\documentclass[twocolumn,showpacs,preprintnumbers,amsmath,amssymb,superscriptaddress]{revtex4-1}
\usepackage{graphicx}% Include Fig.~files
\usepackage{dcolumn}% Align table columns on decimal point
\usepackage{bm}% bold math
\usepackage{amsmath}
\usepackage{xparse}
\usepackage{subfigure}
\usepackage{hyperref}
\usepackage[normalem]{ulem}
\hypersetup{
    colorlinks,
    citecolor=blue,
    filecolor=black,
    linkcolor=blue,
    urlcolor=blue
}
\usepackage{epstopdf}
 \usepackage{relsize}
\usepackage{physics}
\setcitestyle{square}

\begin{document}
\title{A Landauer-B\"uttiker approach for hyperfine mediated electronic transport in the integer quantum Hall regime.}% Force line breaks with \\
%\thanks{A footnote to the article title}%}%
%\author{Aniket Singha}
%\affiliation{Department of Electrical Engineering,\\
%Indian Institute of Technology Bombay, Powai, Mumbai-400076, India\\}
\author{Aniket Singha}
\affiliation{%
Department of Electrical Engineering,\\
Indian Institute of Technology Bombay, Powai, Mumbai-400076, India}
%Visiting Scientist, AIMR Program, Tohoku University, 2-1-1, Aoba-ku, Sendai-980-8577, Japan \\}

 \author{M. H. Fauzi}
 \affiliation{Department of Physics, Graduate School of Science, Tohoku University,
 6-3 Aramaki aza Aoba, Aoba-ku, Sendai, Miyagi-980-8578, Japan
}

\author{Y. Hirayama}
 \affiliation{Department of Physics, Graduate School of Science, Tohoku University,
 6-3 Aramaki aza Aoba, Aoba-ku, Sendai, Miyagi-980-8578, Japan
}

\author{Bhaskaran Muralidharan}
 \email{bm@ee.iitb.ac.in}
\affiliation{%
Department of Electrical Engineering,\\
Indian Institute of Technology Bombay, Powai, Mumbai-400076, India}
\date{\today}% It is always \today, today,
             
\begin{abstract}
The interplay of  spin-polarized electronic edge states with the dynamics of the host nuclei in quantum Hall systems presents rich and non-trivial transport physics. Here, we develop a Landauer-B\"uttiker approach to understand various experimental features observed in the integer quantum Hall set ups featuring quantum point contacts. The approach developed here entails a phenomenological description of spin resolved inter-edge scattering induced via hyperfine assisted electron-nuclear spin flip-flop processes. A self-consistent simulation framework between the nuclear spin dynamics and edge state electronic transport is presented in order to gain crucial insights into the dynamic nuclear polarization effects on electronic transport and in turn the electron-spin polarization effects on the nuclear spin dynamics. In particular, we show that the hysteresis noted experimentally in the conductance-voltage trace as well as in the resistively detected NMR lineshape results from a lack of quasi-equilibrium between electronic transport and nuclear polarization evolution.  In addition, we present circuit models to emulate such hyperfine mediated transport effects to further facilitate a clear understanding of the electronic transport processes occurring around the quantum point contact. Finally, we extend our model to account for the effects of quadrupolar splitting of nuclear levels and also depict the electronic transport signatures that arise from single and multi-photon processes.
\end{abstract}
\maketitle
\section{Introduction}
Nuclear spintronics concerns the manipulation of nuclear spins by means of hyperfine interaction between the host nuclei and the itinerant electrons and their read out using electronic transport \cite{Hirayama_Rev} or optical \cite{Imamoglu} measurements. Quantum Hall geometries in both the integer \cite{Wald_PRL,APL_Edge_Scatt_ORIG,APL_Edge_Scatt,Wieck,PRB_Folk,PRB_West_2009,PRL_SONG_BD,APL_BD_2007,Nano_Hamilton,Hirayama_Rev} and the fractional regime \cite{PRL_Hirayama_FQHE_2002,PRL_WEST_FQHE,APL_Hirayama_FQHE_OPT,PRB_Hirayama_FQHE_OPT,PRB_Hirayama_Diffusion} featuring gated quantum point contacts (QPC) offer a viable method for controlling the spin polarization of the electronic edge channels. This in turn facilitates the manipulation of the nuclear spins via a hyperfine mediated interplay between the spin-polarized edge states and the dynamics of the host nuclei. Such an interplay has revealed rich and non-trivial transport physics in the form of hysteresis in the observed  conductance-voltage traces and non-trivial lineshapes in the resistively detected NMR (RDNMR) traces   \cite{Wald_PRL,APL_Edge_Scatt_ORIG,APL_Edge_Scatt,APL_BD_2007,kawamura_quadru}.  Despite several advancements in the transport experiments involving such set ups, theoretical models for hyperfine interaction mediated edge transport through the QPC in the Hall geometry are clearly missing in the current literature. The object of this work is hence to develop transport models that couple the dynamics of the host nuclei with  edge channel electronic transport as an attempt to fill this gap and theoretically interpret various experiments with specific focus on the conductance-voltage traces \cite{Wald_PRL} and the RDNMR lineshapes \cite{rdnmr1,rdnmr2,Nano_Hamilton,rdnmr4,rdnmr_dis1,rdnmr_dis2,rdnmr_dis3,rdnmr_dis4,rdnmr_dis5} .\\
%\indent  Hyperfine mediated electronic spin flips in the quantum Hall channel result in a change in the channel resistivity simultaneously inducing a dynamic nuclear polarization (DNP) and has been observed consistently in the quantum Hall regime \cite{Wald_PRL,APL_Edge_Scatt_ORIG,APL_Edge_Scatt,PRB_Folk,PRB_West_2009} as well as in the breakdown regime \cite{PRL_SONG_BD,APL_BD_2007,Nano_Hamilton,Hirayama_Rev,PRB_Spatial_BD,NJP_BD_Klitzing}. In the quantum Hall regime, a change in both Hall resistivity \cite{APL_Edge_Scatt_ORIG,APL_Edge_Scatt,kawamura_quadru} as well as longitudinal resistivity \cite{Wald_PRL} is commonly noted in the experiments involving the QPC that selectively filters out the up-spin (down-spin) electrons resulting in an effective spin polarization. \\
\indent We develop our transport models based on a modified Landauer-B\"uttiker formalism that includes a {\it{spin-flip transmission coefficient}},  which is nuclear polarization dependent and describes the rate of electron-nuclear spin flip-flops per unit energy around the QPC region. Using this approach, we show that the hysteresis noted in both the conductance and the RDNMR traces \cite{Wald_PRL,rdnmr1,rdnmr2,Nano_Hamilton,rdnmr4,rdnmr_dis1,rdnmr_dis2,rdnmr_dis3,rdnmr_dis4,rdnmr_dis5} results from a lack of steady state between electronic transport and nuclear polarization evolution and can be explained by taking into account the finite rate of electron-nuclear spin flip-flops in a source limited channel in addition to a finite nuclear spin-lattice relaxation time. The self-consistent simulation framework between the nuclear spin dynamics and the edge state electronic transport developed here offers crucial insights into the dynamic nuclear polarization effects on electronic transport and in turn the electron-spin polarization effects on the nuclear spin dynamics. In addition, we present circuit models to emulate such hyperfine mediated transport effects for a clear understanding of the phenomena occurring near the QPC. Finally, we also address the effects of quadrupolar splitting of the nuclear levels and  depict the electronic transport signatures that arise from single and multi-photon absorption processes \cite{kawamura_quadru}. \\
 \indent This paper is organized as follows. In Sec. \ref{experimentaldetails}, we briefly detail the experimental set up and features that form our current focus after which we spell out the generic formalism. Specifically, in Sec. \ref{SQPC}, a phenomenological model for hyperfine mediated transport through the QPC is developed in detail. Section \ref{simulation_results} elucidates the results from the simulation framework developed with the specific focus on explaining the various experimental trends noted. Specifically, Sec. \ref{subsec:GV} is devoted to the understanding of the hysteritic conductance voltage traces noted for different filling factors and Sec. \ref{subsec:RDNMR} deals with the RDNMR lineshape features in great detail.
\section{Experimental details and theoretical description} \label{experimentaldetails}
\begin{figure*}[]
\begin{minipage}[b]{0.5\textwidth}

\subfigure[]{\includegraphics[scale=.065]{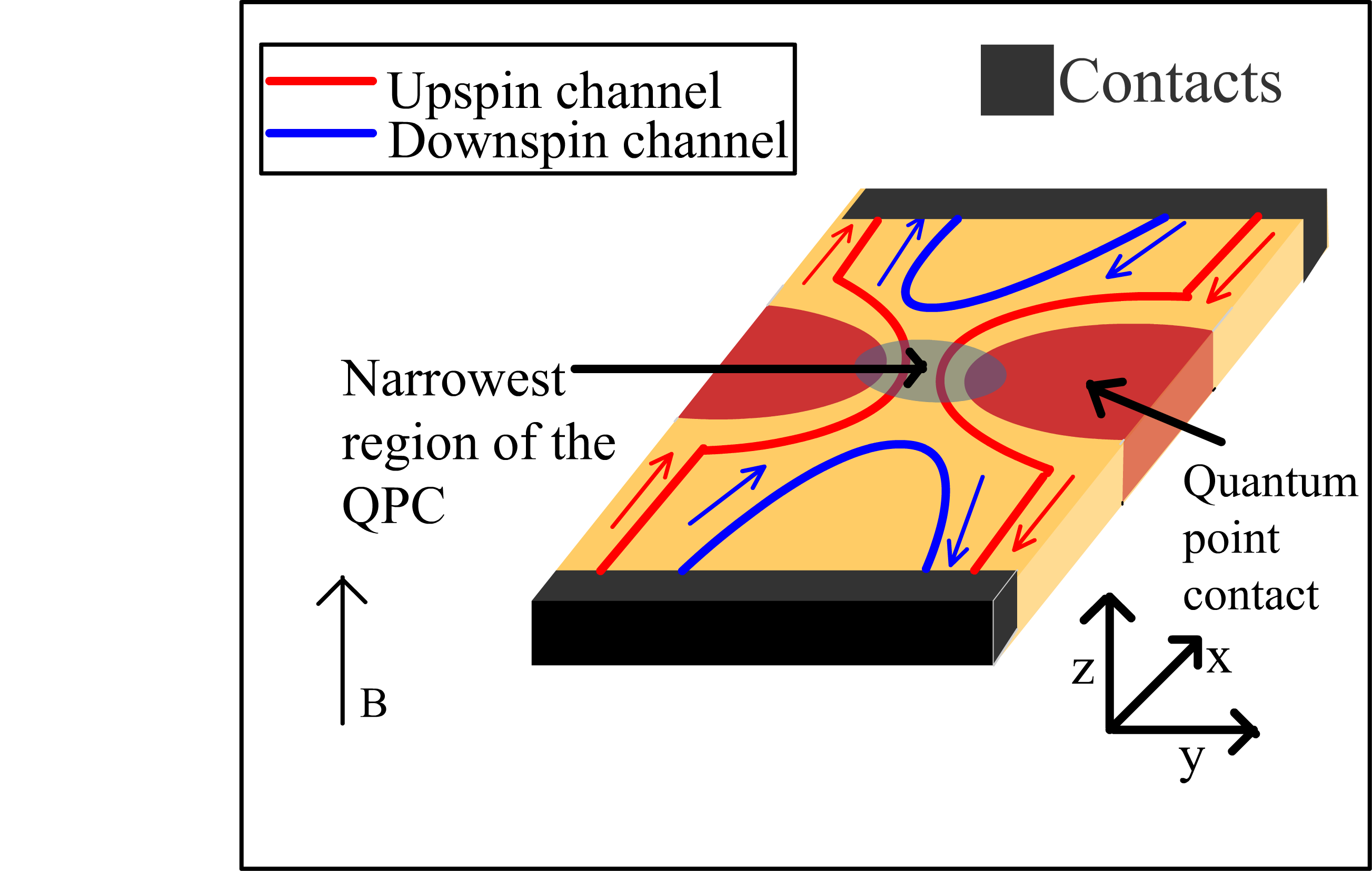}}
 \renewcommand{\thesubfigure}{(c)}% New fixed/manual numbering

\subfigure[]{\includegraphics[scale=.4]{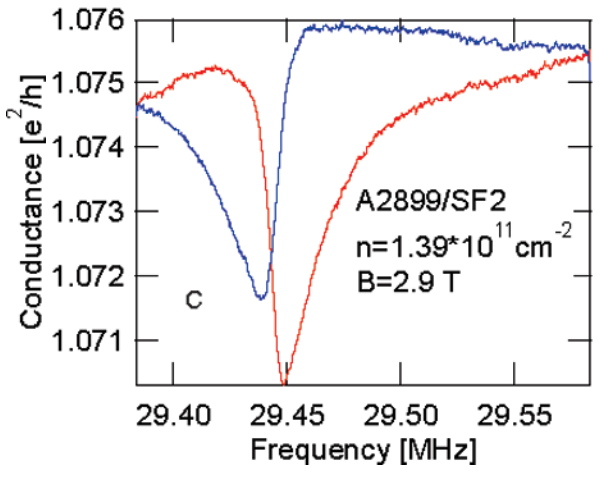}
}
\end{minipage}%
\begin{minipage}[b]{0.5\textwidth}

\renewcommand{\thesubfigure}{(b)}
\subfigure[]{\includegraphics[scale=.4]{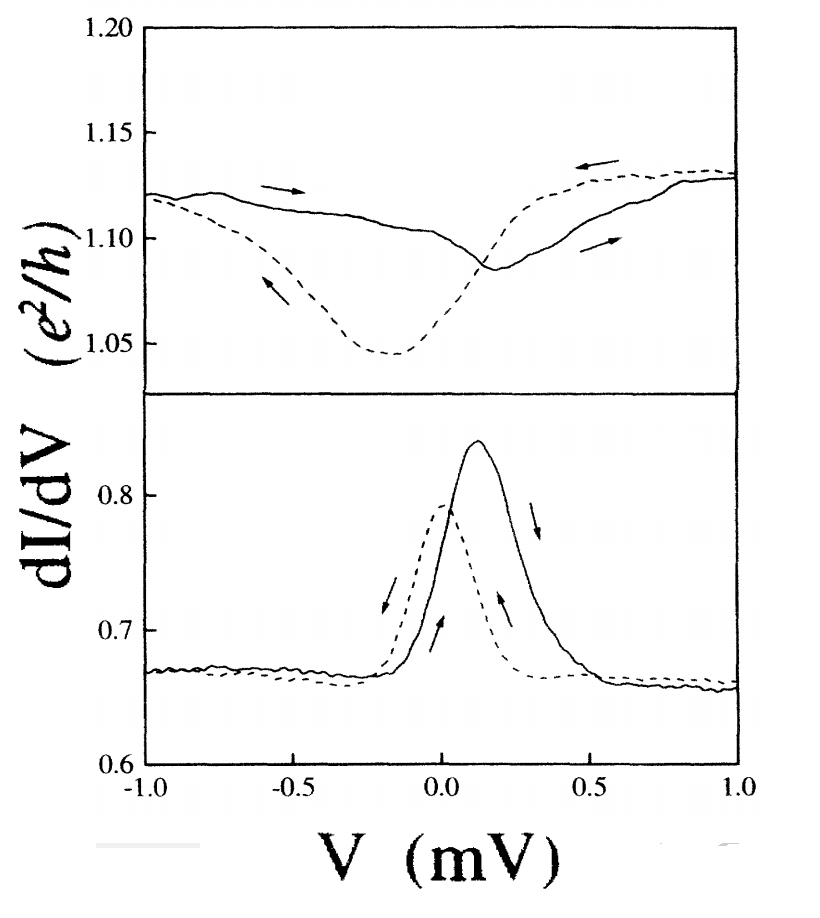}
}
\end{minipage}
\caption{Experimental details.  (a) Schematic of the spin channels propagating through the quantum Hall device with a single QPC for the case $G=\frac{e^2}{h}$.  (b) Experimental traces  (reproduced with permission from Ref. \cite{Wald_PRL} ) of the differential conductance $dI/dV$ versus voltage $V$ for $dI/dV > e^2/h$ (top), $dI/dV < e^2/h$ (bottom) measured on a single QPC  at $T  =50 mK $ and $B=5. 4 T$. Solid curve: forward sweep, dashed curve: backward sweep. Round-trip sweep time for each trace is 200 seconds. (c) Experimental traces  (reproduced with permission from Ref. \cite{Nano_Hamilton} ) of the Hall conductance change and hysteresis in the Hall conductance in a GaAs sample with RF frequency sweep measured at $B=2.9T$ and $T=30mK$.  }
\label{trace}
\end{figure*}
%\begin{figure*}[!htb]
%\subfigure[]{\includegraphics[scale=.22]{single_vs_dbl_qpc/sgl_qpc}
%}
%\hspace{-2cm}\subfigure[]{\includegraphics[scale=.22]{single_vs_dbl_qpc/dbl_qpc}}
%\caption{Schematic of edge state paths in the single and double QPC. (a) For $dI/dV<e^2/h$, any scattering (forward scattering) that increases the net input current to the right of the QPC also increases the net output current and hence the differential conductance $dI/dV$. A decrease in the voltage results in a decrease in the nuclear polarization $F_I$, and hence an increase spin flip-flop scattering probability. For $dI/dV>e^2/h$, electron nuclear spin flip flop scattering generally happens from the forward propagating up channel to the backward propagating down channel. A decrease in the nuclear polarization $F_I$ increases the spin-flip back-scattering and hence decreases the total current. (b) For a double quantum contact, the hyperfine current exits from a different terminal. Hence, an increase in the spin flip-flop scattering actually decreases the net current at the drain terminal.
% }
%\label{fig:qpc}
%\end{figure*}
In the schematic of the experimental set up shown in Fig.~\ref{trace}(a), an appropriately gated single QPC is utilized to selectively filter out a single spin channel into the region beyond the QPC thereby creating an imbalance between the up-spin channel and the down-spin channel.   The principal experimental signature here is the change in conductance with voltage sweep near $V=0$ \cite{Wald_PRL} as shown in Fig.~\ref{trace} (b) as well as  the change in Hall resistance with RF frequency sweep (also known as resistively detected NMR or RDNMR) \cite{Nano_Hamilton} as shown in Fig.~\ref{trace} (c). Along with the change in the conductance, another feature which has attracted significant attention is the hysteresis in the conductance plots during forward and reverse   voltage or RF frequency  sweep as shown in Fig.~\ref{trace} (b) and (c) respectively. A compact theoretical model to elaborate the physics of such a conductance modulation as well as hysteresis occuring in the gated QPC set up forms the primary focus of this work.   \\ %Since both the spin channels terminate at the drain contact,  the spin flip flop scattering from the contact to the device channel in addition to change in Overhauser field due to hyperfine spin exchange  change the conductance of the device. 
\indent  An accurate modeling of such phenomena involves taking into account the details of wavefunction correlations via the density matrix approach \cite{densitymatrix}. Mathematical modeling of such  hyperfine mediated electron transport process self-consistently with evolution of the nuclear polarization  from the density matrix formalism \cite{densitymatrix} is complicated and computationally heavy.  In this paper, we thus adapt a computationally efficient phenomenological model to account for  such hyperfine mediated electronic transport through the QPC.\\
\indent We now provide a theoretical description of the nuclear spin dynamics coupled to the electronic transport following which we focus on how to apply this to our specific set up.  We begin with the description of the nuclear spin dynamics by formulating a master equation in the nuclear spin space followed by the description of the extended Landauer-B\"uttiker formalism for the edge state electronic transport. 
\subsection{Description of scattering processes}
In order to describe the electron-nuclear hyperfine interaction, we start with the Fermi contact hyperfine interaction Hamiltonian for the case with non-varying electronic density of states in space,  given by \cite{Imamoglu}

\begin{widetext}
\begin{equation}
\hat{H}_{HF}(\textbf{r}_n)=\sum_n A_{eff} \psi^{*}(\textbf{r}_n) \psi(\bf{r}_n)  a_0^3 \left [ \hat{S}_z \otimes \hat{I}_z^{n} + \left \{ \frac{\hat{I}_{+}^{n} \otimes \hat{S}_{-}+\hat{S}_{+} \otimes \hat{I}_{-}^{n}}{2} \right \}\right ],
\label{eq:hyperfine_ham1}
\end{equation}
\end{widetext}
where $\psi(\bf{r}_n) $ represents the electron wavefunction at the point $\bf{r}_n$, with $\psi^*(\bf{r}_n)\psi(\bf{r}_n)$ representing the effective electron density  per unit volume at the point $\bf{r}_n$,  $A_{eff}$  is the effective hyperfine coupling constant, $\hat{S}_z,\hat{I}^{n}_z$ are the operators representing the $z$ component of the electronic  spin  and the nuclear spin  respectively, with $a_0^3$ representing a unit cell volume. The operator $`\otimes'$ represents the tensor product between the electron spin  and the nuclear spin spaces. The operators, 
$\hat{S}_{+(-)}$ and $\hat{I}_{+(-)}^{n}$ are respectively the corresponding spin raising (lowering) operators for the electron and the nuclear spins respectively. The above equation assumes  $\psi_{\uparrow}(\textbf{r}_n) = \psi_{\downarrow}(\textbf{r}_n)=\psi(\textbf{r}_n)$, where `$\uparrow$' and  `$\downarrow$' represents the eigen states in the electron spin-space.  In the quantum Hall regime, however, the eigen states are localized in space along the transverse direction. In this case, $\psi_{\uparrow}(\textbf{r}_n) \neq \psi_{\downarrow}(\textbf{r}_n)$ and thus the Hamiltonian should be recast in the form \cite{sakurai_modern,sakurai_advanced,gri}:

\begin{widetext}
\begin{gather}
\hat{H}_{HF}(\textbf{r}_n)=A_{eff}  a_0^3\bf \sum_{\ket{\phi,\beta}\bra{\varphi,\alpha}}\bra{\phi,\beta}\left [ \hat{S}_z \otimes \hat{I}_z^{n} + \left \{ \frac{\hat{I}_{+}^{n} \otimes \hat{S}_{-}+\hat{S}_{+} \otimes \hat{I}_{-}^{n}}{2} \right \}\right ]\ket{\varphi,\alpha} \times \sum_n \Big\{\bra{\psi^{}_{\beta}}\ket{\textbf{r}_n}\ket{\phi,\beta}\bra{\varphi,\alpha} \bra{\textbf{r}_n}\ket{\psi_{\alpha}}\Big\}, \nonumber \\
\label{eq:hyperfine_ham}     
\end{gather}
\end{widetext}
\normalsize
where   $\ket{\phi, \beta}=\ket{\phi} \otimes \ket{\beta}$ and   $\ket{\varphi, \alpha}=\ket{\varphi} \otimes \ket{\alpha}$ with  $\ket{\alpha}$ and $\ket{\beta}$ belonging to   the electron spin-space  and $\ket{\phi}$ and $\ket{\varphi}$ belonging to  the nuclear-spin space in the presence of a magnetic field pointing along the $z-$ direction. 
\indent  When the coupling constant $A_{eff}$ is small,  the effect of the two terms in \eqref{eq:hyperfine_ham1} can be separated. The first term results in an effective magnetic field and the second term within the curly brackets represents the electron-nuclear spin flip-flop processes. The first term in \eqref{eq:hyperfine_ham1}, which also corresponds to  $\ket{\alpha} =\ket{\beta}$ and $\ket{\phi} =\ket{\varphi}$ in \eqref{eq:hyperfine_ham}  introduces an additional shift in the electronic energy levels as well as between states of different nuclear spins.  The electronic energy difference $\Delta$ between the up-spin channel electrons and the down-spin channel electrons  is given by \cite{NMR_book}
\begin{equation}
\Delta=g_{el}\mu_B B_{app} + A_{eff}<I_z>,
\label{eq:electron_energy}
\end{equation}
where $g_{el}$, $\mu_B$ and $B_{app}$ are the effective Lande $g$-factor of the electron in GaAs, the Bohr magneton and the applied magnetic field respectively.  The above expression is obtained by assuming  an isotropic nuclear spin distribution. Similarly, the energy difference $\epsilon$ between the adjacent nuclear spin states differing by a magnetic quantum number of $\Delta s = \pm 1$ is given by 
\begin{equation}
\epsilon=g_{nuc}\mu_B^{nuc} B_{app} + A''<S_z>, 
\label{eq:nucleus_energy}
\end{equation}
where $g_{nuc}$ is the effective Lande $g$-factor of the nuclide in consideration and $\mu_B^{nuc}$ is the nuclear Bohr magneton. The effective coupling constant $A''=A_{eff} a_0^3 n_{el}$ \cite{DMS_DS,NMR_book,Imamoglu}, where $n_{el}$ is the electronic carrier density. The magnetic quantum number for the GaAs nuclei varies from $-3/2$ to $+3/2$ in steps of $\Delta s= +1$. In standard literature, the second terms in \eqref{eq:electron_energy} and \eqref{eq:nucleus_energy} represent the Overhauser shift and the Knight shift respectively. However, for practical purposes, $g_{nuc}<<g_{el}$ and $A''<<A_{eff}$ and hence $\epsilon$ may be neglected with respect to $\Delta$ and the nuclear spin flips may be considered elastic. \\
\indent The electron-nuclear spin flip-flop processes are described by the second term in the Hamiltonian in \eqref{eq:hyperfine_ham1}, which also corresponds to $\ket{\alpha} \neq \ket{\beta}$ and $\ket{\phi} \neq \ket{\varphi}$ in \eqref{eq:hyperfine_ham} and the scattering rates are evaluated via the Fermi's golden rule \cite{SB_BM,F_model}, typically related to the densities of the  initial and the final states. In this case, the electronic wavefunction distribution $\psi_{\alpha}(\bf{r}_n)$ and $\psi_{\beta}(\bf{r}_n)$ overlaps with the nuclear  wavefunction on each site $\bf{r_n}$ differently, and this effect is accounted for via the overlap terms  $\psi^{*}_{\beta}(\bf{r}_n) \psi_{\alpha}(\bf{r}_n)$ for up to down or down to up electronic spin transitions.  The procedure for a self-consistent description of electronic transport coupled to hyperfine spin dynamics then entails the time-dependent simulation of the nuclear spin dynamics, with the electronic transport processes  in steady state. This is because the nuclear spin dynamics are typically slow due to slow relaxation rates, slow diffusion rates as well as longer flip-flop times in comparison with the electronic transport velocities. The nuclear spin dynamics at each point $\bf{r}_n$ are dictated via the electron-nuclear hyperfine flip rates calculated from the Fermi's golden rule \cite{SB_BM} given by:
\begin{eqnarray}
\Gamma_{  \uparrow \downarrow}(\textbf{r}_n)  = \frac{2 \pi}{\hbar} \mid A_{eff} \mid^2 \int dE n_{\uparrow}(\textbf{r}_n, E)  p_{\downarrow}(\textbf{r}_n,E)  \nonumber \\
\Gamma_{ \downarrow\uparrow}(\textbf{r}_n) = \frac{2 \pi}{\hbar} \mid A_{eff} \mid^2 \int dE n_{\downarrow}(\textbf{r}_n, E) p_{\uparrow}(\textbf{r}_n, E), \nonumber\\ 
\label{eq:scattering_rates}
\end{eqnarray}
where  $\Gamma_{ \downarrow \uparrow}(\textbf{r}_n)$ ($\Gamma_{\uparrow \downarrow}(\textbf{r}_n)$) represents the up to down (down to up) nuclear spin transition rate at the nuclear co-ordinate $\bf{r}_n$  between magnetic quantum numbers that differ by $+1$ ($-1$) in the nuclear spin space due to flip-flop transitions of electrons at energy $E$.  The quantities $n_{\uparrow}(\textbf{r}_n,E)$ and $n_{\downarrow}(\textbf{r}_n,E)$ ($p_{\uparrow}(\textbf{r}_n, E)$ and $p_{\downarrow}(\textbf{r}_n,E)$) denote the densities of filled (vacant) states per unit energy per unit area at the point $\textbf{r}_n$. \color{black}  The up to down (down to up) electronic spin transition rate based on \eqref{eq:scattering_rates} depends not only on the the availability of electrons in the up (down) spin density of states $D_{\uparrow(\downarrow)}(\textbf{r}_n,E)$ and the vacancy in the down (up) spin density of states $D_{\downarrow(\uparrow)}(\textbf{r}_n,E)$, but also on the spatial overlap of the corresponding density of states. The total rate of electron-nuclear spin flip-flop now depends on the integral of  $\Gamma_{ \downarrow \uparrow}(\textbf{r}_n)$ ($\Gamma_{\uparrow \downarrow}(\textbf{r}_n)$) over the spatial co-ordinates.
\begin{eqnarray}
\Gamma_{\uparrow \downarrow }  = \frac{2 \pi}{\hbar} \mid A_{eff} \mid^2 \int \int d^3 \textbf{r}_n dE n_{\uparrow}(\textbf{r}_n, E)  p_{\downarrow}(\textbf{r}_n,E)  \nonumber \\
\Gamma_{ \downarrow\uparrow} = \frac{2 \pi}{\hbar} \mid A_{eff} \mid^2 \int \int d^3 \textbf{r}_n dE n_{\downarrow}(\textbf{r}_n, E) p_{\uparrow}(\textbf{r}_n, E) \nonumber\\ 
\label{eq:scattering_rates2}
\end{eqnarray}
The spatial dynamics of the nuclear spins can be described by the following master equation:
\begin{eqnarray}
\frac{d\left [F(\textbf{r}_n) \right ] }{dt} = \left [ \Gamma (\textbf{r}_n) \right ] \left [F(\textbf{r}_n) \right ] - \frac{[F(\textbf{r}_n) -F^0  ]}{\tau_I} \nonumber \\+ D_n {\nabla}^2 [F(\textbf{r}_n) ],\nonumber \\
\label{eq:master2}
\end{eqnarray}
where $\left[{F(\bf{r}_n)}\right]$ is the  probability column vector representing the probability of occupancy of the nuclear spin levels, and $\tau_I$ is a phenomenological nuclear spin relaxation time, which is typically a very slow process.  The matrix $\Gamma(\textbf{r}_n)$ takes into account the transition between the individual nuclear spin levels. The vector $\left [ F^0 \right ]$ denotes the probability of occupation of the nuclear spin levels  in equilibrium. The above equation also includes nuclear spin diffusion described by the last term, where $D_n$ is the phenomenological diffusion constant. In this paper, we neglect the exact spatial distribution of nuclear spins due to diffusion and  approximate the effects of nuclear spin diffusion by incorporating a larger  number of nuclei. The equation governing the dynamics of the nuclear spins is then given by:
\begin{equation}
\frac{d\left [F \right ] }{dt} = \left [ \Gamma \right ] \left [F\right ] - \frac{\left [F-F^0 \right ]}{\tau_I}.
\label{eq:master}
\end{equation}
The transition probability matrix $[\Gamma]$ may be specifically cast for the spin-$3/2$ case in the current study in terms of the spin-flip rates defined in \eqref{eq:scattering_rates2} as:
\begin{eqnarray}
\left [ \Gamma \right ] =\begin{bmatrix}
-\Gamma_{ \downarrow \uparrow} & \Gamma_{\uparrow \downarrow}  & 0 & 0 \\
\Gamma_{ \downarrow \uparrow} & -\left (\Gamma_{ \downarrow \uparrow}+ \Gamma_{\uparrow \downarrow} \right ) & \Gamma_{\uparrow \downarrow}&  0 \\
0 & \Gamma_{ \downarrow \uparrow} & -\left (\Gamma_{ \downarrow \uparrow}+ \Gamma_{\uparrow \downarrow} \right ) & \Gamma_{\uparrow \downarrow}   \\
0 & 0 & \Gamma_{ \downarrow \uparrow} & -\Gamma_{\uparrow \downarrow}
\end{bmatrix}, \nonumber \\ 
\label{eq:matrix}
\end{eqnarray}
where $\Gamma_{\uparrow\downarrow}$ and $\Gamma_{\downarrow\uparrow}$ are defined in \eqref{eq:scattering_rates2}. An additional constraint used to solve \eqref{eq:master} using \eqref{eq:matrix} is that of the normalization of the nuclear state probabilities, i.e.,  $\sum_{s} F_s =1$, where $F_s$ is the occupation probability of the nuclear density of states with spin $s$. \\ 
The temporal evolution of the average electronic polarization $<S_z>$ and the average nuclear polarization $<I_z>$  at the QPC are calculated self-consistently by solving  \eqref{eq:master} and \eqref{eq:electron_energy} via the relations:
\begin{gather}
<S_z>  = \frac{1}{2} \frac{n_{ \uparrow} - n_{ \downarrow}}{n_{ \uparrow} + n_{ \downarrow}} \nonumber \\
F_I  = <I_z>  = \sum_{s} s F_s=\left[s\right]\left[F\right],
\label{eq:pol_nuc}
\end{gather}
where $ F_s $ are obtained by solving the master equations. The matrix $[s]=\left[ \frac{3}{2}~~\frac{1}{2}~~-\frac{1}{2}~~-\frac{3}{2} \right]$ comprises the row vector of the spin magnetic quantum numbers of the GaAs nuclei. The procedure for transport calculations follows solving \eqref{eq:electron_energy}, \eqref{eq:master},  and  \eqref{eq:pol_nuc}  sequentially in a self-consistent loop with the electronic transport to be described now. \\
\begin{figure}
\subfigure[]{\includegraphics[scale=.08]{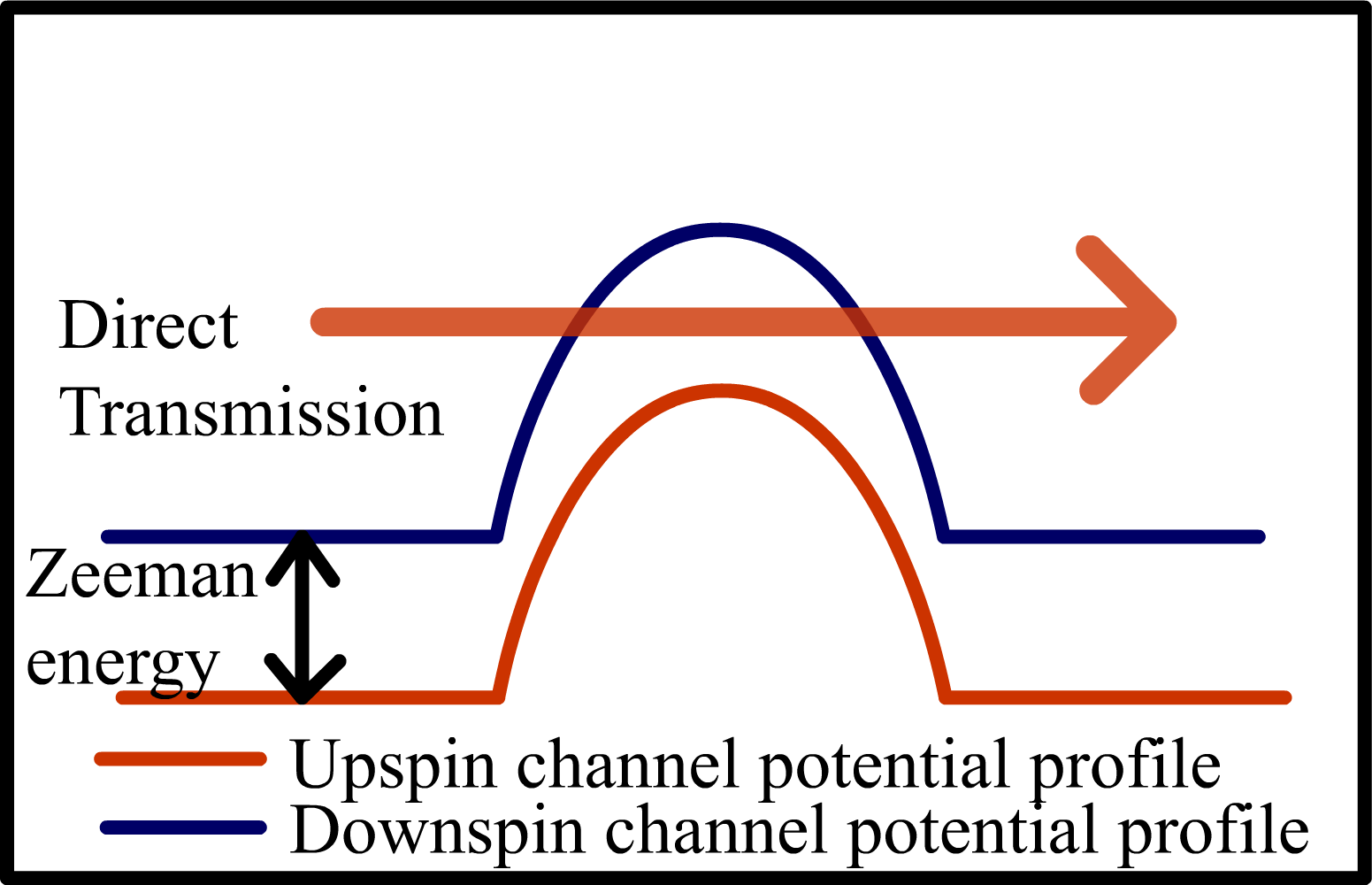}
}\subfigure[]{\includegraphics[scale=.08]{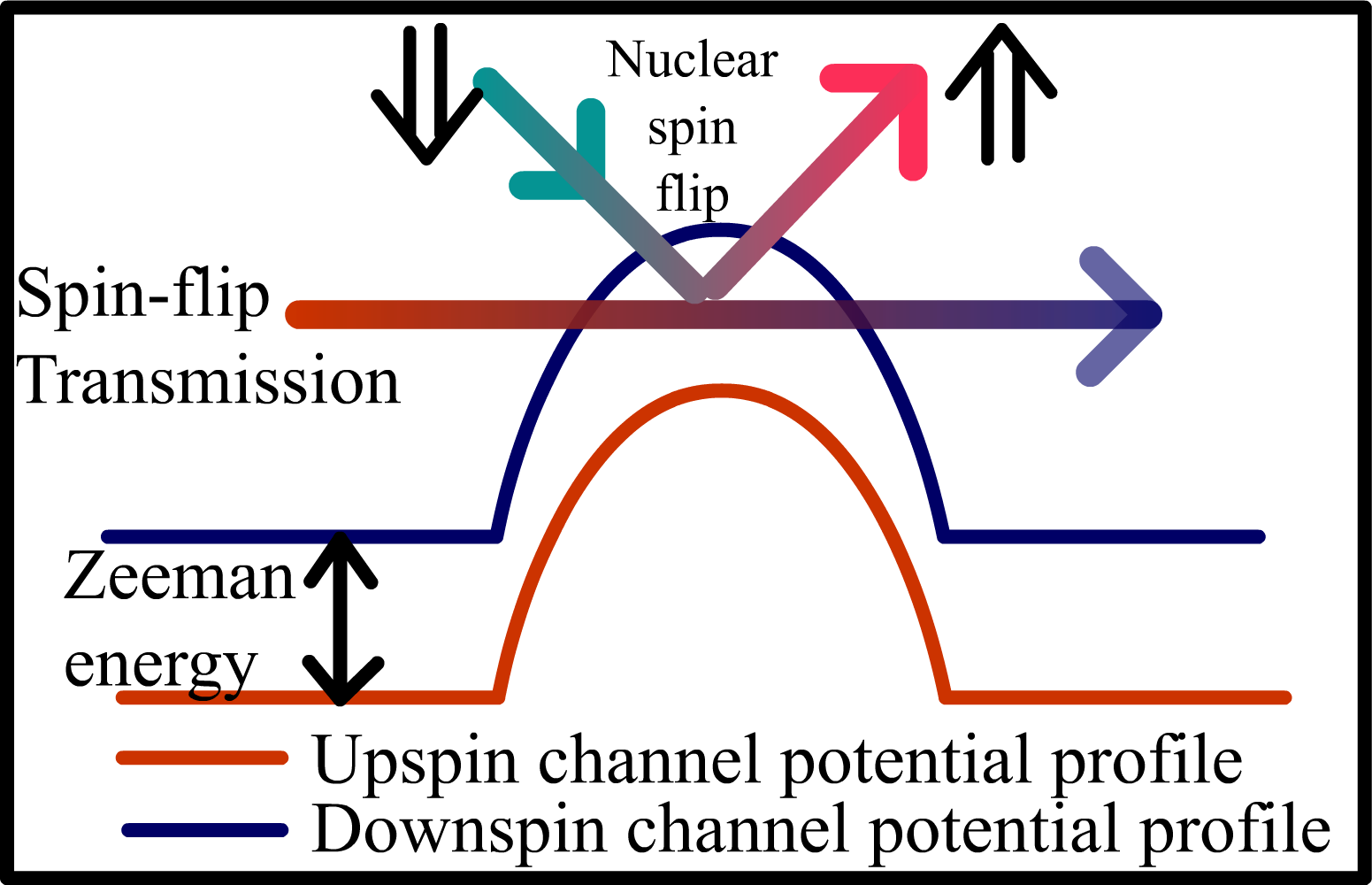}
}
\subfigure[]{\includegraphics[scale=.06]{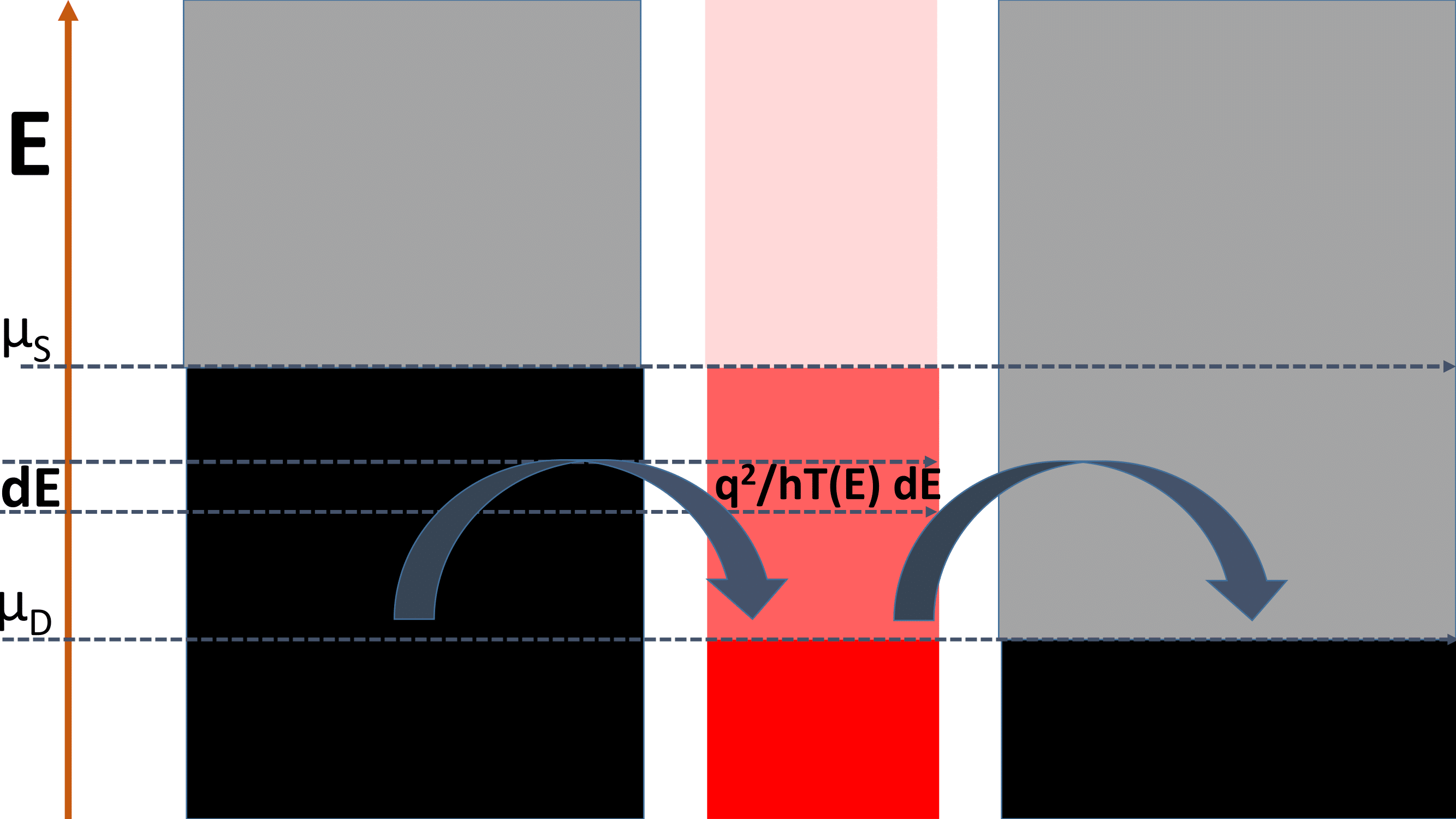}}
\caption{ Schematic of electrons tunneling through a QPC. (a) Direct transmission without spin-flip processes. The up-spin electrons are more likely to be transmitted than the down-spin electrons as a result of Zeeman splitting. (b) Electrons can suffer a spin-flip process around the QPC and transmit from the up-spin channel originating in the source contact to the down-spin channel terminating in the drain contact. (c) Schematic of the model used to simulate a single QPC structure.}
\label{fig:single_qpc}
\end{figure}

\begin{figure*}[!htb]
\begin{minipage}[b]{.5\textwidth}
\subfigure[]{\includegraphics[height=3in]{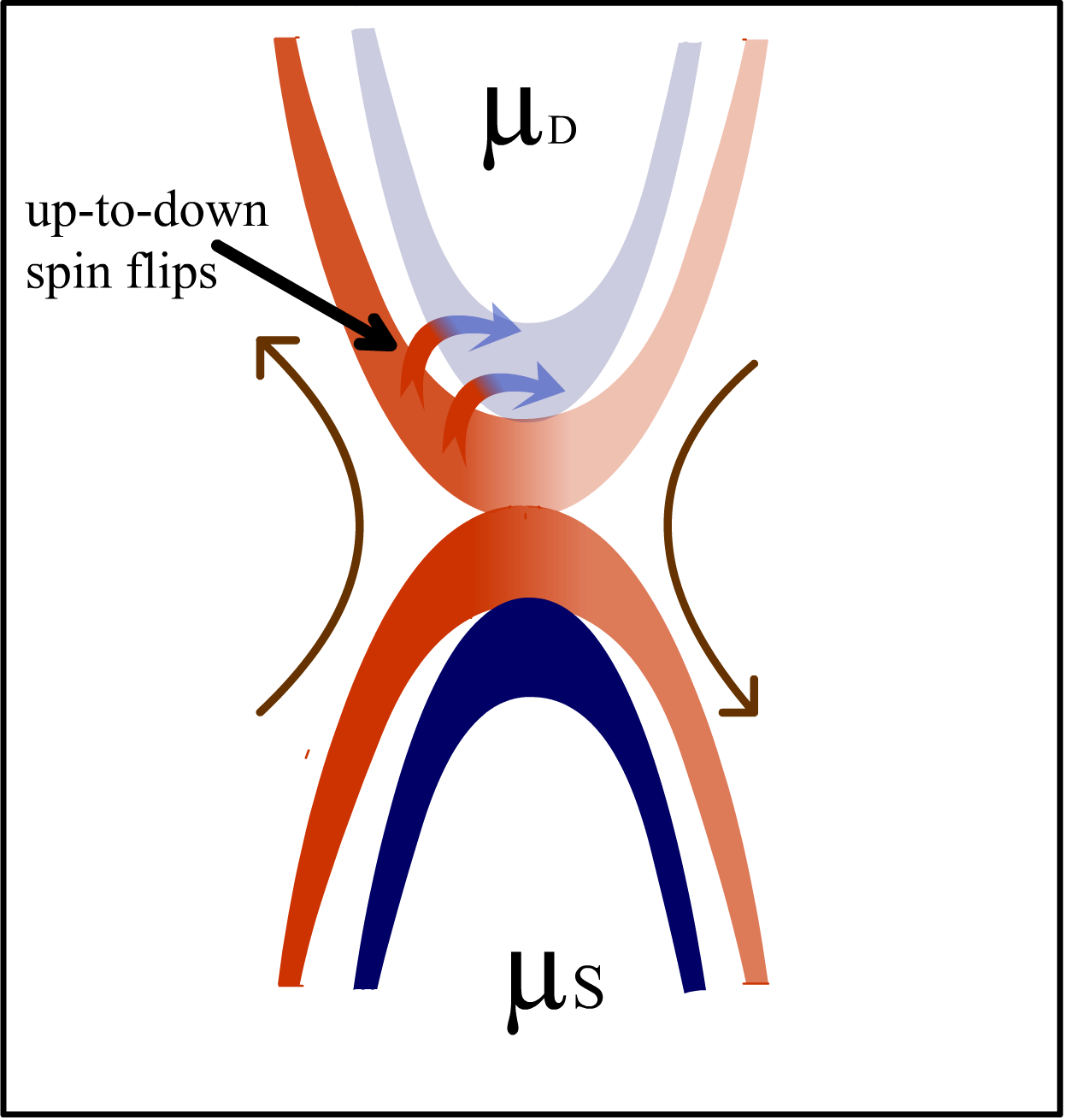}
}
%\subfigure[]{\includegraphics[height=2.5in]{paperfig/ckt_G_less_1}

\end{minipage}%
\begin{minipage}[b]{0.5\textwidth}
\subfigure[]{\includegraphics[height=3in]{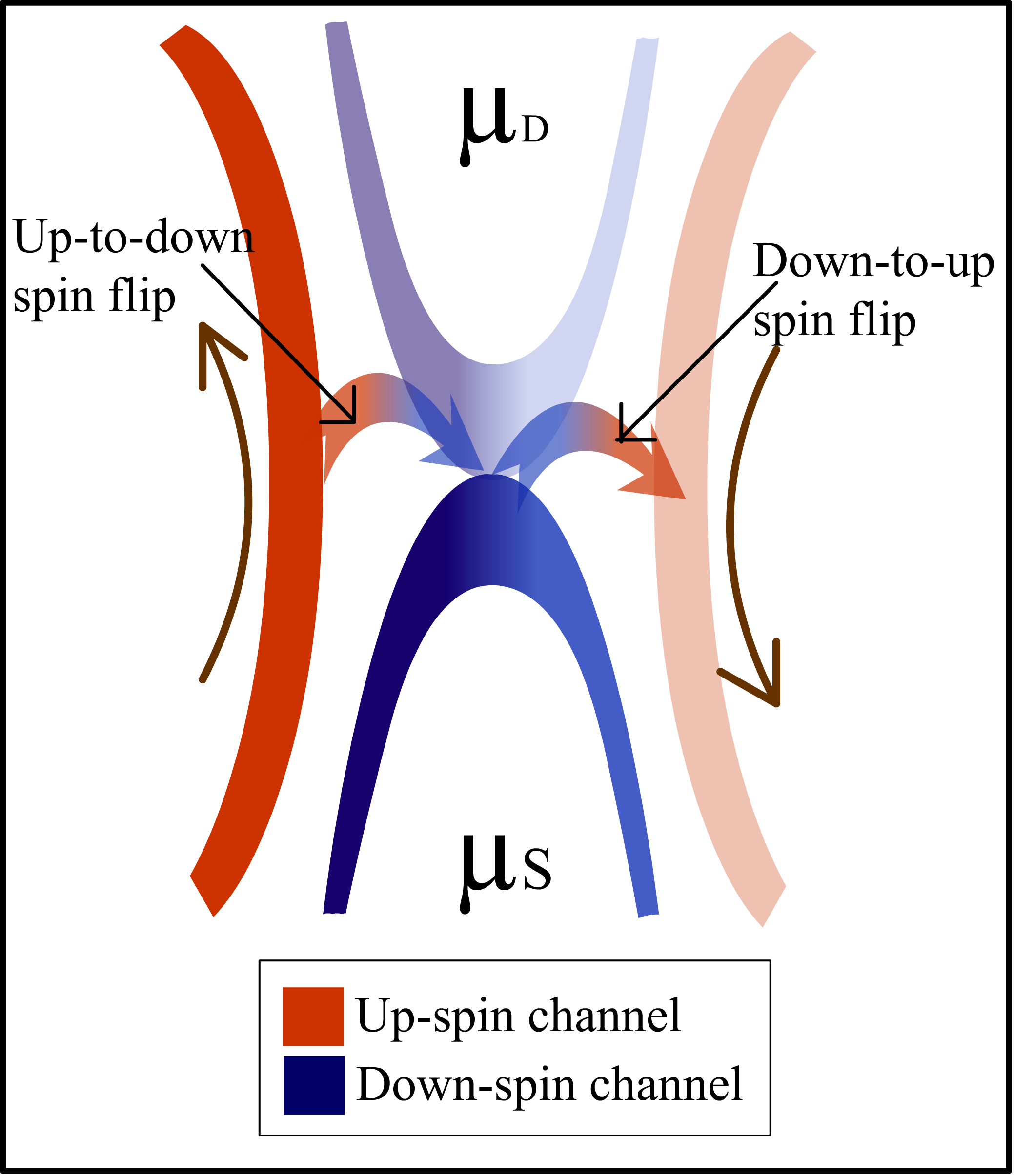}}
%\subfigure[]{\includegraphics[height=2.5in]{paperfig/ckt_G_great_1}

\end{minipage}
\caption{Scattering schematics. (a) Associated scattering phenomena for $dI/dV<\frac{e^2}{h}~(\nu_{QPC}<1)$. Electron-nuclear spin flip-flop scattering at the QPC occurs from a forward propagating up-spin channel terminating in the drain contact to a forward propagating down-spin channel terminating in the drain contact.     (b) Associated scattering phenomena for $dI/dV>\frac{e^2}{h}~\nu_{QPC}>1$. Electron-nuclear flip-flop scattering at the QPC occurs from forward propagating up-spin channel terminating  in the drain contact  to the forward propagating down-spin channel terminating in the drain contact and also from forward propagating down-spin  channel terminating in the drain contact to the backward propagating up-spin channel originating at the drain contact. The later spin flip  scattering process decrease the conductance $dI/dV$ because the scattering occurs to a backward propagating edge channel. A darker colour in (a) and (b) indicates filled channel while a lighter colour indicates an empty channel. The up-spin and down-spin channels in figure (a) and (b) are indicated by brown and blue colour respectively.}
\label{fig:sgl_qpc1}
\end{figure*}
\subsection{Electronic edge-state transport in the QPC region} \label{SQPC}
\begin{figure}[!h]
\includegraphics[scale=.25]{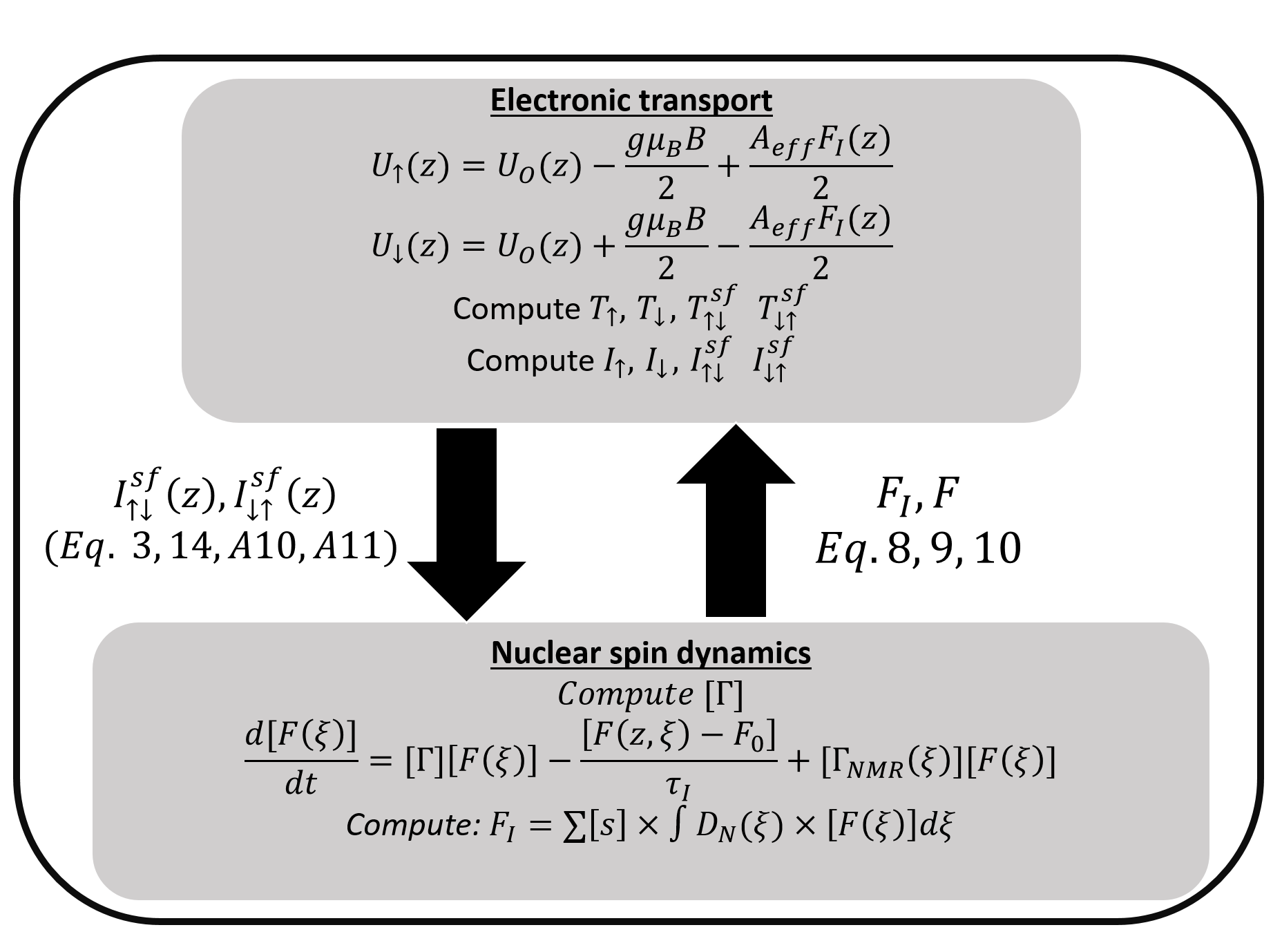}
\caption{Schematic diagram showing the self-consistency involved in solving the electron transport dynamics and the nuclear spin dynamics in time domain.}
\label{fig:loop}
\end{figure}

While the dynamics of the nuclear spins simply follow the master equation \eqref{eq:master} described above, a description of electronic transport involves transport currents due to the source and drain reservoirs held at electrochemical potentials $\mu_S$ and $\mu_D$ respectively. From a Landauer-B\"uttikker perspective, a consistent description of transport currents in our case demands the use of both a) direct transmission and b) spin-flip transmission. The need to include spin flip transmission follows from the interaction between the  edge channels of different spins that gives rise to nuclear polarization  which in close proximity of the QPC region determines the electronic transport. Near the QPC, the forward propagating edge channels and the backward propagating edge channels come in close proximity and hence spin-flip scattering can occur to the forward propagating as well as to the backward propagating edge channels \cite{Wald_PRL}.\\
\indent The Landauer direct transmission $T_{\uparrow (\downarrow)}(E)$ denotes the tunneling probability of the up (down)-spin electrons through the QPC. We model the spin-split edge states in the device by a continuum of density of states as in a ballistic $1-D$ conductor \cite{salahuddin,aono,qpc_model_1,qpc_model_2} with the region of the QPC being represented by a Gaussian potential barrier, as shown in Fig.~\ref{fig:single_qpc} (a) and (b)  along with the model used for simulation of electronic transport  shown in Fig. \ref{fig:single_qpc} (c). The direct transmission coefficients $T_{\uparrow}(E)$ and $T_{\downarrow}(E)$ are then calculated using the non-equilibrium Green's function (NEGF) method applied to the barrier described above using a $1-D$ atomistic tight-binding Hamiltonian \cite{Datta_Green,qtransport}. The pertinent details of the approach used here have been briefly discussed in Appendix \ref{appendix4}.
%Hence, in steady state, the {\it{energy resolved}} currents must satisfy conservation laws, as dictated by the Kirchoff's rules. 
%In the absence of scattering between the two spin channels, with the grounded terminal considered as the drain, the terminal currents originating from the source and drain are given by a simple Landauer type description \cite{Datta_Green} as
%\begin{align}
%I_{S \uparrow}(E) &= q\gamma_{S \uparrow} D_{\uparrow}(E)   f_S(E) \nonumber \\
%I_{S \downarrow}(E) &= q\gamma_{S \downarrow} D_{\downarrow}(E)   f_S(E) \nonumber \\
%I_{D \uparrow}(E) &= -q\gamma_{D \uparrow} D_{\uparrow}(E)   f_D(E)  \nonumber  \\
%I_{D \downarrow}(E) &=-q\gamma_{D \downarrow} D_{\downarrow}(E)   f_D(E)
%\label{eq:Kirchoff2}
%\end{align}
%where the quantities $\gamma_{S(D) \uparrow }$ denote the electronic transfer rates related to the source (drain) up-spin (down-spin) channel. Depending on the geometry of the device, spin flip scattering can occur inside the channel.
In our scheme, we only consider the number of transmitted modes and the filling factor at the QPC for the electronic transport which is related to the geometry of the set up that is ascertained apriori. Near the vicinity of $\nu_{QPC}=1$ at the QPC, the down-spin edge channel at the QPC is almost empty in the energy range between $\mu_S$ and $\mu_D$, resulting in a considerable simplification of the transport equations.\\
\indent  We begin with the case where the filling factor $\nu_{QPC}<1$, i.e., $G<\frac{e^2}{h}$, where only the up-spin edge channel originating from the source contact contributes to the total current terminating in the drain contact. The electrons in the forward propagating up-spin edge channel originating from the source contact can tunnel through the QPC to the up-spin edge channel terminating in the drain contact with a probability $T_{\uparrow}(E)$ while the forward propagating down-spin edge channel originating in the source contact is completely disconnected from the forward propagating down-spin edge channel  terminating in the drain contact, as depicted in Fig.~\ref{fig:single_qpc}(a) and (b) respectively.  \\
\indent A few up-spin electrons at the QPC in the forward propagating edge channel terminating in  the drain contact can however scatter to the forward propagating down-spin edge channel terminating in the drain contact with a spin-flip process  as shown in Fig. \ref{fig:sgl_qpc1} (a). This gives rise to the spin-flip scattering current $I^{sf} _{\uparrow \downarrow (\downarrow \uparrow)}$,  where the superscript `$sf$' denotes the flow of current due to spin-flip scattering at the QPC and the subscript $'\uparrow \downarrow (\downarrow \uparrow)'$ denotes the current flow from the up (down)-spin to down (up)-spin edge channel via electronic spin-flips. Assuming that the direct transmission coefficients ($T_{\uparrow}$ and $T_{\downarrow}$) depend on the nuclear polarization only via the  Overhauser field \eqref{eq:electron_energy}, for a system with four nuclear spin levels, the spin-flip transmission coefficient at the QPC from the forward propagating up-spin channel terminating in the drain contact to the forward propagating down-spin channel terminating in the drain contact is given by (details given in Appendix \ref{appendix1}):
\[
T_{\uparrow\downarrow}^{sff}(E) =T_{\uparrow}(E)T_{\uparrow\downarrow}^f \Big \{1-F_{+\frac{3}{2}} \Big\}.
\] 
Note that $T_{\uparrow\downarrow}^f$ depends on the spatial overlap of the density of states of the up-spin and down-spin edge channel at the QPC between the energy range $\mu_S$ and $\mu_D$  (details given in Appendix \ref{appendix1}, Eq. \ref{eq:J} and \ref{eq:appref1}). We approximate $T_{\uparrow\downarrow}^f$ as a constant. %Since the supply of electrons to the up-spin  channel originating in the source contact is limited to $\frac{1}{h}$ per unit energy range per unit time, a limiting condition must be imposed as:
%\[
%T_{\uparrow }(E)+T_{\uparrow \downarrow}^{sff}(E) \leq 1.
%\]
%
Therefore, the down-spin current recorded just outside the QPC relies entirely on such spin-flip processes    and hence is simply the spin-flip current $I^{sf}_{\uparrow \downarrow}$ while the up-spin current in the edge channel just outside the QPC is reduced by $I^{sf}_{\uparrow \downarrow}$.  Based on the above discussions, the up and down spin channel currents are given by
\[
I = \int dE \left (I_{\uparrow}(E)+I_{\downarrow}(E) \right ) 
\] 
\begin{multline}
 = \frac{q}{h} \int dE  \Big\{[T_{\uparrow}(E)-T^{sff}_{\uparrow\downarrow}(E)]+T^{sff}_{\uparrow\downarrow}(E)\Big\}  \\
 \times\Big \{f_S(E)-f_D(E)\Big \}. 
\label{eq:first}
\end{multline}
The subscripts $'S'$ and $'D'$  the source and drain contacts respectively, with $f_{S(D)}(E)$ denoting the Fermi-Dirac distribution in the source (drain) contact held in quasi-equilibrium at $\mu_{S(D)}$. The parameter $T_{\uparrow \downarrow}^{sff}(E)$ takes into account the spin-flip scattering of electrons at and around the QPC from the forward propagating up-spin edge channel terminating in the drain contact to  the forward propagating down-spin channel terminating in the drain contact, with  the superscript $'sff'$ denoting spin-flip scattering to a forward propagating channel. It must be noted that the edge channels in the quantum Hall arrangement are uni-directional and hence the expressions for the current in  \eqref{eq:first} depend on the factors $f_S(E)$ and $f_D(E)$ only and not on the factors $f_S(E)\{1-f_D(E) \}$ and $f_D(E)\{1-f_S(E) \}$ as expected in a typical Landauer type scattering treatment. \\
\indent Turning our attention to the case when the filling factor $\nu_{QPC}>1$, i.e., $G>\frac{e^2}{h}$,  the down-spin electrons in the edge channel originating in the source contact are partially transmitted  through the QPC to the down-spin edge channel terminating in the drain contact, as depicted in Fig.~\ref{fig:sgl_qpc1}(b). In this case, the spin-flip scattering at and around the QPC can occur from the forward propagating up-spin channel terminating in the drain contact to the forward propagating down-spin channel terminating in the drain contact as well as from the forward propagating down-spin channel terminating in the drain contact to the backward propagating up-spin  channel originating from the drain contact.   Again, assuming that the direct transmission coefficients $T_{\uparrow}$ and $T_{\downarrow}$ depend on the nuclear polarization only via the Overhauser field \eqref{eq:electron_energy}, the spin-flip currents in this case are given by (details in Appendix \ref{appendix1})
\begin{align}
I_{\uparrow \downarrow}^{sf} &\approx \frac{q}{h} \int dE T_{\uparrow \downarrow}^{sff}(E)  \Big ( f_S(E)-f_D(E) \Big ) \nonumber \\
I_{\downarrow\uparrow}^{sf}&\approx \frac{q}{h} \int dE T_{\downarrow \uparrow}^{sfb}(E)  \Big ( f_S(E)-f_D(E) \Big ), \nonumber 
\end{align}
 where superscript $'sfb'$ denote spin-flip scattering to a backward propagating edge channel while the superscript $'sff'$ has the same meaning as described previously. The spin-flip current $I_{\uparrow \downarrow}^{sf}$ flows from the forward propagating up-spin channel terminating in the drain contact to the forward 
propagating down-spin edge channel terminating in the drain contact while the spin-flip current $I_{\downarrow \uparrow}^{sf}$ flows from the forward propagating down-spin edge channel terminating in the drain contact  to the backward propagating up-spin channel originating in the drain contact.  Hence, $I_{\downarrow \uparrow}^{sf}$  causes a change in the total output current since the spin-flip scattering occurs to a backward propagating edge channel. It however does play a role in the nuclei polarization near the QPC.~%
%
%%If $T_{\downarrow \uparrow}^{sf}=0$, $n_{\downarrow}^f=T_{\downarrow}(E)D_{\downarrow}(E)$ and $p_{\uparrow}^{b}=D_{\uparrow}(E)$. For finite value of $T_{\downarrow \uparrow}^{sf}$, $n_{\downarrow}^f$ and $p_{\uparrow}^{b}$ needs to be solved self-consistently. However, for small values of $T_{\downarrow \uparrow}^{sf}$, $n_{\downarrow}^f \approx T_{\downarrow}(E)D_{\downarrow}(E)$ and $p_{\uparrow}^{b} \approx D_{\uparrow}(E)$
The current in the up-spin and down-spin channel terminating in the drain contact just outside the QPC is then given by:
\[
I_{\uparrow} = \int \frac{q}{h}\{T_{\uparrow }(E)- T_{\uparrow \downarrow}^{sff}(E)\}\{f_S(E)-f_D(E)\}dE 
\]
\begin{multline}
= \int \frac{q}{h}\{T_{\uparrow }(E) - T_{\uparrow}(E)T_{\uparrow \downarrow}^f(1-F_{\frac{3}{2}})\} \\
\times \{f_S(E)-f_D(E)\}dE 
\label{eq:second}
\end{multline}
\[
I_{\downarrow} = \int \frac{q}{h}\{T_{\downarrow }(E)+ T_{\uparrow \downarrow}^{sff}(E)- T_{\downarrow\uparrow }^{sfb}(E)\}\{f_S(E)-f_D(E)\}dE 
\]
\begin{multline}
= \int \frac{q}{h}\{T_{\downarrow }(E) + T_{\uparrow}(E)T_{\uparrow \downarrow}^f(1-F_{\frac{3}{2}})- T_{\downarrow}(E)T_{\downarrow\uparrow }^b(1-F_{-\frac{3}{2}})\} \\
\times \{f_S(E)-f_D(E)\}dE 
\label{eq:second}
\end{multline}
From the above discussion, the generalized equations for the up-spin, down-spin and spin-flip currents through the QPC are given by:
\begin{widetext}
\begin{align}
I_{\uparrow} &=\frac{q}{h}\int \Big\{ \underbrace{T_{\uparrow}(E)}_{\frac{Direct}{Transmission}}+~~~\underbrace{T^{sff}_{\downarrow\uparrow}(E)-T^{sff}_{\uparrow\downarrow}(E)}_{\frac{spin~flip}{forward~transmission}}~~~~~-\underbrace{T^{sfb}_{\uparrow\downarrow}(E)}_{\frac{spin~flip}{backward~transmission}} \Big\}    \times\{f_S(E)-f_D(E)\}dE  \nonumber \\
&=\frac{q}{h}\int \Big\{{T_{\uparrow}(E)}+{T_{\downarrow\uparrow}^{f}T_{\downarrow}(E)\{1-F_{-\frac{3}{2}}\}-T_{\uparrow\downarrow}^{f}T_{\uparrow}(E)\{1-F_{\frac{3}{2}}\}-T_{\uparrow\downarrow}^{b}T_{\uparrow}(E)\{1-F_{\frac{3}{2}}\}}\Big\}    \times\{f_S(E)-f_D(E)\}dE \nonumber  \\
I_{\downarrow} &=\frac{q}{h}\int \Big\{ \underbrace{T_{\downarrow}(E)}_{\frac{Direct}{Transmission}}+~~~\underbrace{T^{sff}_{\uparrow\downarrow}(E)-T^{sff}_{\downarrow\uparrow}(E)}_{\frac{spin~flip}{forward~transmission}}~~~~-\underbrace{T^{sfb}_{\downarrow\uparrow}(E)}_{\frac{spin~flip}{backward~transmission}} \Big\}    \times\{f_S(E)-f_D(E)\}dE  \nonumber \\
&=\frac{q}{h}\int \Big\{{T_{\downarrow}(E)}+{T_{\uparrow\downarrow}^fT_{\uparrow}(E)\{1-F_{\frac{3}{2}}\}-T_{\downarrow\uparrow}^{f}T_{\downarrow}(E)\{1-F_{-\frac{3}{2}}\}-T_{\downarrow\uparrow}^{b}T_{\downarrow}(E)\{1-F_{-\frac{3}{2}}\}}\Big \}    \times\{f_S(E)-f_D(E)\}dE \nonumber \\
I^{sf} &= |I_{\uparrow \downarrow}^{sf}|-|I_{\downarrow\uparrow}^{sf}| \nonumber \\
&= \frac{q}{h}\int \{ T^{sff}_{\uparrow\downarrow}(E)+T^{sfb}_{\uparrow\downarrow}(E)-T^{sff}_{\downarrow\uparrow}(E)-T^{sfb}_{\downarrow\uparrow}(E) \} \times\{f_S(E)-f_D(E)\}dE
\label{eq:sqpct}
\end{align}
\end{widetext}
where $T_{\uparrow}(E)$ and $T_{\downarrow}(E)$ are the direct transmission coefficients between the forward propagating edge channels originating and terminating in the source and drain contacts respectively through the QPC in the absence of electron-nuclear spin flip-flop scattering. As already discussed, the term $T^{sff}_{\downarrow\uparrow}(E)$ and $T^{sff}_{\uparrow \downarrow}(E)$ characterize spin-flip scattering from a forward propagating edge channel terminating in the drain contact  to a forward propagating edge channel terminating in the drain contact at the QPC, while $T^{sfb}_{\uparrow\downarrow}(E)$ and $T^{sfb}_{\downarrow\uparrow}(E)$ characterize spin-flip scattering  at the QPC from a  forward propagating edge state terminating in the drain contact to a backward propagating edge channel  originating in the drain  contact at the QPC. %Other processes like scattering from a forward  to backward propagating edge states near the QPC  without  electron-nuclear spin flip-flop scattering  can be incorporated in the model by changing the direct transmission coefficient $T_{\uparrow}(E)$. 
The terms $T^{sff}_{\downarrow\uparrow}(E)$, $T^{sff}_{\uparrow\downarrow}(E)$, $T^{sfb}_{\uparrow\downarrow}(E)$ and $T^{sfb}_{\downarrow\uparrow}(E)$, being the probability of  electron-nuclear spin flip-flop processes, are dependent on the nuclear polarization (details given in Appendix \ref{appendix1}).
%\indent The spin-flip current is related to net spin-flip rate that the electron feels due to the bed of $N$ nuclei treated as point particles. For a system with four nuclear spin levels, $I_{sf}$ is given as:
%\begin{align}
%I^{sf} &=|I_{\uparrow\downarrow}^{sf}|-|I_{\downarrow\uparrow}^{sf}| \nonumber \\
%&= qN_I\left[ S \right] \left[ \frac{dF}{dt}\right]_{conserving} \nonumber \\
%&=qN_I \left[S\right] \left[\Gamma\right] \left[ F\right]
%\label{eq:spin_flip_curr}
%\end{align}
%where $\left[S\right]$ is the row matrix denoting the nuclear  spin levels in GaAs.
%In this paper, we are interested in the spin flip-flop transitions near the QPC which changes the conductivity of the device via the Overhauser field and the spin flip-flop scattering process from a forward propagating edge channel originating in the source contact to a forward propagating edge channel terminating in the drain contact. 
The spin-flip currents $I^{sf}_{\uparrow \downarrow}$ and $I^{sf}_{\downarrow \uparrow }$ give rise to nuclear polarization at and around the QPC region. \\ 
\indent Turning our attention to the self-consistent solution of the electronic transport and the temporal evolution of the nuclear polarization, the electronic transport is influenced by the nuclear polarization via the Overhauser field while the evolution of nuclear polarization is determined by the spin-flip current and the nuclear spin-lattice relaxation time ($\tau_I$). The matrix $\left[\Gamma\right]$, which determines the temporal evolution in nuclear polarization is hence related to the spin-flip currents $I^{sf}_{\uparrow \downarrow}$ and $I^{sf}_{\downarrow \uparrow }$. A schematic diagram on self-consistency involved in the temporal evolution of nuclear polarization and electronic transport phenomena is shown in Fig. \ref{fig:loop}.  
For a system with quad nuclear spin levels as in GaAs,  it can be shown that $\Gamma_{\downarrow\uparrow}=C_2|I_{\downarrow\uparrow}^{sf}|$ and $\Gamma_{\uparrow\downarrow}=C_1|I_{\uparrow\downarrow}^{sf}|$ with  $C_1=\frac{1}{qN_I\{1-F_{\frac{3}{2}}\}}$  and $C_2=\frac{1}{qN_I\{1-F_{-\frac{3}{2}}\}}$ (details given in Appendix \ref{appendix2}), $N_I$ being the number of nuclei that are being influenced by spin flip-flop processes at the QPC. We can hence rewrite the expression for $[\Gamma]$ as (details given in Appendix \ref{appendix2}):
\begin{widetext}
\begin{equation}
\left [ \Gamma \right ] =\begin{bmatrix}
-C_2|I_{\downarrow\uparrow}^{sf}| & C_1|I_{\uparrow\downarrow}^{sf}|  & 0 & 0 \\
C_2|I_{\downarrow\uparrow}^{sf}| & -\left (C_2|I_{\downarrow\uparrow}^{sf}|+ C_1|I_{\uparrow\downarrow}^{sf}| \right ) & C_1|I_{\uparrow\downarrow}^{sf}|&  0 \\
0 & C_2|I_{\downarrow\uparrow}^{sf}| & -\left (C_2|I_{\downarrow\uparrow}^{sf}|+ C_1|I_{\uparrow\downarrow}^{sf}| \right ) & C_1|I_{\uparrow\downarrow}^{sf}|   \\
0 & 0 & C_2|I_{\downarrow\uparrow}^{sf}| & -C_1|I_{\uparrow\downarrow}^{sf}|.
\end{bmatrix}
\label{eq:rewrite}
\end{equation}
\end{widetext}

\begin{figure*}[!htb]
\begin{minipage}[b]{.5\textwidth}
\subfigure[]{{\includegraphics[scale=.16]{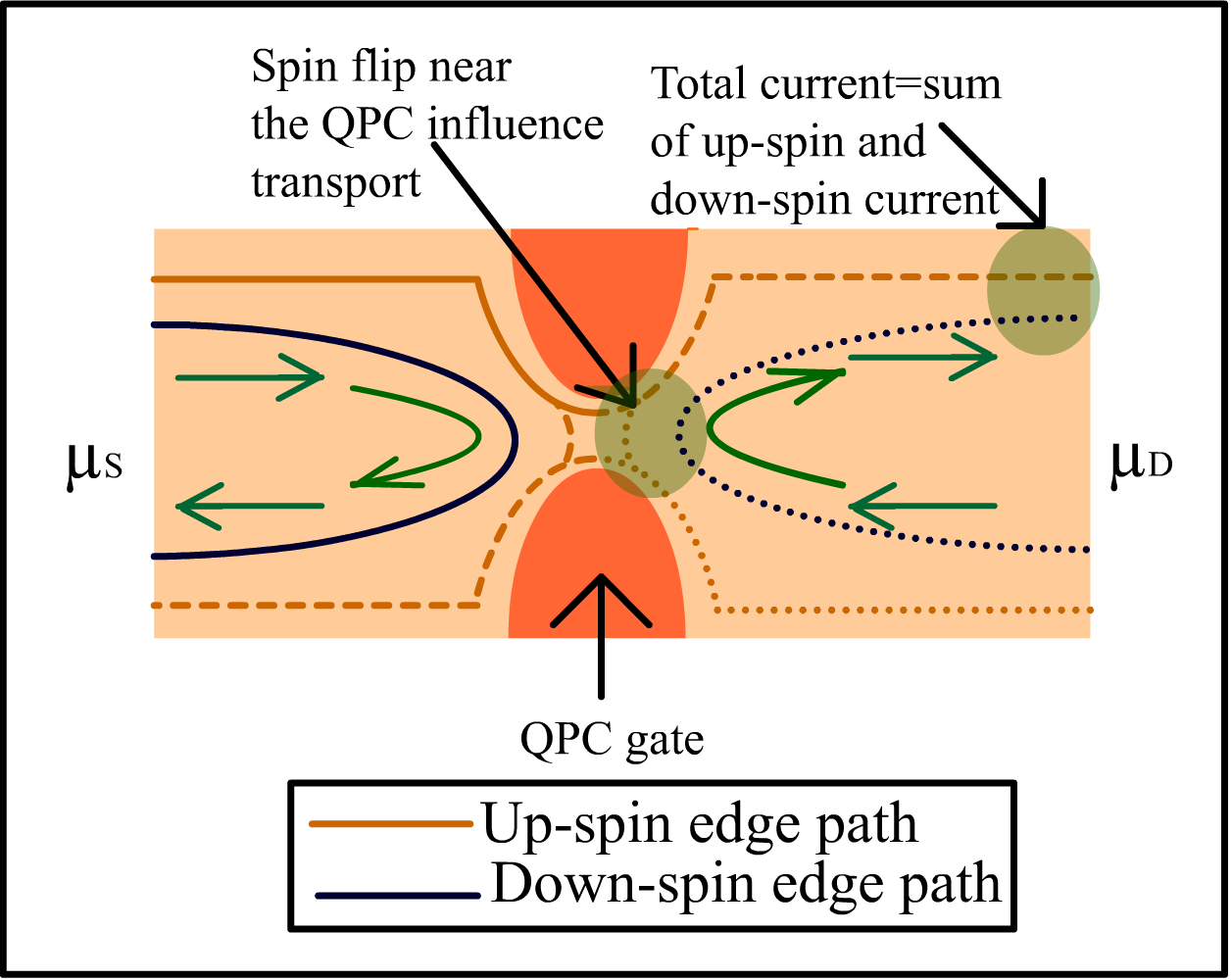}}
}
\subfigure[]{\includegraphics[height=2.2in]{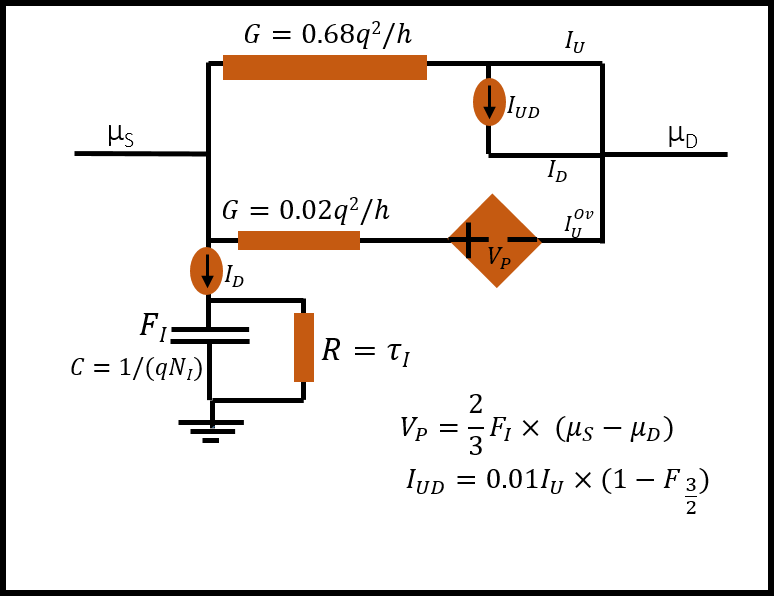}
}

\end{minipage}%
\begin{minipage}[b]{0.5\textwidth}
\subfigure[]{{\includegraphics[scale=.16]{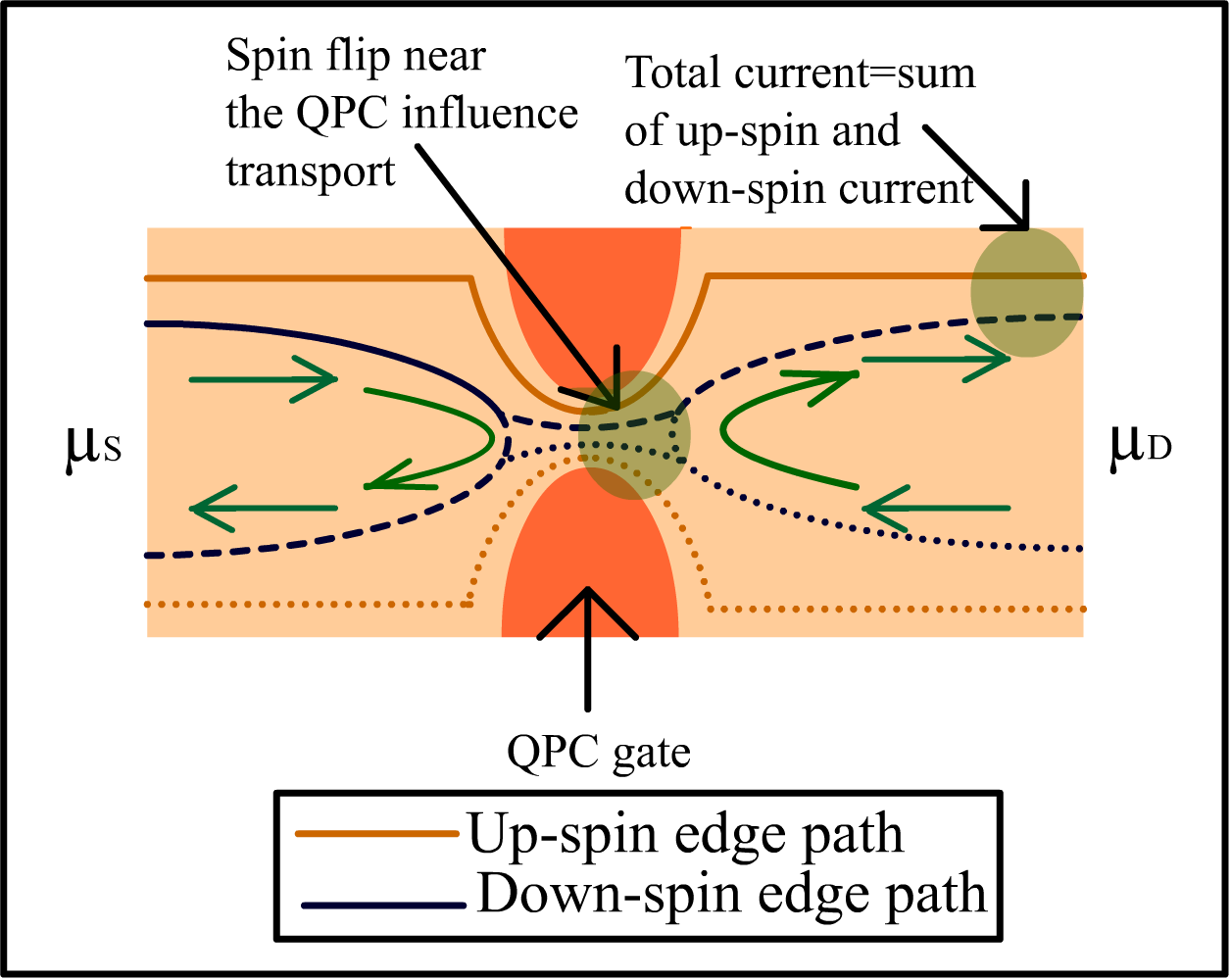}}}
\subfigure[]{\includegraphics[height=2.2in]{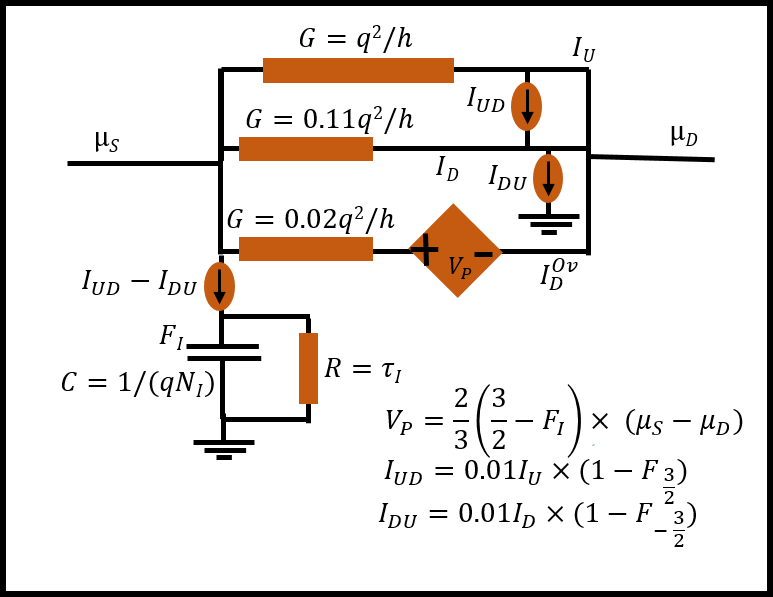}
}

\end{minipage}
\caption{Schematic of current path and equivalent circuit models. (a) Associated up-spin and down-spin current path  for $dI/dV<\frac{e^2}{h}~(\nu_{QPC}<1)$.    (b) Equivalent lumped circuit diagram for (a).  (c) Associated up-spin and down-spin current path  for $dI/dV>\frac{e^2}{h}~(\nu_{QPC}>1)$.  (d) Equivalent lumped circuit diagram for (c). The up-spin and down-spin channels in figure (a) and (c) are indicated by brown and blue colour respectively. In the energy range between $\mu_S$ and $\mu_D$, the filled, partially filled end empty edge states are denoted by solid line, dashed lines and dotted lines respectively.}
\label{fig:sgl_qpc_schem}
\end{figure*}
Let us now consider the experimental features on a case by case basis. 
\section{Results}\label{simulation_results}
\subsection{Conductance hysteresis with voltage sweep} \label{subsec:GV}
We first reproduce  some trends noted in the conductance plots of a recent experiment \cite{Wald_PRL} where a change in the conductance along with hysteresis in the conductance was noted in the vicinity of $V=0$  with positive and negative source to drain voltage sweep. We explain the possible phenomena giving rise to such experimental trends. \\
 \textbf{Case I: $dI/dV < e^2/h$}\\
A schematic of the scattering processes in this regime is shown in Fig.~\ref{fig:sgl_qpc1}(a) while the up-spin and down-spin edge current paths and equivalent circuit models for the phenomena occuring around the QPC are shown in Fig.~\ref{fig:sgl_qpc_schem}(a) and (b) respectively. In this case, the following points are to be noted:
\begin{enumerate}
\item Only the up-spin channel is transmitted through the QPC.
\item The down-spin channel originating in the source contact is totally reflected at the QPC.
\item Some up-spin electrons in the edge channel terminating in the drain contact can scatter at the QPC to the down-spin edge channel terminating in the drain contact   via electron-nuclear spin flip-flop scattering. Such a scattering decreases the current in the up-spin channel just outside the QPC and increases the current in the down-spin edge channel outside the QPC. However, the total current remains proportional to $T_{\uparrow}(E)$.
\end{enumerate}
 \textbf{Reason for an increase in $dI/dV$ near $V=0$.}
\begin{enumerate}
\item Near V=0, the nuclear polarization cannot be maintained.
\item Nuclear polarization drops due to spin lattice relaxation. 
\item A drop in nuclear polarization results in an increase in  the direct transmission coefficient $T_{\uparrow}$ of the up-spin channel due to a decrease in the Overhauser field as well as an  increase in  the spin-flip transmission coefficient $T_{\uparrow\downarrow}^{sff}$. 
\end{enumerate}
\textbf{Equivalent circuit model:}\\
A schematic of the edge channel path in this case is shown in Fig. \ref{fig:sgl_qpc_schem} (a) while the  equivalent circuit in this case is detailed in Fig.~\ref{fig:sgl_qpc_schem} (b) to aid a visualization of the various transport phenomena inside the device. The circuit model can be described as follows:
\begin{enumerate}
\item The up-spin edge channel is represented by a conductance $G=0.68\frac{e^2}{h}$.
\item The up-to-down spin-flip scattering can be represented by an equivalent current source from the up-spin channel. $I_{UD}=0.01I_U \times \{1-F_{\frac{3}{2}}\}$.
\item The change in transmissivity of the up-spin channel due to the  Overhauser field is represented by a by an equivalent conductor $(G=0.02\frac{e^2}{h})$ in series with a voltage dependent voltage source $V_p=\frac{2}{3}F_I \times \{\mu_S-\mu_D\}$. The current change due to the Overhauser field is represented by $I_U^{Ov}$.
\item The nuclear polarization is represented by the voltage across the capacitor.
\item The resistance in parallel with the capacitor represents nuclear spin-lattice relaxation. 
\end{enumerate}
\textbf{Case II $dI/dV>e^2/h$} \\
A schematic of the scattering processes in this regime is shown in Fig.~\ref{fig:sgl_qpc1}(b) while the up-spin and down-spin edge current paths and equivalent circuit models for the phenomena occuring around the QPC is shown in Fig.~\ref{fig:sgl_qpc_schem} (c) and (d) respectively. In this case, the following points are to be noted:
\begin{enumerate}
\item The up-spin electrons in the edge channel originating in the source contact are fully transmitted through the QPC to the up-spin edge channel terminating in the drain contact.
\item The down-spin electrons in the edge channel originating in the source contact  are partially transmitted through the QPC to the down-spin edge channel terminating in the drain contact.
\item Two kinds of electron-nuclear spin-flip  scattering dominate at the QPC in this case:
\begin{enumerate}
\item Electrons from the forward propagating up-spin channel terminating in the drain  contact can undergo  spin-flip  scattering to the forward propagating down-spin channel terminating in the drain contact which is almost empty in the energy range between $\mu_S$ and $\mu_D$. Such scattering at and around the QPC results in a positive nuclear polarization.
\item Electrons in forward propagating down-spin channel propagating through the QPC  can undergo a spin-flip scattering  to the backward propagating up-spin channel (which is totally empty in the energy range between $\mu_S$ and $\mu_D$) terminating in the source contact. Such scattering results in a negative nuclear polarization in addition to decreasing the total current through the QPC.
\item Out of these two processes, the former process dominates at the QPC  due to the presence of more up-spin electrons compared to down-spin electrons resulting in a net positive nuclear polarization at the QPC. 
\end{enumerate}
\end{enumerate}
 \textbf{Reason for a decrease in $dI/dV$ near $V=0$:}
\begin{enumerate}
\item Near $V=0$, the nuclear polarization cannot be maintained.
\item Nuclear polarization drops due to spin-lattice relaxation. 
\item A drop in polarization results in an increase in the up-to-down spin-flip rate as well as a decrease in down-to-up spin-flip rate in addition to a decrease in the direct transmission coefficient $T_{\downarrow}$ of the down-spin channel due to decrease in the Overhauser field.
\item The decrease in transmission coefficient of the down-spin channel decreases the conductance of the QPC in the vicinity of $V=0$.
\end{enumerate}
\textbf{Equivalent circuit model:}\\
The equivalent circuit in this case is detailed in Fig.~\ref{fig:sgl_qpc_schem}(d). The circuit model can be described as follows:
\begin{enumerate}
\item The up spin edge channel originating in the source contact  is fully transmitted through the QPC  and hence is represented by a conductance $G=e^2/h$.
\item The down-spin edge channel originating in the source contact is partially transmitted through the QPC to the down-spin edge channel terminating in the drain contact and hence  is represented by a conductance $G=0.11\frac{e^2}{h}$. 
\item The up-to-down spin-flip current at the QPC from the forward propagating up-spin channel terminating  in the drain  contact  to the forward propagating down-spin channel  terminating in the drain contact is represented by a current dependent current source  $I_{UD}=0.01I_U \times \{1-F_{\frac{3}{2}}\}$.
\item The down-to-up spin-flip current from the forward propagating down-spin channel originating in the source contact to the backward propagating up-spin channel terminating in the source contact   is represented by a current dependent current source  $I_{DU}=0.01I_D \times \{1-F_{-\frac{3}{2}}\}$.
\item The change in the transmission coefficient of the down-spin channel due to the  Overhauser field is represented by a by an equivalent conductor $(G=0.02\frac{e^2}{h})$ in series with a voltage dependent voltage source ($V_p=\frac{2}{3}\{\frac{3}{2}-F_I\} \times \{\mu_S-\mu_D\}$). The current change due to  the Overhauser field is represented by $I_D^{Ov}$.
\item The nuclear polarization is represented by the voltage across the capacitor.
\item The resistance in parallel with the capacitor represents   nuclear spin-lattice relaxation by causing charge leakage from the capacitor. \\%
\end{enumerate}

\begin{figure}[!htb]

\begin{minipage}[b]{0.25\textwidth}
\subfigure[]{\includegraphics[scale=.175]{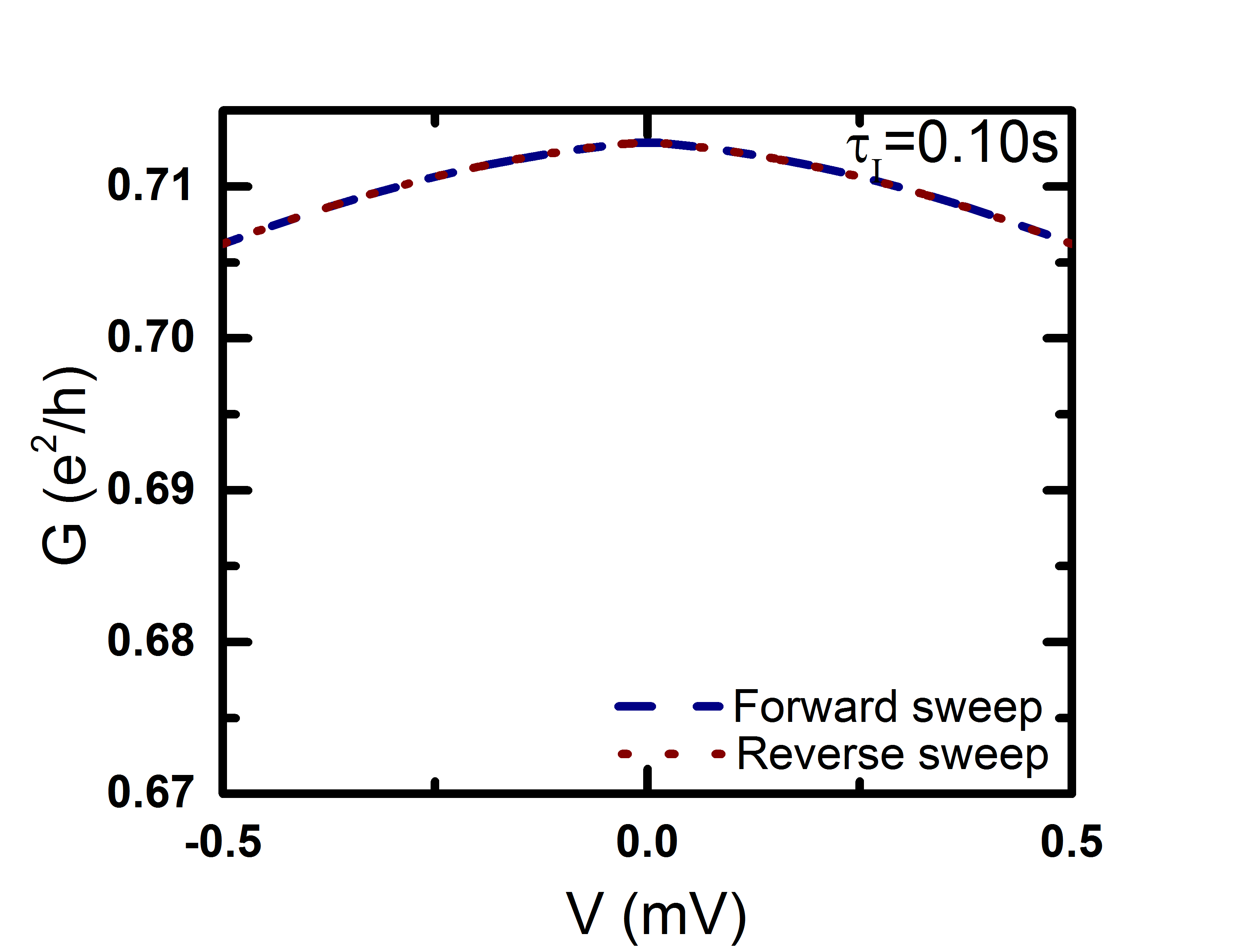}
}

\subfigure[]{\includegraphics[scale=.175]{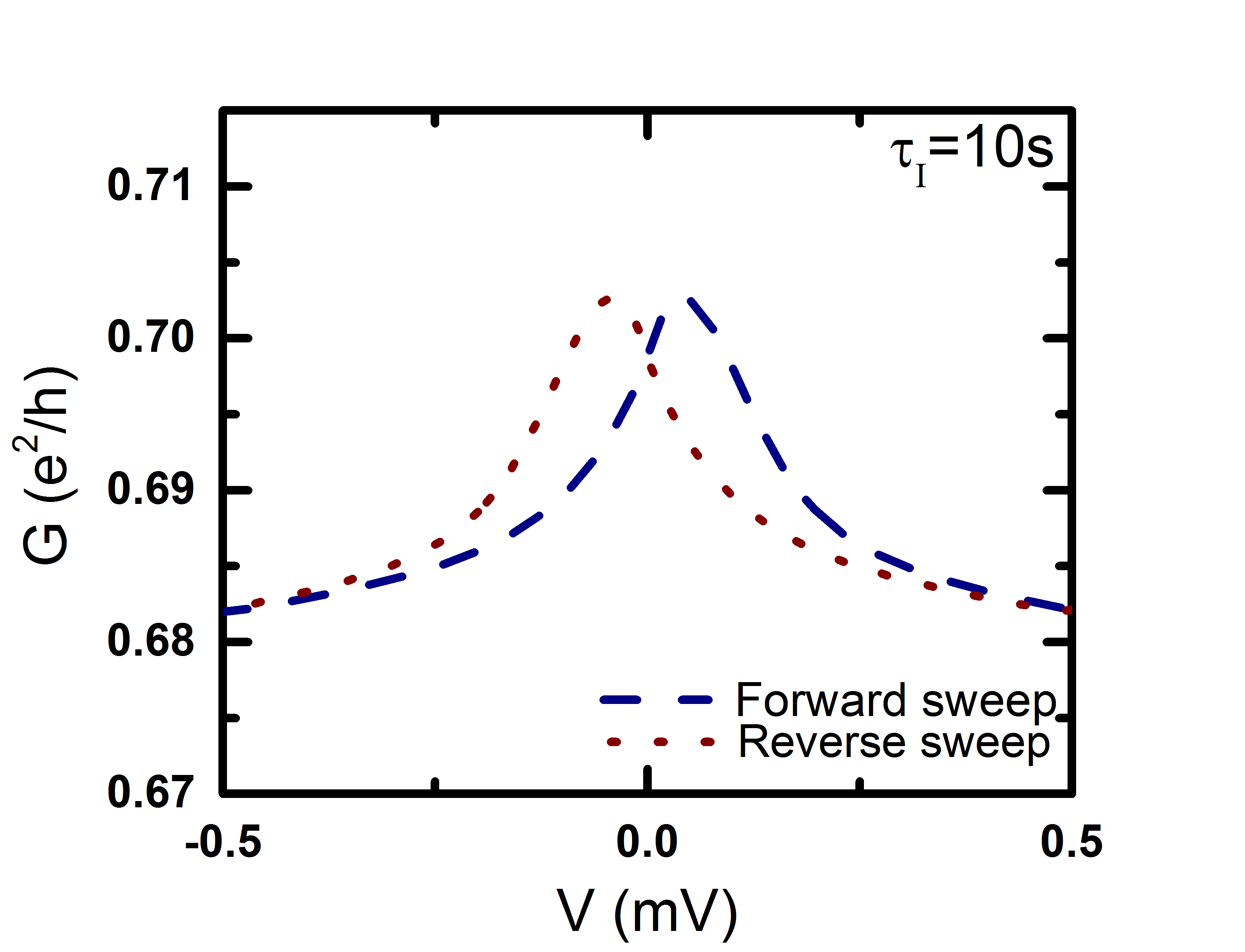}}
\subfigure[]{\includegraphics[scale=.175]{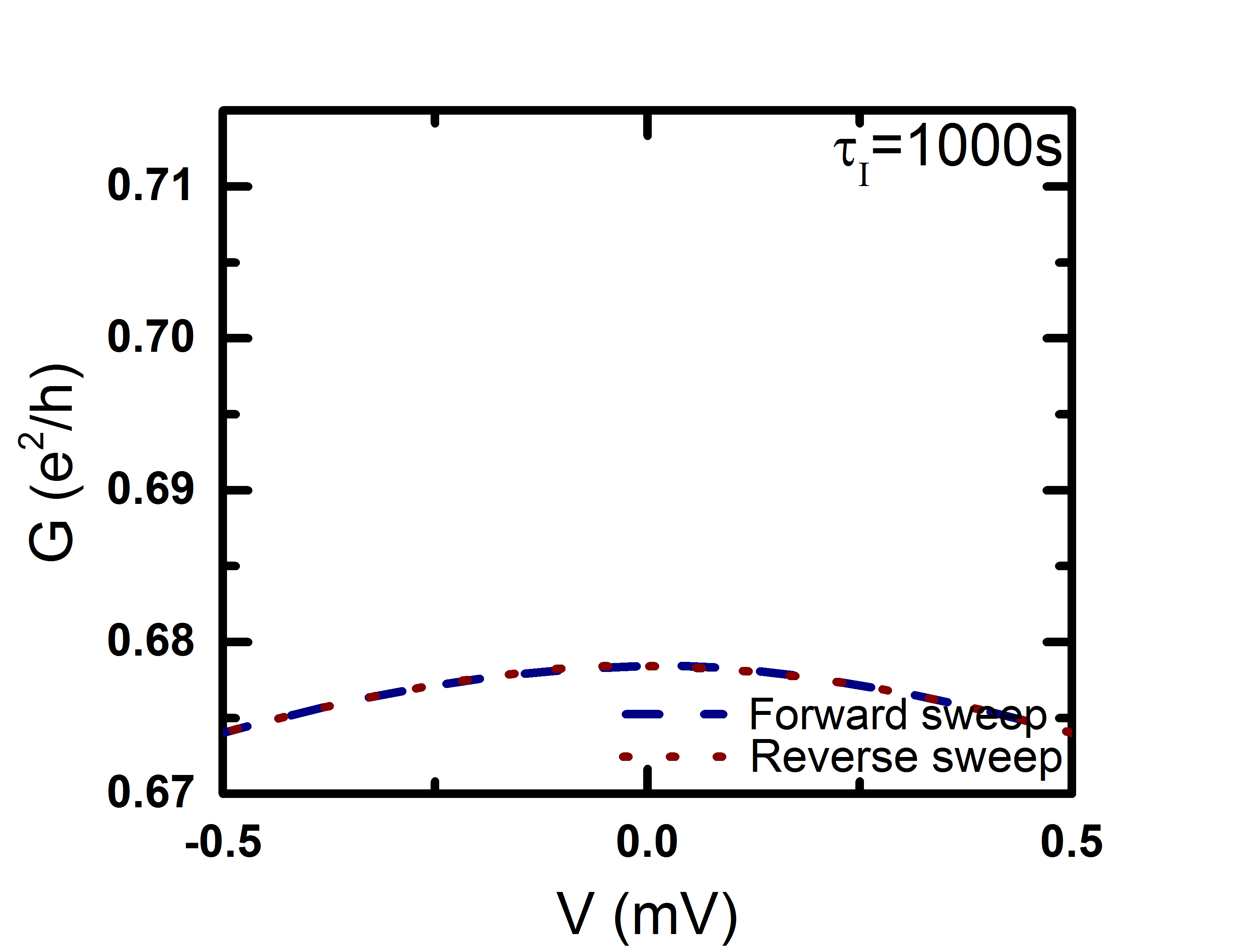}}

\end{minipage}%
\begin{minipage}[b]{0.25\textwidth}
\subfigure[]{\includegraphics[scale=.175]{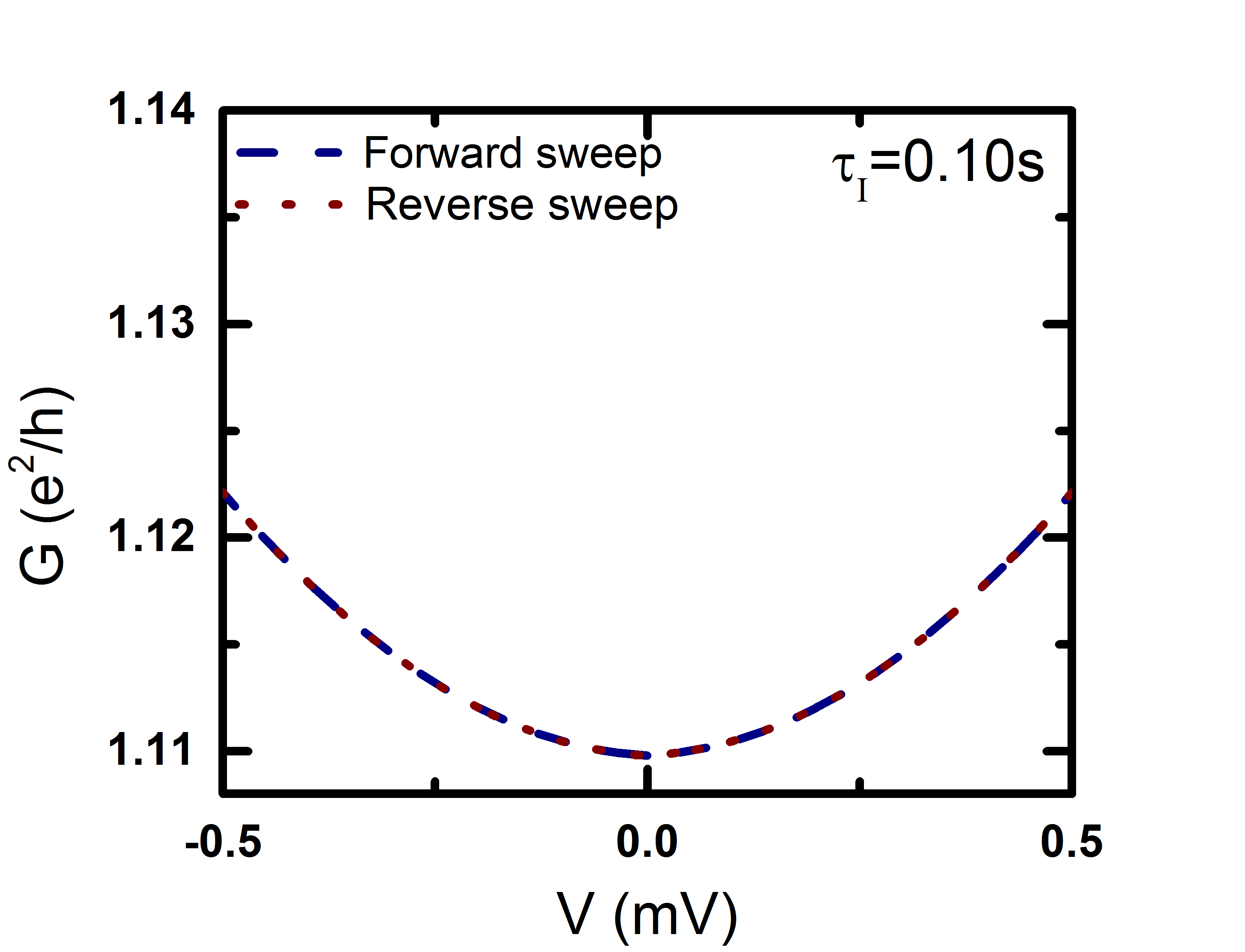}}

\subfigure[]{\includegraphics[scale=.175]{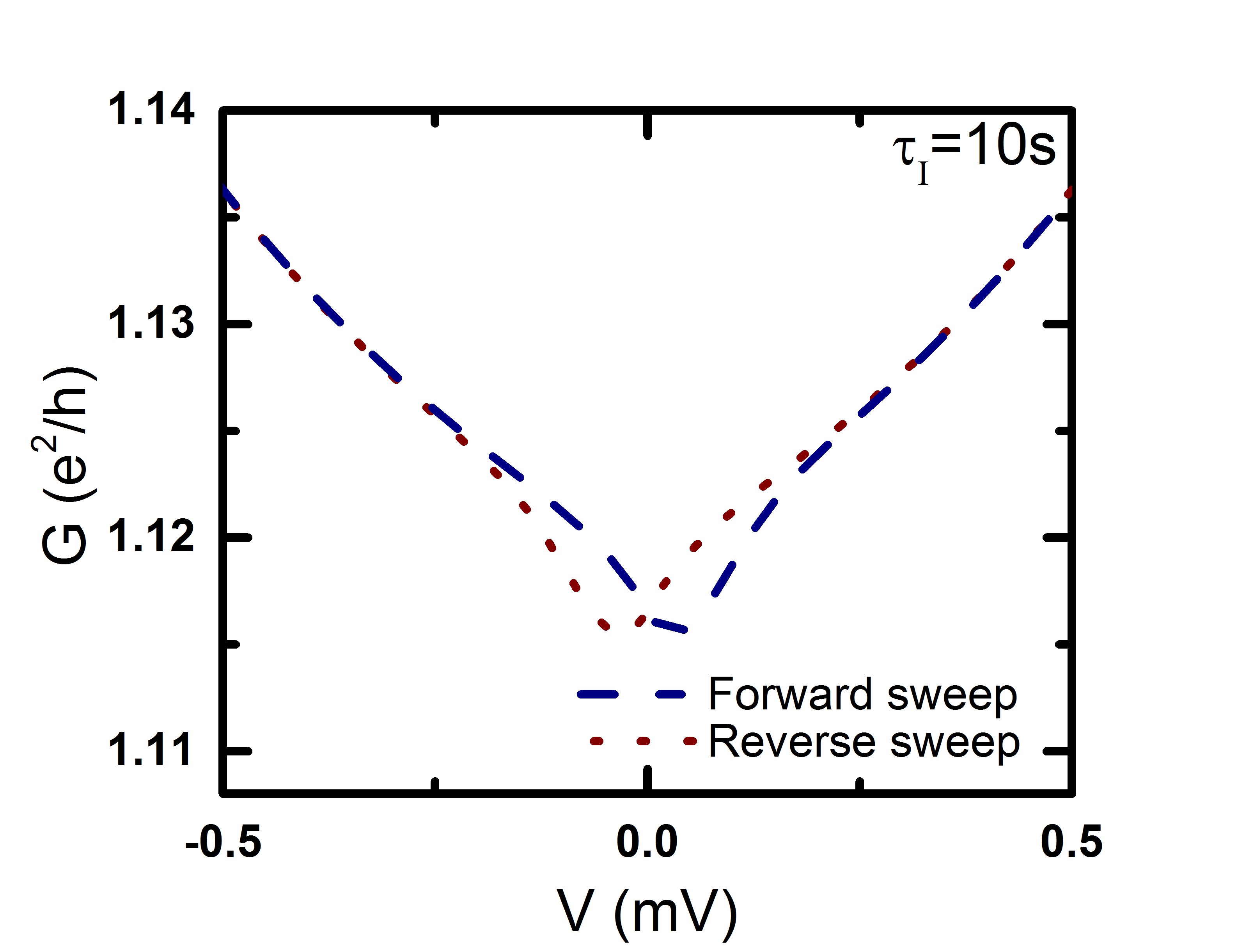}
}
\subfigure[]{\includegraphics[scale=.175]{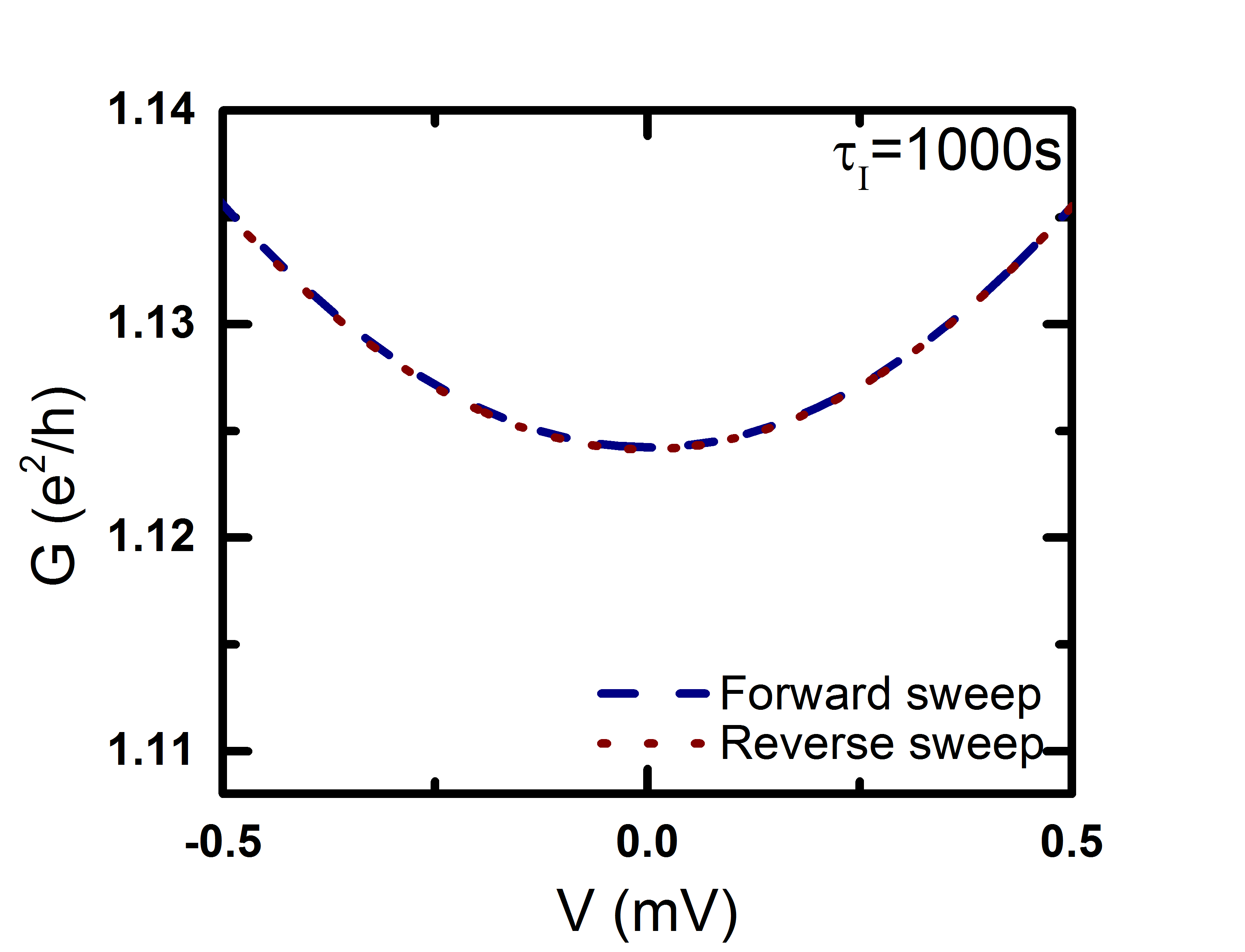}
}

\end{minipage}

\caption{ Simulated plots of conductance-voltage ($G-V$) traces during a voltage sweep. \emph{Left panel (a,b,c):} for the case $G<\frac{e^2}{h}$, \emph{Right panel (d,e,f):} for the case $G>\frac{e^2}{h}$\emph{Top panel (a,d):} the case when $\tau_I=0.1s$, \emph{Middle panel (b,e):}  the case when $\tau_I=10s$, \emph{Bottom panel (c,f):} the case when $\tau_I=1000s$. The parameters used in the simulations are: $B=4T$, $N_I=10^{8},~T_{\uparrow \downarrow}^f=0.01,~T_{\downarrow \uparrow}^f=0,~T_{\uparrow \downarrow}^b=0,~T_{\downarrow\uparrow}^b=0.01$. Total sweep time $=200s$.}
\label{fig:sgl_qpc2}
\end{figure}

%\begin{widetext}
%\begin{figure}[!htb]
%\begin{minipage}[b]{1\textwidth}
%\subfigure[]{\includegraphics[scale=.22]{Voltage_sweep/dI_dV_less_1_low_tau}
%}\subfigure[]{\includegraphics[scale=.22]{Voltage_sweep/dI_dV_less_1}
%}\subfigure[]{\includegraphics[scale=.22]{Voltage_sweep/dI_dV_less_1_high_tau}}

%\subfigure[]{\includegraphics[scale=.22]{Voltage_sweep/dI_dV_great_1_low_tau}
%}\subfigure[]{\includegraphics[scale=.22]{Voltage_sweep/dI_dV_great_1}
%}\subfigure[]{\includegraphics[scale=.22]{Voltage_sweep/dI_dV_great_1_high_tau}
%}

%\caption{Plots of   $dI/dV-V$ curve during voltage sweep with our simple model. \it{Top panel:} $\tau_I=0.1s$, \it{Middle panel:} $\tau_I=10s$, \it{Bottom panel:} $\tau_I=200s$, \it{Left panel:} $G<\frac{e^2}{h},~T_{\uparrow}=0.65,~T_{\uparrow\downarrow}=0.2$, \it{Right panel:} $G>\frac{e^2}{h},~T_{\downarrow}=1,~T_{\downarrow\uparrow}=0.2$. Simulations done with $N_I=10^{10}$ and total sweep time $=200s$}
%\label{fig:sgl_qpc2}
%\end{minipage}
%\end{figure}
%\end{widetext}

\indent The simulated results of the change in conductance with source to drain voltage sweeps are shown in Fig. \ref{fig:sgl_qpc2}. The parameters $T_{\uparrow}$ and $T_{\downarrow}$ in the simulations are calculated directly via a $1-D$ non-equilibrium Green's function (NEGF) method using an atomistic tight-binding Hamiltonian \cite{Datta_Green,qtransport,LNE} while the parameters $T_{\uparrow \downarrow}^{sff}$ and $T_{\downarrow \uparrow }^{sfb}$ are calculated using \eqref{eq:appref1}.  The parameters in the above illustration for the circuit diagrams in Fig. \ref{fig:sgl_qpc1} are chosen to match the simulated result of the change in conductance with source to drain voltage sweep.  The maximum change in the conductance due to a difference in the Overhauser field between the fully polarized nuclei and the  non-polarized nuclei is less than $G_{Ov}=0.02\frac{e^2}{h}$. However this maximum change can be enhanced due to spin flip-flop tunneling as noted experimentally \cite{Wald_PRL}. \color{black} The hysteresis in the $G-V$ curves in  Fig. \ref{fig:sgl_qpc2} near $V=0$ occurs only when the   nuclear spin relaxation time ($\tau_I$) is of the order of the voltage sweep time. This results in a lag between the applied voltage and nuclear polarization near $V=0$ thereby resulting in the hysteresis. The hysteresis in $G-V$ plots disappear when the $\tau_I$ is very large such that the change in nuclear polarization is negligible during the time of voltage sweep.   The hysteresis in $G$ vs $V$ plots also disappear when $\tau_I$   is very small compared to the  voltage sweep time because the nuclear polarization is always in a steady state  with the applied voltage.
\subsection{Resistively detected nuclear magnetic resonance (RDNMR)} \label{subsec:RDNMR}
\begin{figure}[!htb]
\begin{minipage}[b]{0.25\textwidth}
\subfigure[]{\includegraphics[scale=.175]{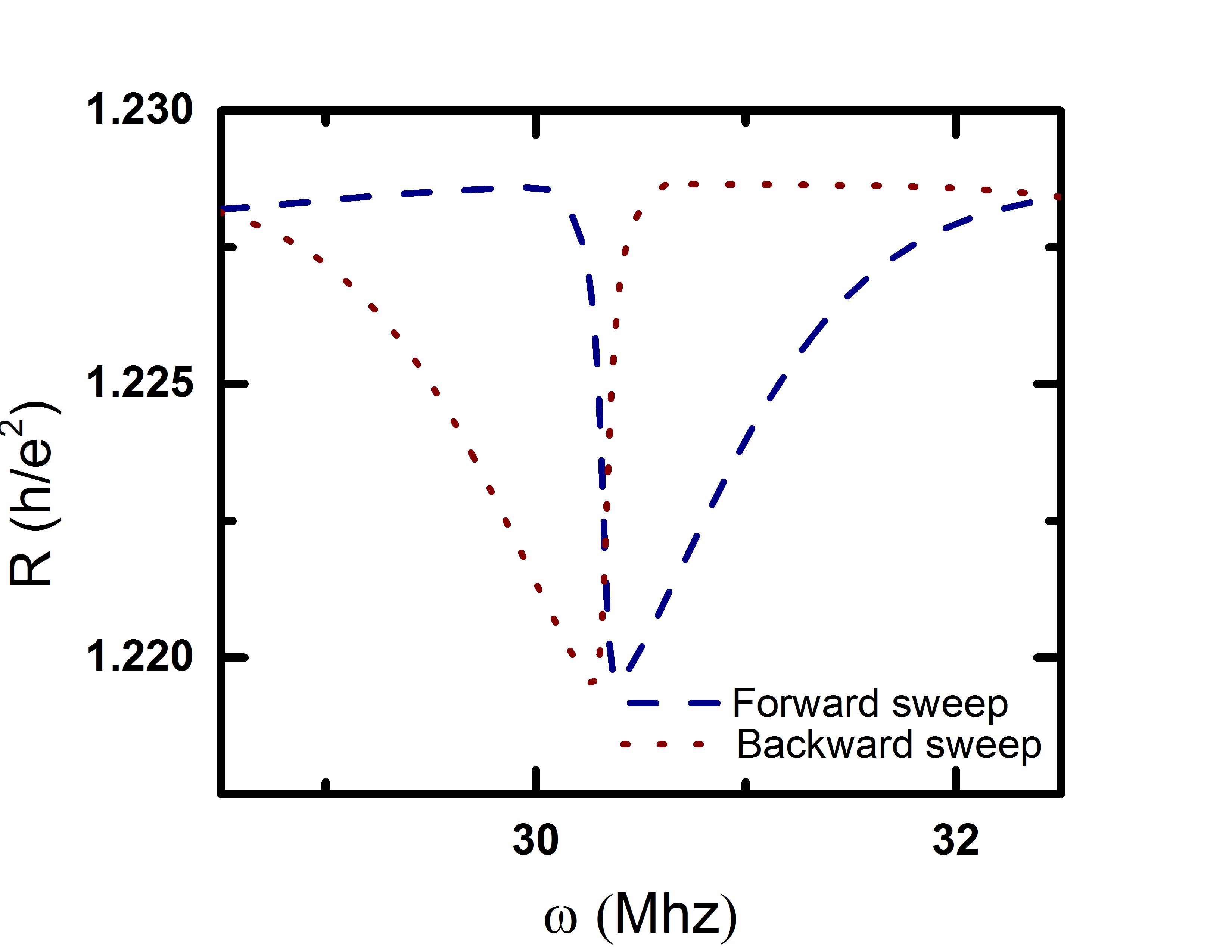}}
\subfigure[]{\includegraphics[scale=.175]{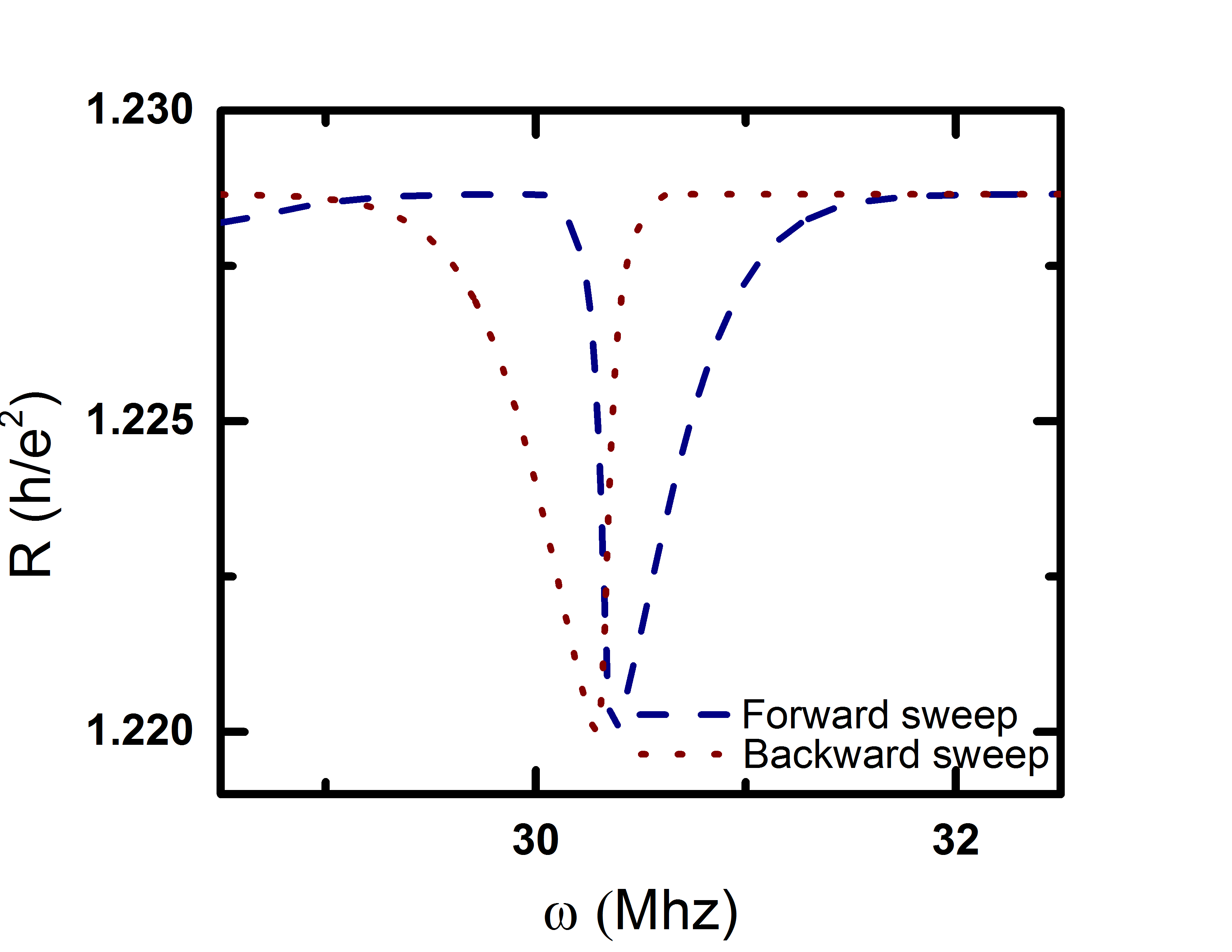}}
\subfigure[]{\includegraphics[scale=.175]{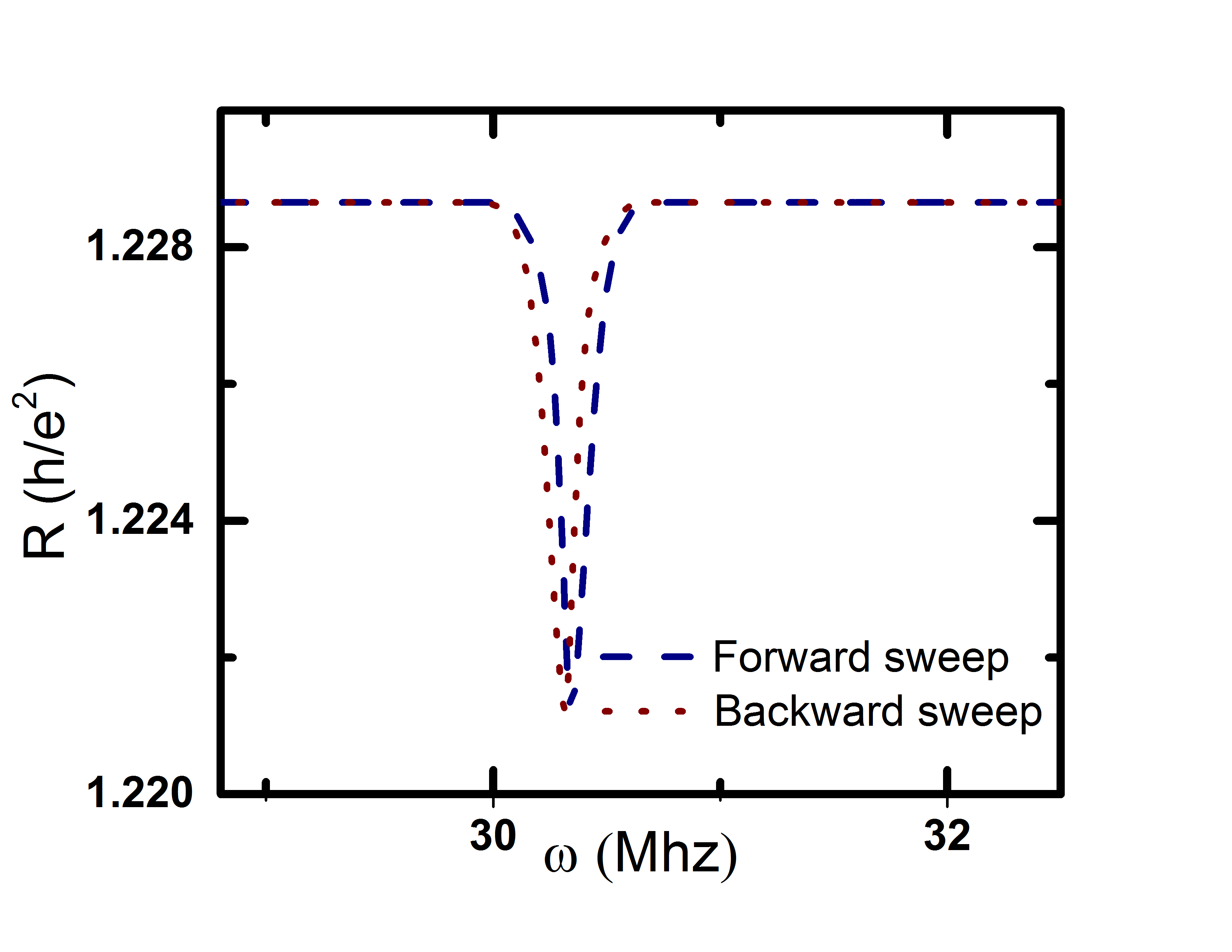}}
\end{minipage}%
\begin{minipage}[b]{0.25\textwidth}
\subfigure[]{\includegraphics[scale=.183]{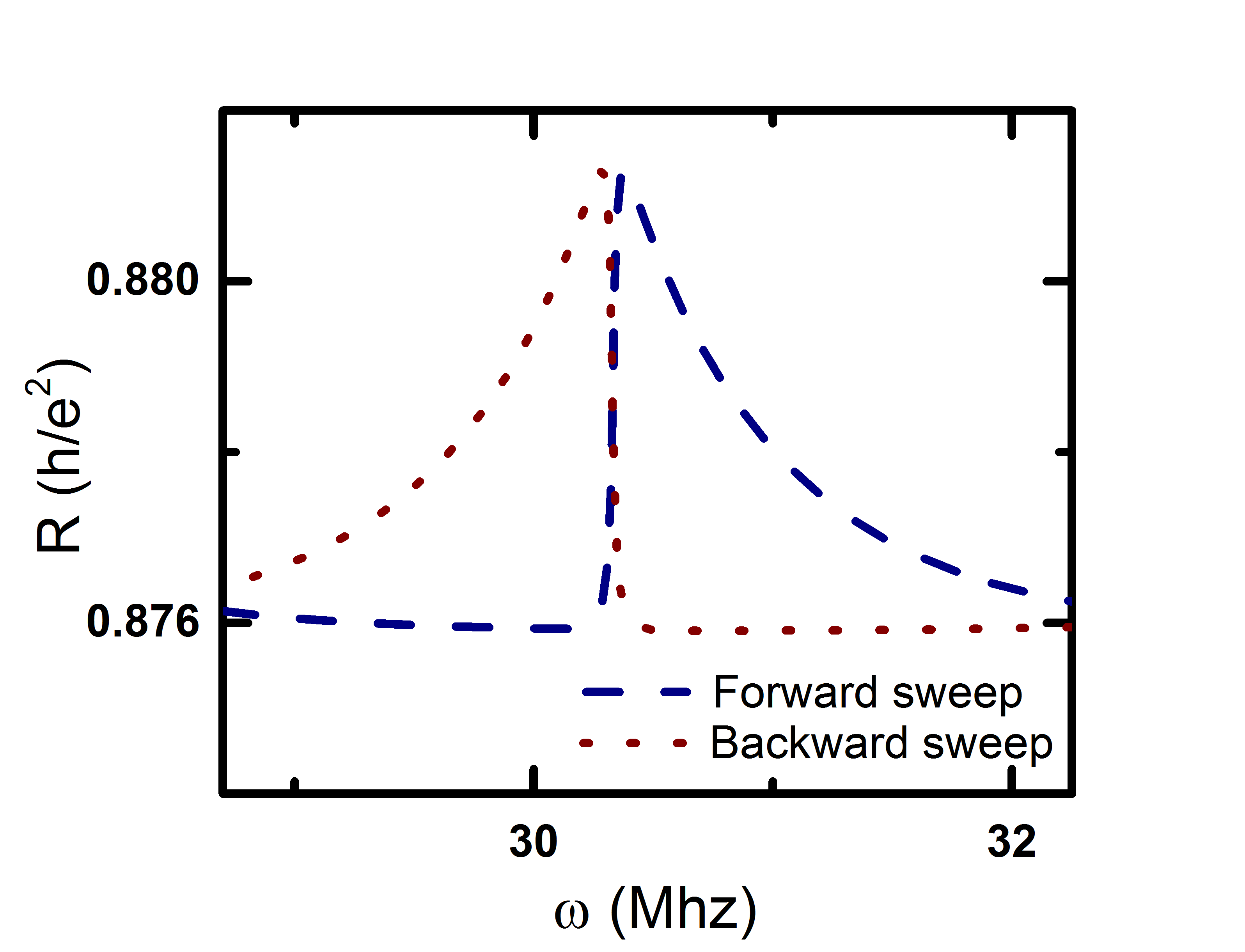}}
\subfigure[]{\includegraphics[scale=.183]{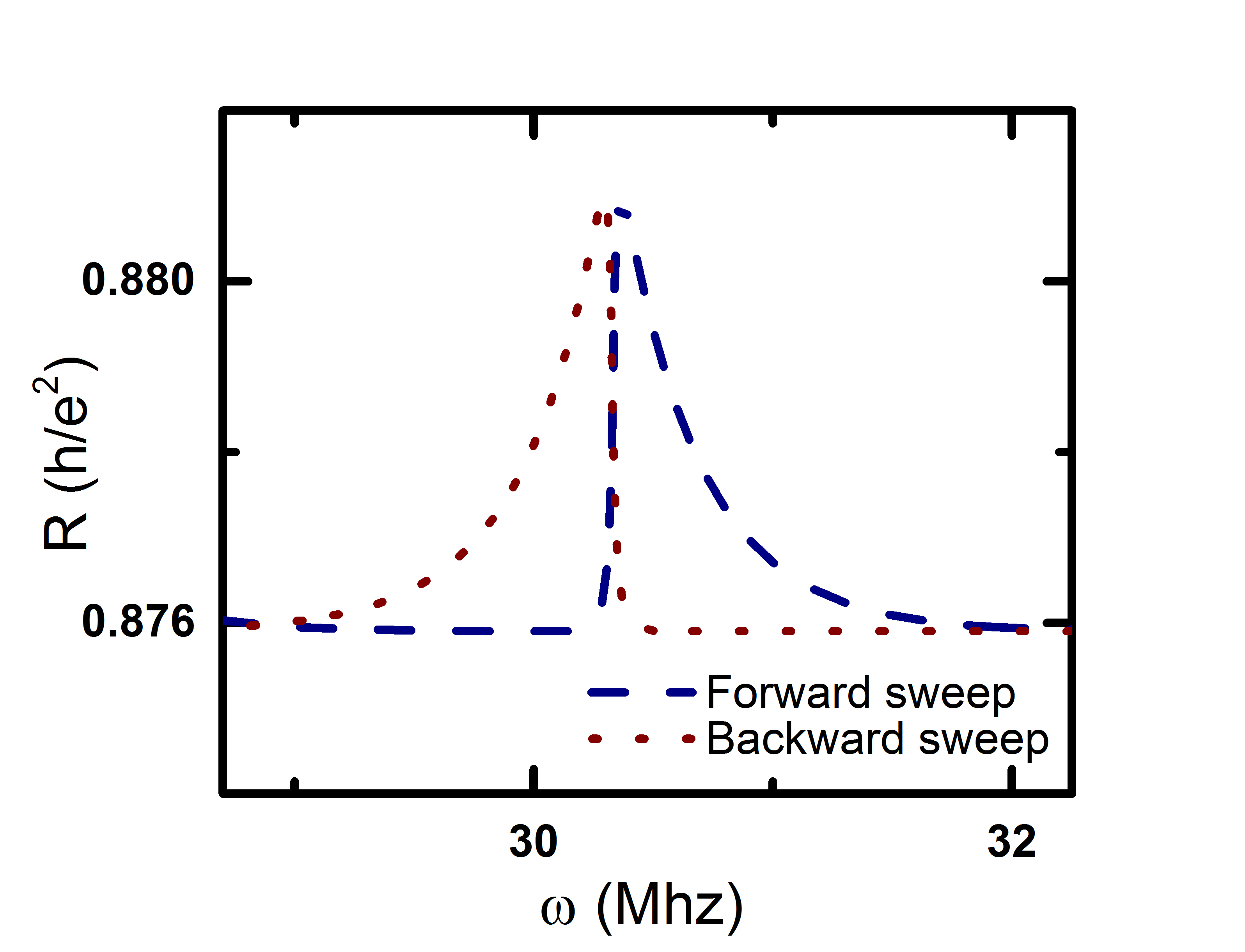}}
\subfigure[]{\includegraphics[scale=.183]{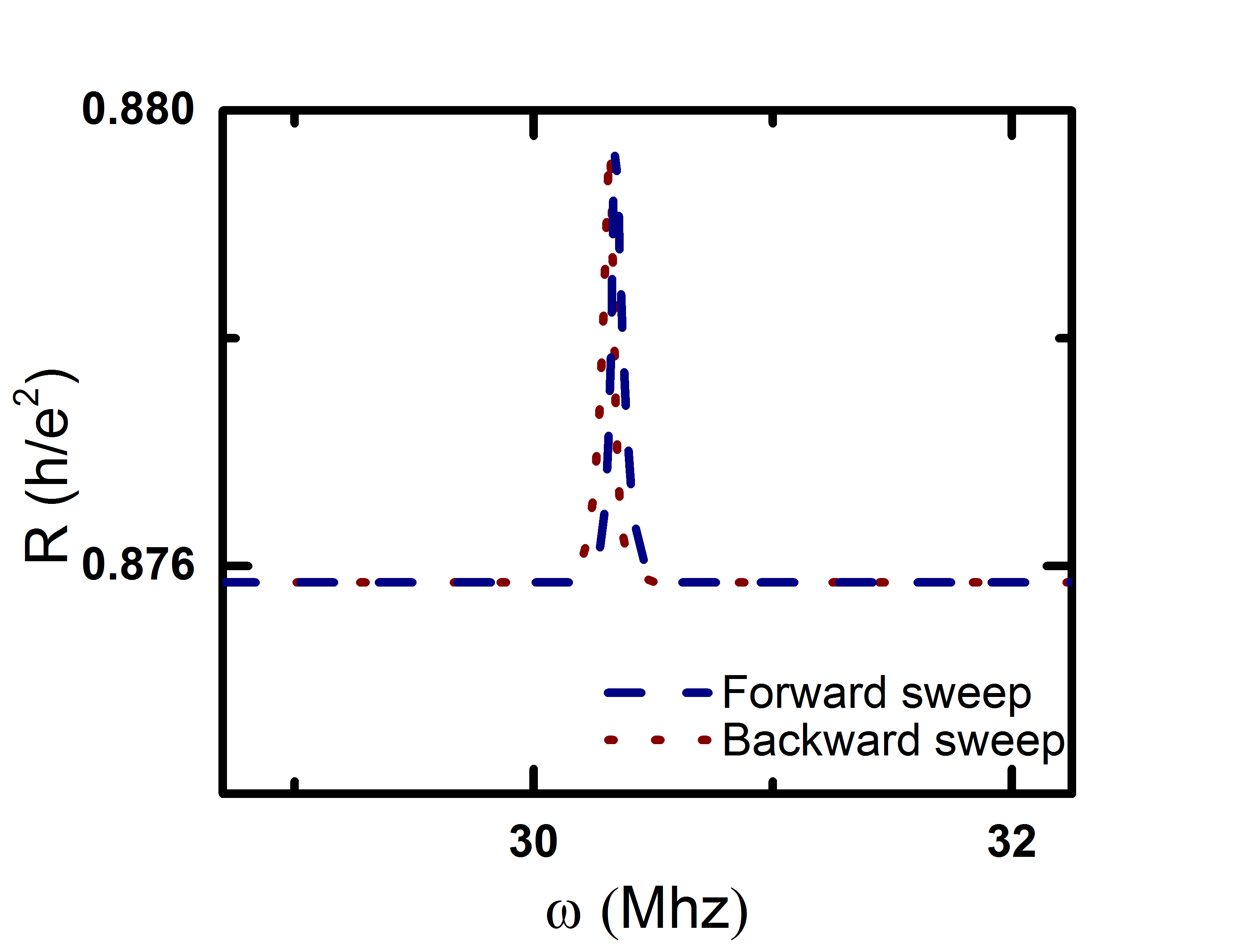}}
\end{minipage}
\caption{{Plots of resistance variation with NMR frequency sweep (RDNMR traces) at various sweep rates. Top panel: Plots for total sweep time=$50s$, Middle panel: Plots for total sweep time=$100s$, Bottom  panel: Plots for total sweep time=$500s$. Left panel (a, b, c): Plots for $R>\frac{h}{e^2}~(\nu_{QPC}<1)$, Right panel (d, e, f): Plots for $R<\frac{h}{e^2} ~(\nu_{QPC}>1)$ }. Simulations  are done for  $B=4T$, $N_I=10^8,~T^f_{\uparrow\downarrow}=0.01,~T^b_{\uparrow\downarrow}=0,~T^f_{\downarrow\uparrow}=0,~T^b_{\downarrow\uparrow}=0.01$. $T_{\uparrow}$ and $T_{\downarrow}$ are calculated using NEGF formalism as elaborated in the text.}
\label{fig:nmr1}
\end{figure}

%\begin{widetext}
%\begin{align}
%I_{\uparrow} &=\frac{q}{h}\int \Big\{ \underbrace{T_{\uparrow}(E)}_{\frac{Direct}{Transmission}}+\underbrace{T^{sff}_{\downarrow\uparrow}(E)}_{\frac{spin~flip}{forward~transmission}} \Big\}    \times\{f_S(E)-f_D(E)\}dE  \nonumber \\
%&=\frac{q}{h}\int \Big\{{T_{\uparrow}(E)}+{T_{\downarrow\uparrow}^{f}T_{\downarrow}(E)\{1-F_{-\frac{3}{2}}\}}\Big\}    \times\{f_S(E)-f_D(E)\}dE \nonumber 
%\end{align}
%\begin{align}
%I_{\downarrow} &=\frac{q}{h}\int \Big\{ \underbrace{T_{\downarrow}(E)}_{\frac{Direct}{Transmission}}+\underbrace{T^{sff}_{\uparrow\downarrow}(E)}_{\frac{spin~flip}{forward~transmission}} \Big\}    \times\{f_S(E)-f_D(E)\}dE  \nonumber \\
%&=\frac{q}{h}\int \Big\{{T_{\downarrow}(E)}+{T_{\uparrow\downarrow}^b(E)T_{\uparrow}(E)\{1-F_{\frac{3}{2}}\}}\Big \}    \times\{f_S(E)-f_D(E)\}dE \nonumber 
%\end{align}
%\begin{align}
%I^{sf} &= |I_{\uparrow \downarrow}|-|I_{\downarrow\uparrow}| \nonumber \\
%&= \frac{q}{h}\int \{ T^{sff}_{\uparrow\downarrow}(E)+T^{sfb}_{\uparrow\downarrow}(E)-T^{sff}_{\downarrow\uparrow}(E)-T^{sfb}_{\downarrow\uparrow}(E) \} \times\{f_S(E)-f_D(E)\}dE
%\label{eq:sqpct}
%\end{align}
%\end{widetext}
We begin our analysis with \eqref{eq:sqpct}, where the nuclear polarization in the vicinity of the QPC is perturbed by an externally applied alternating magnetic field in the radio frequency (RF) range resulting in the Zeeman split nuclear levels to interact with each other. Near the frequency corresponding to the difference in energy between the two spin split nuclear energy levels ($\hslash \omega=\xi_{s+1}-\xi_{s}$), precession of the nuclear spins accompanied by a rapid decay in the nuclear polarization occurs. 
 %\cite{rdnmr1,rdnmr2,Nano_Hamilton,rdnmr4,rdnmr_dis1,rdnmr_dis2,rdnmr_dis3,rdnmr_dis4,rdnmr_dis5}
%\indent One of the features of quantum Hall experiments is that a change in nuclear polarization is accompanied by change in conductance \cite{rdnmr1,rdnmr2,Nano_Hamilton,rdnmr4,rdnmr_dis1,rdnmr_dis2,rdnmr_dis3,rdnmr_dis4,rdnmr_dis5}. Such effects arise due to a change the  in Overhauser field accompanied by a change  in the direct transmission coefficient  as well as a change in the rate of spin-flip tunneling through the QPC which is accompanied by electron-nuclear spin flip flop. The change in conductance mainly manifests itself in the vicinity of $dI/dV=\frac{e^2}{h}$, where the filled and empty conducting channels are spatially close to each other.  
To model such processes, we model the spin split nuclear energy levels by a broadened normalized density of states \cite{Datta_Green,qtransport,NMR_book}.
\[
D_{s}(\xi)=\frac{1}{2\pi}\frac{\eta}{(\xi-\epsilon_{s})^2+(\frac{\eta}{2})^2},
\]
where $\xi$ is the free variable denoting energy in the nuclear spin space, $\eta$ is related to the amount of broadening of the nuclear spin levels and $\epsilon_{s}$ is the  energy level of the $s^{th}$ nuclear spin  in the absence of broadening. Broadening might be a result of thermal motion of the nucleus \cite{NMR_book}, hyperfine interaction mediated electron-nuclear spin exchange \cite{NMR_book} as well as nuclear dipole-dipole exchange interaction \cite{NMR_book} which is the causative agent for nuclear spin diffusion\cite{NMR_book}. We take broadening to be $\eta=10^{-4}\mu eV$. We simulate the case of quad nuclear spin levels, as in GaAs, separated in energy due to Zeeman splitting. The rate equations in this case are given by:

  \begin{align}
  \left[ \frac{dF(\xi)}{dt}\right]=\left[ \frac{dF(\xi)}{dt}\right]_{flip-flop}&+\left[ \frac{dF(\xi)}{dt}\right]_{relaxation} \nonumber \\
  &+\left[ \frac{dF(\xi)}{dt}\right]_{NMR}  \nonumber \\
\label{eq:nmr2}    
  \end{align}

  \begin{equation}
  F_I=\int^{\infty}_{-\infty}\left[s\right] \times [D_N(\xi)] \times \left[ F(\xi) \right]d\xi
\label{eq:nmr3}    
  \end{equation}

  \begin{align}
  T_{\downarrow\uparrow}^{sff}(E) &= T_{\downarrow\uparrow}^f T_{\downarrow}(E) \{1-\int D_{-\frac{3}{2}}(\xi) F_{-\frac{3}{2}}(\xi)d\xi\} \nonumber \\
  T_{\uparrow\downarrow}^{sff}(E) &=T_{\uparrow\downarrow}^f T_{\uparrow}(E) \{1-\int D_{\frac{3}{2}}(\xi)F_{\frac{3}{2}}(\xi)d\xi\} \nonumber \\
   T_{\downarrow\uparrow}^{sfb}(E) &= T_{\downarrow\uparrow}^b T_{\downarrow}(E) \{1-\int D_{-\frac{3}{2}}(\xi)F_{-\frac{3}{2}}(\xi)d\xi\} \nonumber \\
  T_{\uparrow\downarrow}^{sfb}(E) &=T_{\uparrow\downarrow}^b T_{\uparrow}(E) \{1-\int D_{\frac{3}{2}}(\xi)F_{\frac{3}{2}}(\xi)d\xi\} \nonumber \\
\label{eq:nmr4}    
  \end{align}
  
  \begin{eqnarray}
  I_{  \uparrow\downarrow}^{sf}=\int\frac{e}{h} \{T_{\uparrow\downarrow}^{sff}(E)+T_{\uparrow\downarrow}^{sfb}(E)\} \{ f_S(E)-f_D(E)\}dE \nonumber \\
  I_{\downarrow  \uparrow}^{sf}=\int \frac{e}{h} \{ T_{\downarrow\uparrow}^{sff}(E)+T_{\downarrow\uparrow}^{sfb}(E)\} \{ f_S(E)-f_D(E)\}dE, \nonumber \\
\label{eq:nmr5}    
  \end{eqnarray}
  where $\left[s\right]=\left[\frac{3}{2}~~\frac{1}{2}~~-\frac{1}{2}~~-\frac{3}{2}\right]$ is the row vector denoting four nuclear spin levels in GaAs and $[F(\xi)]=\left[F_{\frac{3}{2}}(\xi)~~F_{\frac{1}{2}}(\xi)~~F_{-\frac{1}{2}}(\xi)~~F_{-\frac{3}{2}}(\xi)\right]^{\dagger}$. $F_s(\xi)$ is the probability of occupancy of the density of states of the $s^{th}$ nuclear spin level $D_s$ at energy $\xi$. The matrix $D_N(\xi)$ is the diagonal  matrix representing the nuclear density of states at energy $\xi$ given by:
\[
D_N(\xi)=\begin{bmatrix}
 D_{\frac{3}{2}}(\xi) & 0 & 0 & 0 \\ 
 0 & D_{\frac{1}{2}}(\xi) & 0 & 0 \\ 
 0 & 0 & D_{-\frac{1}{2}}(\xi) & 0 \\ 
 0 & 0 & 0 & D_{-\frac{3}{2}}(\xi)
\end{bmatrix}
\] 
  At low temperatures, $\{ f_S(E)-f_D(E)\}$ is a boxcar function. Assuming that the average value of $T_{\uparrow}$ and $T_{\downarrow}$ in the energy range between $\mu_S$ and $\mu_D$ are $T_{\uparrow}^{avg}$ and $T_{\downarrow}^{avg}$ respectively and the spin-flip transmission coefficients, $T_{\uparrow\downarrow}^{f(b)}$ and $T_{\downarrow\uparrow}^{f(b)}$, are constant in the range of energy between $\mu_S$ and $\mu_D$, the above equation can be simplified to, 
  
  \begin{eqnarray}
  I_{  \uparrow\downarrow}^{sf}=\frac{e}{h} T_{\uparrow}^{avg}\{ T_{\uparrow\downarrow}^{f}+ T_{\uparrow\downarrow}^{b}\}\times\{1-\int D_{\frac{3}{2}}(\xi)F_{\frac{3}{2}}(\xi)d\xi\} \Delta \mu \nonumber \\ =\frac{e^2}{h} T_{\uparrow}^{avg} T_{\uparrow\downarrow}^{avg}\times\{1-\int D_{\frac{3}{2}}(\xi)F_{\frac{3}{2}}(\xi)d\xi\} V \nonumber \\
  I_{\downarrow  \uparrow}^{sf}=\frac{e}{h} T_{\downarrow}^{avg}\{ T_{\downarrow\uparrow}^{f}+T_{\downarrow\uparrow}^{b} \}\times\{1-\int D_{-\frac{3}{2}}(\xi)F_{-\frac{3}{2}}(\xi)d\xi\} \Delta \mu \nonumber \\=\frac{e^2}{h} T_{\downarrow}^{avg} T_{\downarrow\uparrow}^{avg}\times\{1-\int D_{-\frac{3}{2}}(\xi)F_{-\frac{3}{2}}(\xi)d\xi\} V, \nonumber \\
\label{eq:nmr5p1}    
  \end{eqnarray}
where $\Delta \mu=\mu_S-\mu_D$. The set of equations  \eqref{eq:nmr2}-\eqref{eq:nmr5} have to be solved self-consistently to calculate the temporal evolution of the nuclear polarization.  We now turn our attention towards  \eqref{eq:nmr2}. The first term on the right hand side of \eqref{eq:nmr2} is given by:
    
    \begin{multline}
  \left[ \frac{dF(\xi)}{dt}\right]_{flip-flop}=\left[\Gamma_{out}\right]\times [F(\xi)] \nonumber \\
 + \{[I_4]-[F_{diag}(\xi)]\} \times[P_{diag}]^{-1}\left[\Gamma_{in}\right] [N] , 
\label{eq:fllip}    
  \end{multline}
where $I_4$ is the identity matrix of the fourth  order and [$N_{diag}$], [$P_{diag}$], $\left[\Gamma_{in}\right]$ and $\left[\Gamma_{out}\right]$ are given by:
\begin{widetext}
\begin{equation} 
[F_{diag}(\xi)]=\begin{bmatrix}
 F_{\frac{3}{2}}(\xi) & 0 & 0 & 0 \\ 
 0 & F_{\frac{1}{2}}(\xi) & 0 & 0 \\ 
 0 & 0 & F_{-\frac{1}{2}}(\xi) & 0 \\ 
 0 & 0 & 0 & F_{-\frac{3}{2}}(\xi)
\end{bmatrix} 
 \end{equation}
 \begin{equation}
 [N]=\int\left[  D_{\frac{3}{2}}(\xi)F_{\frac{3}{2}}(\xi)~~D_{\frac{1}{2}}(\xi)F_{\frac{1}{2}}(\xi)~~D_{-\frac{1}{2}}(\xi)F_{-\frac{1}{2}}(\xi)~~D_{-\frac{3}{2}}(\xi)F_{-\frac{3}{2}}(\xi)\right]^{\dagger}d\xi
 \end{equation}
 \begin{equation}
P_{diag}=\begin{bmatrix}
\int D_{\frac{3}{2}}(\xi)\{1-F_{\frac{3}{2}}(\xi)\}d\xi & 0 & 0 & 0 \\ 
 0 & \int D_{\frac{1}{2}}(\xi)\{1-F_{\frac{1}{2}}(\xi)\}d\xi & 0 & 0 \\ 
 0 & 0 & \int D_{-\frac{3}{2}}(\xi)\{1-F_{-\frac{3}{2}}(\xi)\}d\xi & 0 \\ 
 0 & 0 & 0 & \int D_{-\frac{3}{2}}(\xi)\{1-F_{-\frac{3}{2}}(\xi)\}d\xi
\end{bmatrix} 
 \end{equation}
\begin{equation}
\left [ \Gamma_{in} \right ] =\begin{bmatrix}
0 & C_1|I_{\uparrow\downarrow}^{sf}|  & 0 & 0 \\
C_2|I_{\downarrow\uparrow}^{sf}| & 0 & C_1|I_{\uparrow\downarrow}^{sf}|&  0 \\
0 & C_2|I_{\downarrow\uparrow}^{sf}| & 0 & C_1|I_{\uparrow\downarrow}^{sf}|   \\
0 & 0 & C_2|I_{\downarrow\uparrow}^{sf}| & 0.
\end{bmatrix}
\end{equation}

\begin{equation}
\left [ \Gamma_{out} \right ] =\begin{bmatrix}
-C_2|I_{\downarrow\uparrow}^{sf}| & 0  & 0 & 0 \\
0 & -\left (C_2|I_{\downarrow\uparrow}^{sf}|+ C_1|I_{\uparrow\downarrow}^{sf}| \right ) & 0 &  0 \\
0 & 0 & -\left (C_2|I_{\downarrow\uparrow}^{sf}|+ C_1|I_{\uparrow\downarrow}^{sf}| \right ) & 0  \\
0 & 0 & 0 & -C_1|I_{\uparrow\downarrow}^{sf}|.
\end{bmatrix}
\end{equation}

\end{widetext}
The second term on the right hand side of \eqref{eq:nmr2} arises  due to nuclear spin lattice relaxation and is given by:
\begin{equation}
  \left[ \frac{dF(\xi)}{dt}\right]_{relaxation}=\frac{[F(\xi)]-[F^0]}{\tau_I} \nonumber
\end{equation}
The third term on the right hand side of \eqref{eq:nmr2} arises due to perturbation via an externally applied RF field.   \color{black} If the coherence between the nuclear spins is neglected, then the rate of decay of nuclear polarization with time due  to perturbation via an externally applied RF field can be characterized phenomenologically by a time constant $\tau_{NMR}$. The rate of decay of nuclear polarization with RF frequency sweep without taking into account the correlation between nuclear spins is given by the equation:
\scriptsize
\begin{equation}
\left [ \frac{dF(\xi)}{dt} \right ]_{NMR} =diag \left(\frac{1}{\tau_{NMR}}\frac{\pi \eta}{2} \times [\Gamma_{NMR}(\xi) ]\times [F_{NMR}(\xi,\hslash \omega)]\right). \nonumber
\label{eq:nuc_eqn}
\end{equation}
\normalsize
where $\frac{\pi \eta}{2}$ acts as a normalization constant, $\eta$ being the broadening of the nuclear density of states (details given in Appendix D). The matrix $\Gamma_{NMR}(\xi)$ takes into account the net rate of transition between consecutive nuclear spin levels depending on the energy of the RF frequency photons ($\hslash \omega$) and is given by (details given in Appendix \ref{appendix3}): 
\begin{widetext}
\scriptsize
\begin{eqnarray}
\Gamma_{NMR}(\xi)=\begin{bmatrix}
-{D_{\frac{1}{2}}(\xi-\hslash \omega)} & {D_{\frac{1}{2}}(\xi-\hslash \omega)}  & 0 & 0 \\
{D_{\frac{3}{2}}(\xi+\hslash \omega)} & -\{D_{\frac{3}{2}}(\xi+\hslash\omega)+D_{-\frac{1}{2}}(\xi-\hslash \omega)\} & {D_{-\frac{1}{2}}(\xi-\hslash \omega)} &  0 \\
0 & {D_{\frac{1}{2}}(\xi+\hslash\omega)} & -\{D_{\frac{1}{2}}(\xi+\hslash \omega)+D_{-\frac{3}{2}}(\xi-\hslash\omega)\} & {D_{-\frac{3}{2}}(\xi-\hslash\omega)}  \\
0 & 0 & {D_{-\frac{1}{2}}(\xi+\hslash\omega)} & -{D_{-\frac{1}{2}}(\xi+\hslash\omega)}
\end{bmatrix} \nonumber \\ \nonumber
\end{eqnarray}
\begin{eqnarray}
F_{NMR}(\xi,\hslash\omega)=
\begin{bmatrix}
F_{\frac{3}{2}}(\xi) & F_{\frac{3}{2}}(\xi+\hslash \omega) & 0 & 0 \\
F_{\frac{1}{2}}(\xi-\hslash \omega) & F_{\frac{1}{2}}(\xi) & F_{\frac{1}{2}}(\xi+\hslash \omega) &  0 \\
0 &  F_{-\frac{1}{2}}(\xi-\hslash \omega) & F_{-\frac{1}{2}}(\xi) & F_{-\frac{1}{2}}(\xi+\hslash \omega) \\
0 & 0 & F_{-\frac{3}{2}}(\xi-\hslash \omega) & F_{-\frac{3}{2}}(\xi)
\end{bmatrix} \nonumber \\
\end{eqnarray}
 
\end{widetext}
with  $\tau_{NMR}$ being of the order of  $100\mu s$ \citep{APL_Edge_Scatt_ORIG,APL_Edge_Scatt}.  It can be shown that when $\eta \rightarrow 0$, in the absence of RF field perturbation, \eqref{eq:nmr2} becomes identical to \eqref{eq:master}. \color{black} We now present some simulation results based on the above model. Specifically, we present results for three different sweep times in Fig. \ref{fig:nmr1} and show how the rate of sweep of RF frequency  influence the hysteresis observed in the $R_H$ traces.

  \begin{figure}[!htb]

\subfigure[]{\includegraphics[scale=0.15]{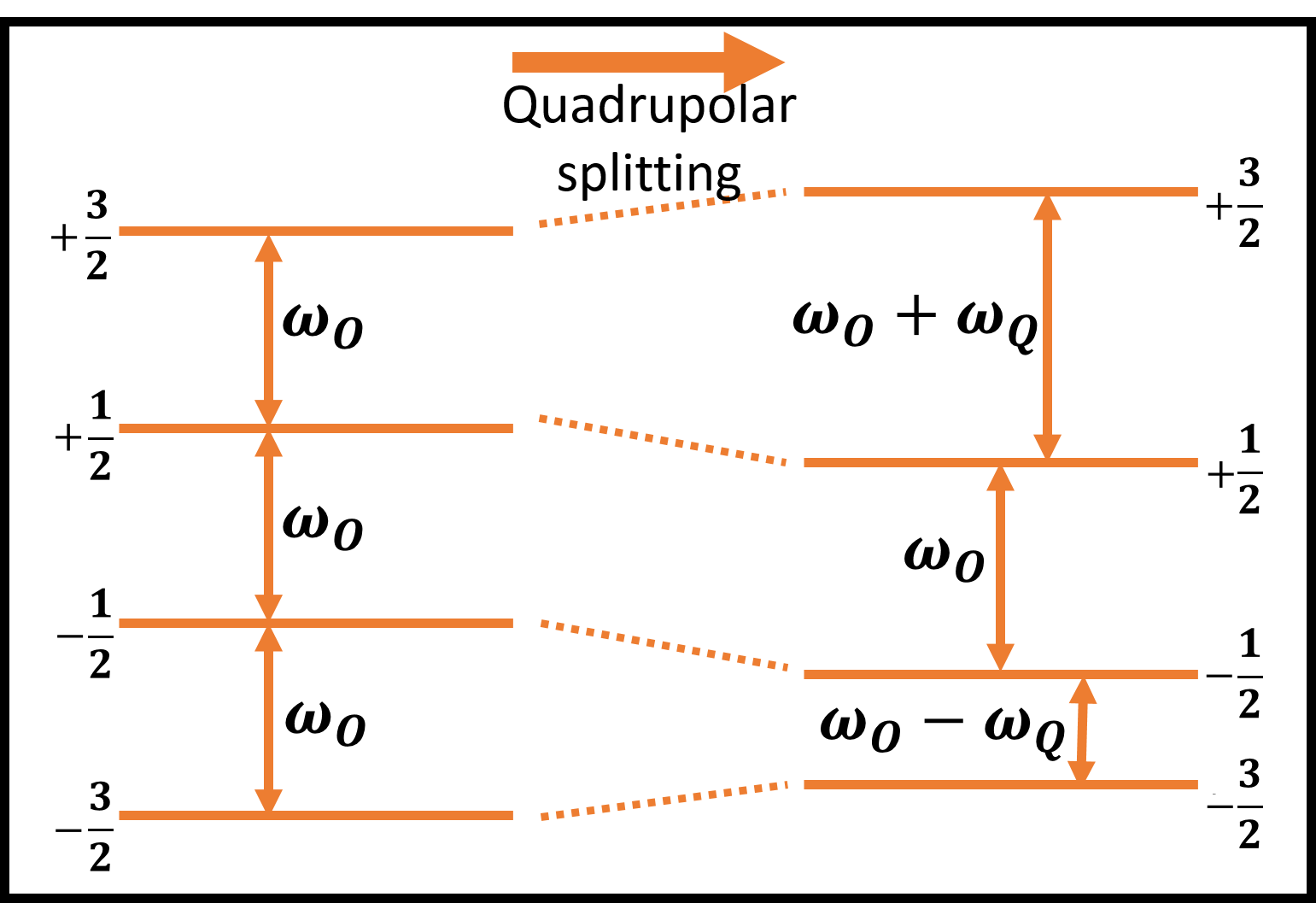}}

\begin{minipage}[b]{0.25\textwidth}

\subfigure[]{\includegraphics[scale=.06]{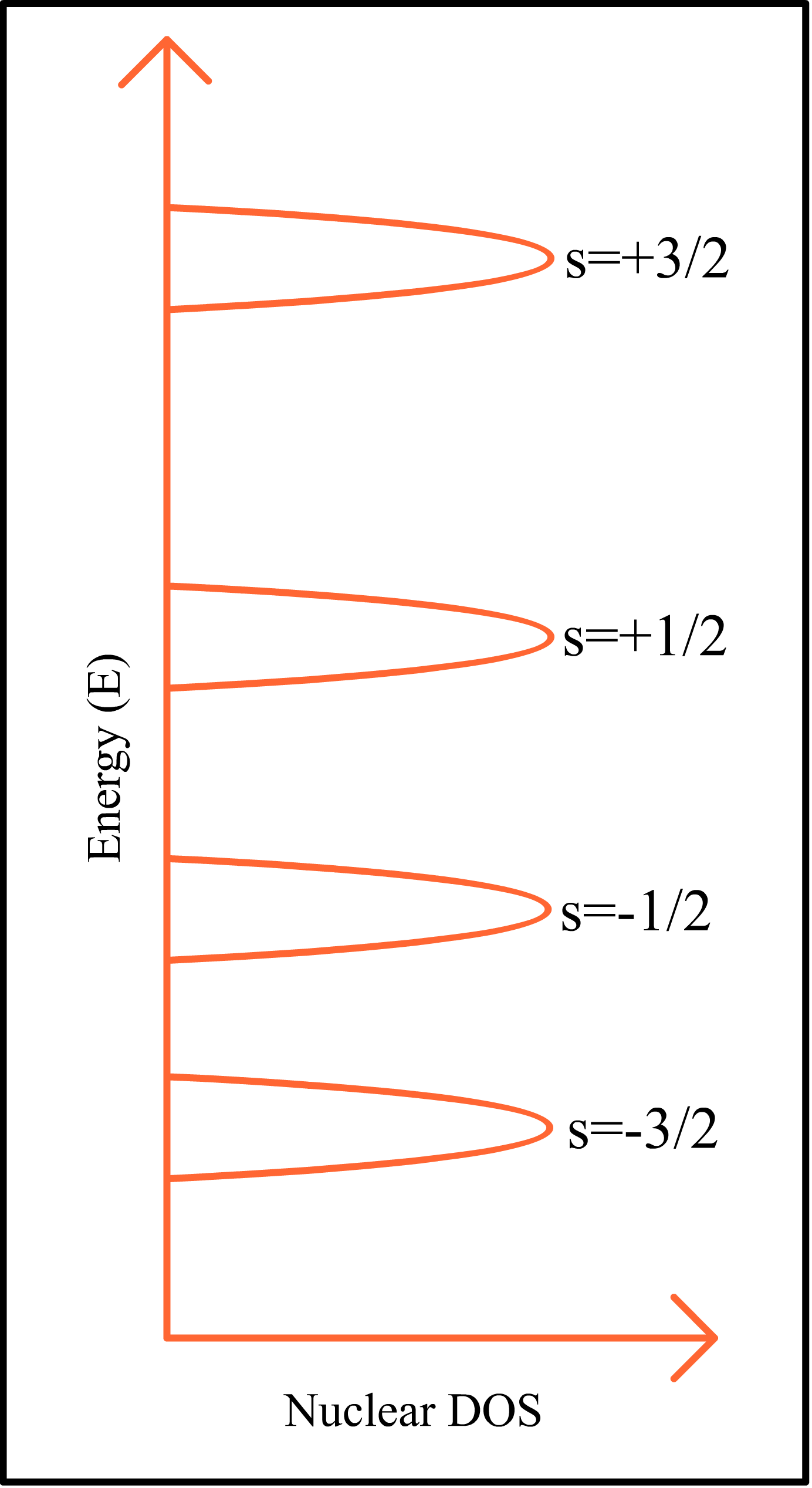}
}

\subfigure[]{\includegraphics[scale=.174]{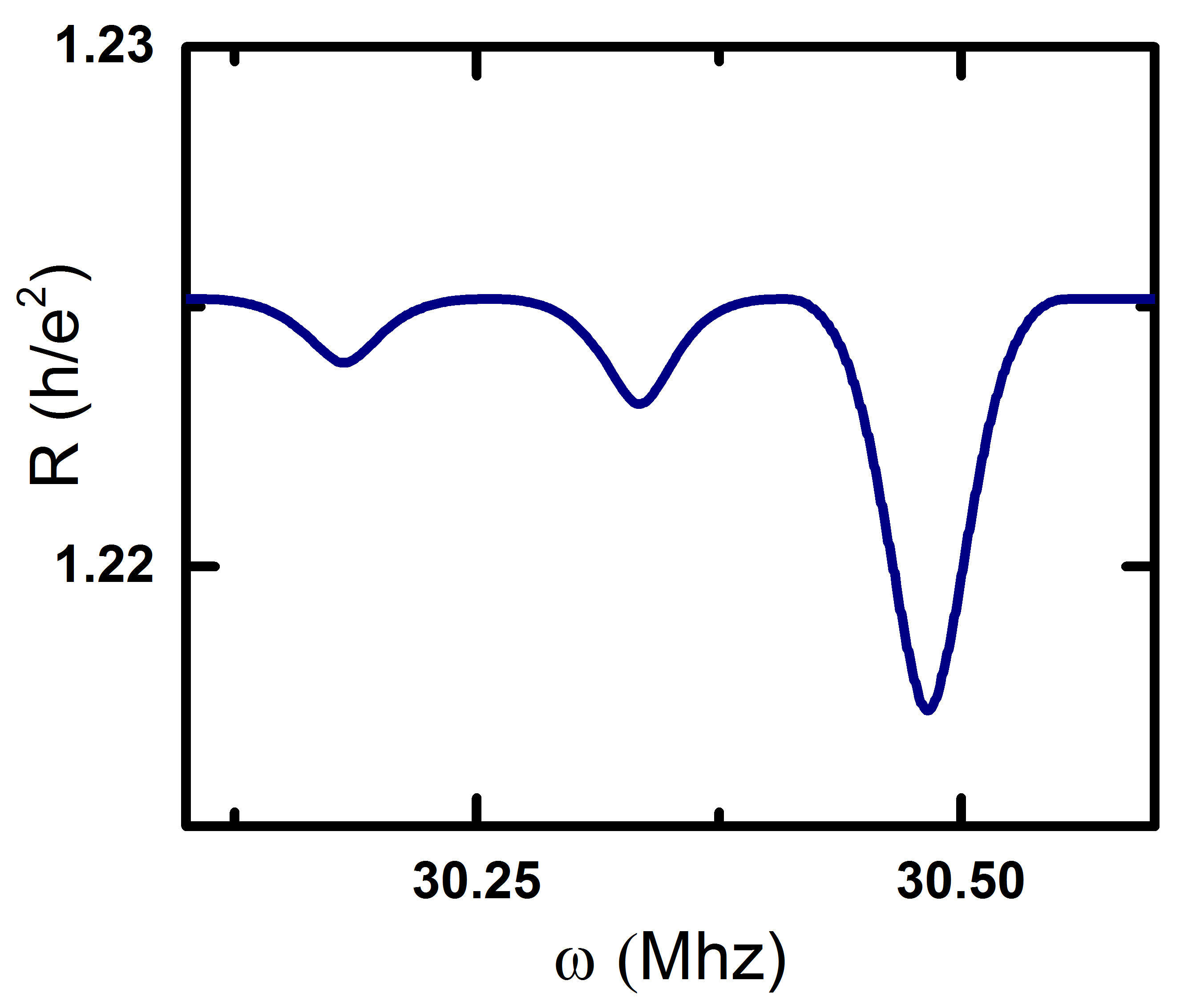}
}

\end{minipage}%
\begin{minipage}[b]{0.25\textwidth}

\subfigure[]{\includegraphics[scale=.06]{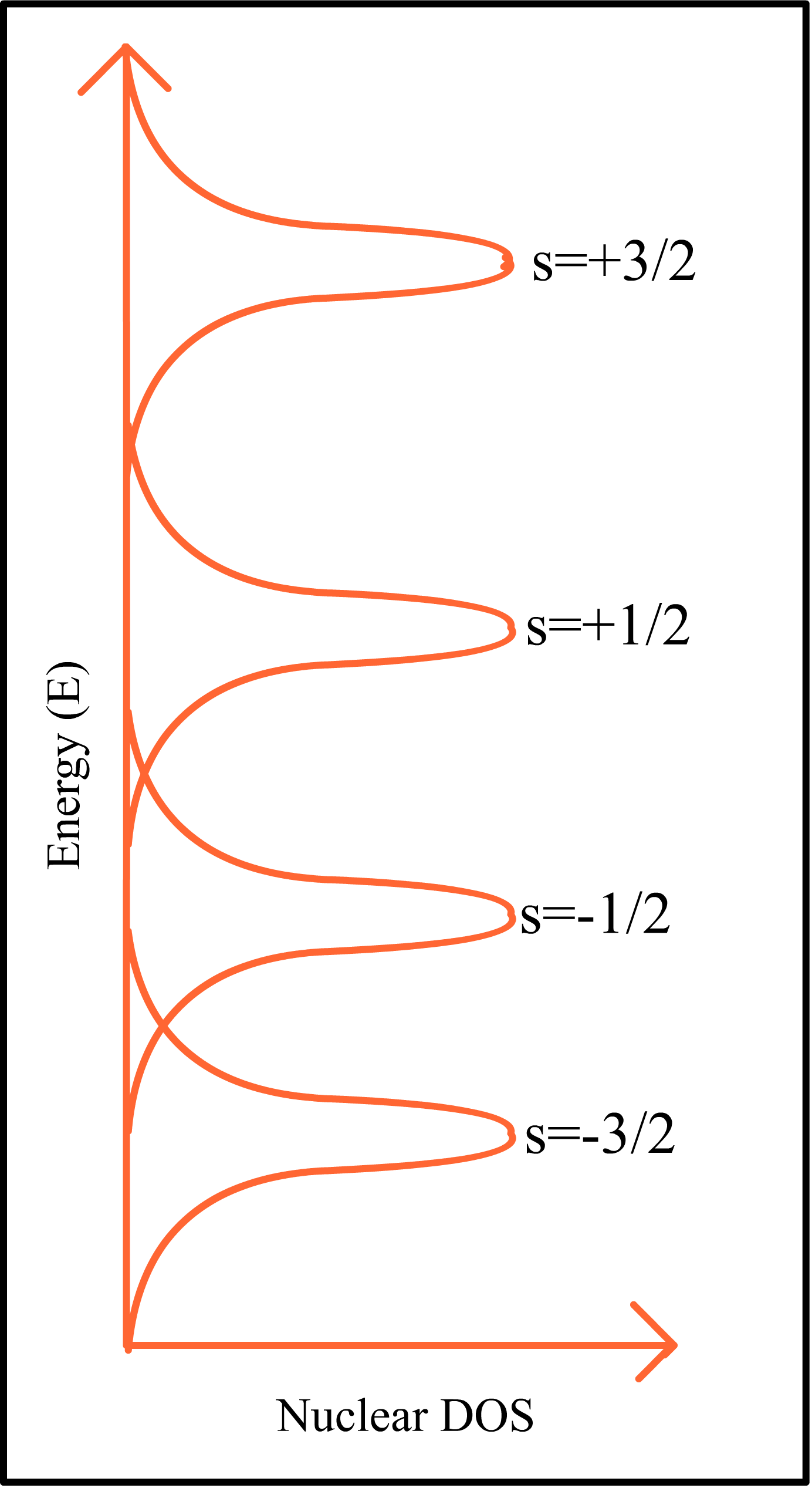}
}

\subfigure[]{\includegraphics[scale=.174]{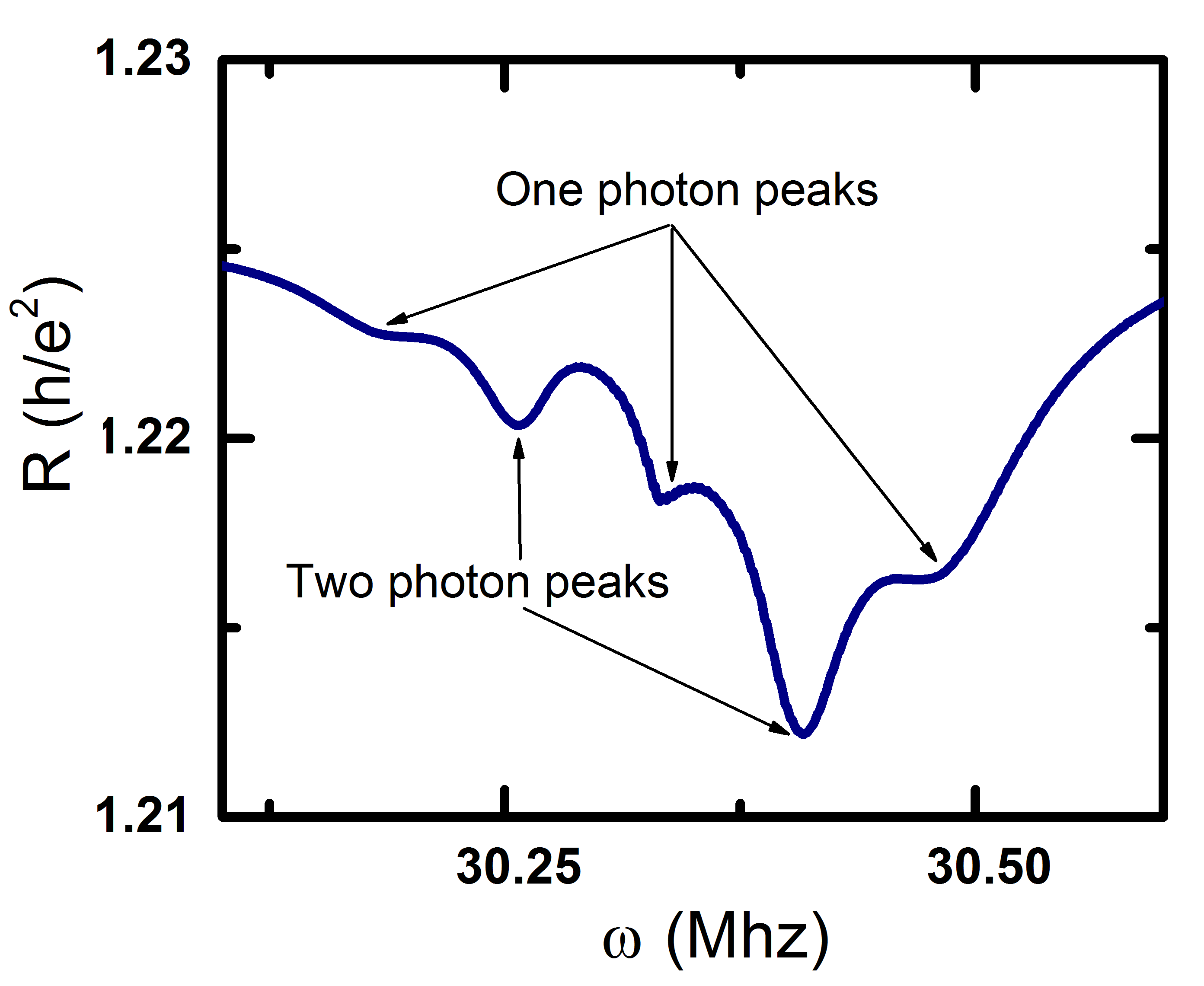}
}
\end{minipage}%

\caption{Signatures of quadrupolar splitting on RDNMR traces. (a) Schematic diagram showing  nuclear spin levels unevenly split due to quadrupolar splitting.  (b) Schematic of broadened nuclear density of states without any intermediate state for multi-photon absorption (c) Plot showing  one photon peaks in the conductance  with RF frequency sweep in case of nuclear quadrupolar splitting  for $R>\frac{h}{e^2}~(\nu_{QPC}<1)$.  \color{black}  (d) Schematic broadened nuclear density of states with  infinitesimally small nuclear density of states added between the peaks in the density of states to aid  multiphoton transitions. (e)  Plot showing  single photon and multi photon  peaks in the conductance  with RF frequency sweep in case of strain induced quadrupolar splitting  for $R>\frac{h}{e^2}~(\nu_{QPC}<1)$ . The two photon processes in our simulation can be accounted for by incorporating infinitesimally small nuclear density of states between the peaks of the nuclear density of states. Simulations are done for $B=4T$, $N_I=10^9$, $~T^f_{\uparrow\downarrow}=0.01,~T^b_{\uparrow\downarrow}=0,~T^f_{\downarrow\uparrow}=0,~T^b_{\downarrow\uparrow}=0.01$. $T_{\uparrow}$ and $T_{\downarrow}$ are calculated using NEGF formalism as elaborated in the text}
\label{fig:nmr2}
\end{figure}
We note that the hysteresis is a result of the slow buildup of nuclear polarization. The hysteresis disappears if the sweep rate is much slower compared to the rate of buildup of nuclear polarization. On the other hand, hysteresis is pronounced  if the sweep rate is much faster compared to the rate of buildup of nuclear polarization.  %To simulate the effect of Overhauser field on conductance during NMR frequency sweep experiments, we use model the QPC as per Eq. \eqref{eq:overhausereffect}. 
 The change in conductance during NMR frequency sweep is mainly due to the effect of the Overhauser field on the transmission coefficient of the spin channels at the QPC as well as the spin-flip scattering at the QPC. For $R>\frac{h}{e^2}~(\nu_{QPC}<1)$, the up-spin electrons in the forward propagating edge channel terminating in the drain contact can scatter at the QPC to the forward propagating down-spin edge channel terminating in the drain contact  with a spin-flip scattering. Such electron-nuclear spin flip-flop processes create a positive nuclear polarization which opposes the effect of the magnetic field on the up-spin channel by enhancing it's potential barrier  at the QPC,  resulting in a decrease in the transmission coefficient
of the up-spin channel.  Destruction of nuclear polarization via NMR frequency perturbation hence results in a  decrease in the up-spin channel resistivity  due to an increase in the transmission coefficient 
  (decrease in the Overhauser field).\\
 \indent For $R<\frac{h}{e^2}~~(\nu_{QPC}>1)$, the electrons in the forward propagating down-spin edge channel originating in the source contact  can transmit partially to the down-spin edge channel terminating in the drain contact. The forward propagating down-spin edge channel originating in the source contact  being partially transmitted  through the QPC, two spin flip-flop scattering mechanism may be dominant  at and around the QPC-(i) A net spin-flip scattering from the forward propagating up-spin edge channel terminating in the drain contact to the forward propagating down-spin edge channel terminating in the drain contact. Such spin-flip scattering creates positive nuclear polarization. (ii) A net spin-flip  scattering from forward propagating down-spin edge channel terminating in the drain contact to backward propagating up-spin edge channel terminating in the source contact (the forward propagating up-spin channel is full). Such spin-flip scattering creates negative nuclear polarization and also decreases the total current through the QPC. At $R\approx \frac{h}{e^2}-\epsilon$, most of the electron  spin flipping at the QPC occurs from the forward propagating up-spin edge channel terminating in the drain contact  to the forward propagating down-spin edge  channel terminating in the drain contact (since the majority of electrons  at the QPC occupy the up-spin channel) which creates net positive nuclear polarization. Although positive nuclear polarization influences the transmission coefficient of both down spin and up-spin channel, the relative feedback on the up-spin channel is much less compared to that on the  down-spin channel. Positive nuclear polarization enhances the transmission coefficient of the down-spin edge channel through the QPC. Destruction of nuclear spin polarization during NMR frequency perturbation hence results in an increase in the  resistivity of the down-spin edge channel due to decrease in  transmission coefficient of the down-spin channel through the QPC. The plots of Hall resistance vs. RF frequency sweep for the two cases (a) $R>\frac{h}{e^2}~(\nu_{QPC}<1)$ (b) $R<\frac{h}{e^2}~(\nu_{QPC}>1)$ are shown in Fig. \ref{fig:nmr1}. % An interesting case occurs when both positive and negative nuclear polarization are present but are localized in different regions near the QPC. This situation is shown schematically in Fig. . We speculate that a net positive nuclear polarization occurs in 'region A' where the filling factor is $\nu=1$, while a net negative nuclear polarization is more likely to occur near 'region B' where $\nu=2$. 'Region A' hence experiences a net decrease in nuclear magnetic resonance frequency due to a negative 'Knight field'. The electronic wavefunctions being spatially delocalized might be influenced by the nuclear polarization in both 'region A' and 'region B'. Assuming such scenario, we get a dispersive $dI/dV ~vs.~\omega$ plot which has been a characteristic feature of several experimental measurements. The peak and dip of the dispersive plot swaps position as the conductance of the QPC is changed from $G>\frac{e^2}{h}$ to $G<\frac{e^2}{h}$. Such phenomenon have been noted in experiments. Currently the concrete explanation of the causative factors of such dispersive feature is still in infancy and lots of experiments are being carried out for overall understanding of such features. Our model is however useful for explaining the basic features of such plots. 
\subsection*{Capturing quadrupolar peaks}\label{subsec:QP}

%It can be seen that the conductance plot in Fig. near the resonance frequency initially increases with frequency sweep followed by a sharp decrease . Such nature of the conductance vs. frequency curve arises from the interplay between increase in the direct transmission coefficient due to decrease in Overhauser field and increase in spin flip transition to backward propagating channel due to destruction in nuclear polarization (see section \ref{appendix1}). Initially, an increase in spin flip transition due to backward propagating up-spin channel decreases the overall current resulting in an increase in resistance. On further destruction of nuclear polarization, the overall transmission function increases due to decrease in Overhauser field leading to a decrease in resistance. \subsection{Extension to quad nuclear basis}
%The above set of simulations can be easily extended to a system with quad nuclear spin basis. Here we carefully include the interaction between each set of energy levels and self consistently calculate the change in spin flip current due to drop/build up of nuclear polarization. Lets now see the results for simulation with quad nuclear spin levels for very slow frequency sweeps. Fig. \ref{fig:quad1}, \ref{fig:quad2}, \ref{fig:quad3} shows the result of our simulations for three different cases. We see that the change in conductance during the NMR sweeps is dependent on the nuclear polarization because of Overhauser field.

\indent A current focus area in quantum computing is the possibility of information processing via independent manipulation of the individual nuclear spin levels. Materials with four nuclear spin levels (GaAs) and nine nuclear spin levels (InAs) offer attractive options since multi-bits of information can be stored and manipulated. An attractive method to manipulate the individual spin levels is introducing  quadrupolar splitting which creates a slight difference in the splitting energy of consecutive nuclear spin levels \cite{kawamura_quadru,Yusa,multibits1,multibits2}.\\ % A suitable method of introducing strain based quadrupolar splitting is to coat a thin film polyimide  on GaAs quantum Hall device at higher temperature \cite{kawamura_quadru}. When such polyimide coated devices are cooled to lower temperatures, the difference in thermal shrinkage coefficient of polyimide and GaAs films introduces a strain in GaAs devices. \\ 
\indent The aforesaid effects are taken into account by carefully incorporating transitions between the individually broadened nuclear spin levels using \eqref{eq:nuc_eqn}. A schematic of the broadened nuclear density of states and simulated conductance variation with RF frequency sweep is shown in Fig. \ref{fig:nmr2} (b) and (c) respectively.  The simulation model can be further extended to show peaks corresponding to multi-photon absorption \cite{kawamura_quadru,Yusa}. Such peaks occur at higher RF powers which induce higher rates of scattering and consequently larger broadenings of nuclear levels. This creates a density of states at energies where nuclear population is otherwise absent. Transitions corresponding to two consecutive photon absorptions can be included in our model by adding an infinitesimally small nuclear density of states  in between the peaks of the nuclear density of states. Such an infinitesimally small density of states does not influence the total nuclear polarization but provides intermediate levels with extremely small lifetimes to facilitate an intermediate transition before absorbing the second photon. \\
\indent A schematic diagram of nuclear density of states is shown in  Fig. \ref{fig:nmr2} (d) where a small nuclear density of states has been added in between the main peaks to account for the multi-photon transitions. The plot of conductance versus RF frequency sweep with single photon and multi-photon peaks is shown in \ref{fig:nmr2}(e). Although the magnitude of conductance change during the multi-photon processes should ideally depend on the amount of RF power being supplied, taking into account such variation requires book keeping of the details of photon density in space and the interaction between photons and nuclear spin levels. We leave these considerations for a future work.
\section{Conclusion}
In this paper, we have developed a Landauer-B\"uttiker approach to understand various experimental features observed in integer quantum Hall set ups featuring QPCs. Starting from the Fermi contact hyperfine Hamiltonian, we have developed physics based models to incorporate electron-nuclear spin flip-flops in to an extended Landauer-B\"uttiker formalism to describe the edge state electronic transport near the QPC region. This self-consistent simulation framework between the nuclear spin dynamics and edge state electronic transport aided a theoretical investigation of the hysteresis in the conductance-voltage and RDNMR lineshapes noted in certain experiments \cite{Wald_PRL,rdnmr1,rdnmr2,Nano_Hamilton,rdnmr4}.
In particular, we demonstrated that the hysteresis noted experimentally results from a lack of quasi-quilibrium between electronic transport and nuclear polarization evolution. In addition, we presented circuit models to emulate such hyperfine mediated transport effects to further facilitate a clear understanding of the electronic transport processes occurring near the QPC. Finally, we extended our models to account for the effects of quadrupolar splitting of nuclear levels and also depict the electronic transport signatures that arise from single and multi-photon processes. We believe that this work sets stage for a more rigorous approach which will include a self-consistent solution of the potential profile\cite{Glazman,poisson1,poisson3,poisson4,poisson2,poisson5,poisson6,poisson7,poisson8,poisson9,poisson10} of the channel along with the spatial distribution the nuclear spin profile. \\
{\it{Acknowledgements:}}  The authors AS and BM have benefitted from various discussions with Kantimay Das Gupta and would like to acknowledge funding from the Department of Science and Technology, India under the Science and Engineering Board (SERB) grant no. SERB/F/3370/2013-2014. The authors YH and BM acknowledge support from WPI-AIMR, Tohoku University. The authors AS and YH acknowledge support from GP-Spin, Tohoku University. The author MHF acknowledges funding from MD program, Tohoku University and the author YH acknowledges funding from KAKENHI (Grant Nos. 26287059 and 15H05867) and CSRN, Tohoku University.
 
\appendix \label{appendix}

%\section{Appendix} \label{appendix}

\section{Derivation of the spin-flip transmission coefficient.} \label{appendix1}
In this section, we show that  the rate of spin-flip scattering from the forward propagating up-spin (down-spin) channel originating in the source contact to the forward propagating  down-spin (up-spin) channel terminating in the drain contact per unit energy at the QPC can indeed be approximated by a transmission coefficient that is dependent on the nuclear polarization.
The region around the QPC can be modeled by a smooth Gaussian/bell shaped potential barrier \cite{qpcmodel} which varies along the transport direction as well as along the direction perpendicular to transport. However, to model transport through a QPC, we need to know the minimum potential  along the transport direction \cite{aono,qpcmodel}.  If both $\mu_D$ and $\mu_S$ lie above the top of the barrier, $T_{\uparrow (\downarrow)}(E)\approx 1$ in the range of energy over which electronic transport takes place. On the other hand, if the top of the barrier lie between $\mu_S$ and  $\mu_D$  the channel is partially transmitted. 

Let us assume that the forward propagating down-spin edge channel originating in the source contact is totally isolated from forward propagating down-spin edge channel terminating in the drain contact. A major part of the up-spin edge channel originating  in the source contact is however transmitted through the QPC to the up-spin edge channel terminating in the drain contact. A very small portion of the electrons in the  forward propagating  up-spin edge channel terminating in the drain contact however suffers a spin-flip scattering at the QPC and is  transmitted to the  forward propagating  down-spin edge channel terminating in the drain contact via electron-nuclear spin flip-flop process. Such processes at the QPC gives rise to nuclear polarization which in turn influences electronic transport via Overhauser field. 

The forward propagating up-spin edge  channel near the QPC is characterized by  $n_{\uparrow}^{in}(E)$  and $v_{\uparrow}(E)$ which are  the density of electrons per unit energy per unit length and velocity of electrons  respectively at the forward propagating up-spin edge channel near the QPC. Therefore,
\[
n_{\uparrow}^{in}(E)v_{\uparrow}(E)=\frac{1}{h}.
\]
A portion of the electrons in the forward propagating  up-spin edge channel is reflected while the rest is transmitted with or without a spin-flip. Since the total current is conserved,
\begin{eqnarray}
n_{\uparrow}^{in}(E)v_{\uparrow}(E)=n_{\uparrow}^{t}(E)v_{\uparrow}^{ch}(E)
+n_{\uparrow}^{R}(E)v_{\uparrow}(E) \nonumber 
\end{eqnarray}
\begin{gather}
\frac{1}{h}=\frac{1}{h}T_{\uparrow}(E)+\frac{1}{h}{R}(E) \nonumber \\
T_{\uparrow}(E)=1-R(E) \nonumber \\
\end{gather}
In the above set of equations $n_{\uparrow}^{t}(E)$ and $n_{\uparrow}^R(E)$ denote the density of up-spin electrons in the forward propagating edge channel  at the QPC at energy $E$,   and density of  electrons that are reflected  to the backward propagating edge-channel terminating in the source contact. In a similar fashion, $v_{\uparrow}^{ch}(E)$ and $v_{\uparrow}(E)$ are the group velocity of up-spin electrons in the forward propagating edge channel (terminating in the drain contact) at the QPC and the group velocity of up-spin electrons in the forward or backward propagating up-spin edge channel far away from the QPC respectively.  
Assuming the QPC to be a single point, let $n_{\downarrow}^{ch}(E)v_{\downarrow}^{ch}(E)$ be the total number of electrons transmitted per unit time per unit energy at the QPC from the forward propagating up-spin edge channel terminating in the drain contact to the forward propagating down-spin edge channel terminating in the drain contact. So,
\[
n_{\uparrow}^t(E)v_{\uparrow}^{ch}(E)=n_{\uparrow}^{ch}(E)v_{\uparrow}^{ch}(E)+n_{\downarrow}^{ch}(E)v_{\downarrow}^{ch}(E).
\]
If the rate of electron-nuclear spin flip-flop at the QPC is much less compared to the rate at which electrons are transmitted from the up-spin edge channel originating in the source contact to the up-spin edge channel terminating in the drain contact  through the QPC, then, $n_{\uparrow}^{ch}(E) \approx n_{\uparrow}^{t}(E)$.
Our intention is to derive the approximate form for $T_{\uparrow\downarrow}^{sff}$. \\
\indent The flow of electrons between the up-spin edge channel originating  in the source contact to the down-spin edge channel terminating in the drain contact at the QPC  is
\begin{eqnarray}
I^{sf}(E)=I_{\uparrow \downarrow}^{sf}(E)-I_{\downarrow \uparrow}^{sf}(E)
=qN_I\times \left[\frac{3}{2}~~~ \frac{1}{2}~ -\frac{1}{2}~ -\frac{3}{2}\right]~  \nonumber \\\times \bigg\{\frac{dF(E)}{dt}\bigg\}_{conserving}  \nonumber \\
 =qN_I\times \Big[-\frac{3}{2} \Gamma_{\downarrow \uparrow}(E) F_{\frac{3}{2}}+\frac{3}{2} \Gamma_{\uparrow \downarrow}(E) F_{\frac{1}{2}} +\frac{1}{2} \Gamma_{\downarrow \uparrow}(E) F_{\frac{3}{2}} \nonumber \\  -\frac{1}{2} \Big \{\Gamma_{\downarrow \uparrow}(E) 
 + \Gamma_{\uparrow \downarrow}\Big \} F_{\frac{1}{2}}(E) 
+\frac{1}{2}\Gamma_{\uparrow \downarrow}(E) F_{-\frac{1}{2}}-\frac{1}{2} \Gamma_{\downarrow \uparrow}(E) F_{\frac{1}{2}}\nonumber \\ +\frac{1}{2}\Big \{\Gamma_{\downarrow \uparrow}(E)  
+ \Gamma_{\uparrow \downarrow}(E)\Big \}F_{-\frac{1}{2}}
-~\frac{1}{2}\Gamma_{\uparrow \downarrow}(E)F_{-\frac{3}{2}}\nonumber \\-\frac{3}{2} \Gamma_{\downarrow \uparrow}(E)F_{-\frac{1}{2}} 
+\frac{3}{2}\Gamma_{\uparrow \downarrow}(E)F_{-\frac{3}{2}} \nonumber \Big]
\end{eqnarray}
\begin{eqnarray}
=\Big [ -\Gamma_{\downarrow\uparrow}(E)F_{\frac{3}{2}}+\Big \{\Gamma_{\uparrow\downarrow}(E)-\Gamma_{\downarrow\uparrow}(E)\Big \}F_{\frac{1}{2}}+\Big \{\Gamma_{\uparrow\downarrow}(E) \nonumber \\ 
-\Gamma_{\downarrow\uparrow}(E)\Big\}F_{-\frac{1}{2}} 
+\Gamma_{\uparrow\downarrow}(E)F_{-\frac{3}{2}} \Big] \times q N_I. \nonumber \\
\label{eq:flip1}
\end{eqnarray}

$\Gamma_{\uparrow \downarrow}(E)$ and $\Gamma_{\downarrow \uparrow}(E)$ are given by:

\begin{eqnarray}
\Gamma_{\uparrow \downarrow}(E)={\frac{2 \pi}{h}J_{eff}^2n_{\uparrow}^t(E)p_{\downarrow}^{ch}(E)} \nonumber \\
\Gamma_{\downarrow \uparrow}(E)={\frac{2 \pi}{h}J_{eff}^2n_{\downarrow}^{ch}(E)p_{\uparrow}^{t}(E)},
\end{eqnarray}
where
\begin{gather}
p_{\downarrow}^{ch}(E)=D_{\downarrow}(E)-n_{\downarrow}^{ch}(E) \nonumber \\
p_{\uparrow}^{t}(E)=D_{\uparrow}(E)-n_{\uparrow}^{t}(E).
\label{eq:p}
\end{gather}
$D_{\uparrow}(E)$ and $D_{\downarrow}(E)$ are the density of states of the up-spin electrons and down-spin electrons respectively at  the QPC. The Overhauser field at the QPC and the electron spin-flip tunneling is determined by the nuclear polarization which in turn is determined by the history of electron-nuclear spin flip-flop scattering at the QPC. While electron-nuclear  spin flip-flop  scattering can also occur between the channels far away from the QPC terminating in the drain contact, such spin flip-flop scattering hardly causes any change in the output current as well as the Overhauser field at the QPC. We are only interested in spin-flip scattering in the vicinity of the QPC that determines the transmission coefficient 
of the spin-split channels. The parameter $J_{eff}$ takes into account the spatial overlap of the density of states of the up-spin channel and the down-spin channel at the QPC. It is quite different from the parameter $A_{eff}$ used in \eqref{eq:scattering_rates}. 
\begin{equation}
J_{eff}^2=A_{eff}^2 \frac{\int d^3\textbf{r}_ndE D_{\uparrow}(\textbf{r}_n, E)D_{\downarrow}(\textbf{r}_n, E)}{\int d^3\textbf{r}_ndE D_{\uparrow}(\textbf{r}_n, E)\int d^3\textbf{r}_ndED_{\downarrow}(\textbf{r}_n, E)}
\label{eq:J}
\end{equation}
Since we are mainly interested in the spin flip-flop scattering that occur at  the QPC, the range of $d^3\textbf{r}$  includes the region at the QPC which determines the transmission coefficients of the  spin-split edge channels (narrowest region of the QPC).\\

\begin{gather}
I^{sf}(E)=I^{sf}_{\uparrow \downarrow}(E)=qn_{\downarrow}^{ch}(E)v_{\downarrow}^{ch}(E). \nonumber
\end{gather}
Substituting $I^{sf}(E)$ from \eqref{eq:flip1} and $p_{\uparrow}(E)$ and $p_{\downarrow}(E)$ from \eqref{eq:p} and doing some algebraic manipulations, we arrive at the equation,
\begin{eqnarray}
n_{\downarrow}^{ch}(E)v_{\downarrow}^{ch}(E)=2\pi\frac{J_{eff}^2N_I}{h}\Bigg\{n_{\uparrow}^{t}(E)D_{\downarrow}(E) \Big\{F_{\frac{1}{2}}+F_{-\frac{1}{2}} \nonumber \\ 
+F_{-\frac{3}{2}}\Big\} -n_{\downarrow}^{ch}(E)D_{\uparrow}(E)   \times \Big\{F_{\frac{3}{2}} 
+F_{\frac{1}{2}}+F_{-\frac{1}{2}}\Big\}  \nonumber \\ 
+n_{\uparrow}^{t}(E)n_{\downarrow}^{ch}(E)\Big\{F_{\frac{3}{2}}-F_{-\frac{3}{2}}\Big\} \Bigg\} \nonumber 
\end{eqnarray}
\begin{widetext}
\begin{eqnarray}
=2\pi\frac{J_{eff}^2N_I}{h}\Bigg\{n_{\uparrow}^{t}(E)D_{\downarrow}(E) \Big\{1-F_{\frac{3}{2}}\Big\} -n_{\downarrow}^{ch}(E)D_{\uparrow} (E)   \times \Big\{1-F_{-\frac{3}{2}} 
\Big\}+n_{\uparrow}^{t}(E)n_{\downarrow}^{ch}(E)\Big\{F_{\frac{3}{2}}-F_{-\frac{3}{2}}\Big\} \Bigg\} \nonumber 
\end{eqnarray}
%\scriptsize
%\begin{eqnarray}
%=2\pi\frac{J_{eff}^2N_I}{h}\Bigg\{n_{\uparrow}^{t}(E)D_{\downarrow}(E) \Big\{1-F_{\frac{3}{2}}\Big\} -n_{\downarrow}^{ch}(E)D_{\uparrow} (E) \nonumber \\  \times \Big\{1-F_{-\frac{3}{2}} 
%\Big\}+n_{\uparrow}^{t}(E)n_{\downarrow}^{ch}(E)\Big\{F_{\frac{3}{2}}-F_{-\frac{3}{2}}\Big\} \Bigg\} \nonumber \\
%\end{eqnarray}
\begin{gather}
\Rightarrow n_{\downarrow}^{ch}(E)
=\frac{2\pi\frac{J_{eff}^2N_I}{h}n_{\uparrow}^{t}(E)D_{\downarrow} (E)\Big\{1-F_{\frac{3}{2}}\Big\}}{2\pi\frac{J_{eff}^2N_I}{h}\Bigg \{n_{\uparrow}^{ch}(E) \Big\{F_{-\frac{3}{2}}-F_{\frac{3}{2}}\Big\}+D_{\uparrow}(E) \Big\{1-F_{-\frac{3}{2}}\Big\}\Bigg\}+v_{\downarrow}^{ch}(E)}.
\end{gather}
\end{widetext}
\normalsize
If the rate of spin flip-flop scattering at the QPC is slower compared to the rate of transfer of electrons between the up-spin channel originating in the source contact and the up-spin edge channel terminating in the drain contact through the QPC, that is, if $2\pi\frac{J_{eff}^2N_I}{h}D_{\uparrow}(E)<<v_{\downarrow}^{ch}(E)$, the first two factors in the denominator can be neglected.
\begin{equation}
n_{\downarrow}^{ch}(E)=\frac{2\pi\frac{J_{eff}^2N_I}{h}n_{\uparrow}^{t}(E)D_{\downarrow} (E)\Big\{1-F_{\frac{3}{2}}\Big\}}{v_{\downarrow}^{ch}(E)}.
\label{eq:app2}
\end{equation}
We assume that $n_{\uparrow}^{t}(E)$ is independent of the nuclear polarization. So,
\begin{equation}
n_{\uparrow}^{ch}(E)=\frac{T_{\uparrow}(E)}{hv_{\uparrow}^{ch}(E)}.
\label{eq:app1}
\end{equation}
Eliminating $n_{\uparrow}^{ch}(E)$ from \eqref{eq:app1} and \eqref{eq:app2} we get
\begin{align}
n_{\downarrow}^{ch}(E) & =\frac{2\pi\frac{J_{eff}^2N_I}{h}T_{\uparrow}(E)D_{\downarrow} (E)\Big\{1-F_{\frac{3}{2}}\Big\}}{hv_{\downarrow}^{ch}(E)v_{\uparrow}^{ch}(E)} \nonumber \\
T_{\uparrow\downarrow}^{sff} (E) & =h\times n_{\downarrow}^{ch}(E)v_{\downarrow}^{ch}(E) \nonumber \\
&=\frac{2\pi\frac{J_{eff}^2N_I}{h}T_{\uparrow}(E)D_{\downarrow}(E) \Big\{1-F_{\frac{3}{2}}\Big\}}{v_{\uparrow}^{ch}(E)} \nonumber \\
 & = k T_{\uparrow}(E)\{1-F_{\frac{3}{2}}\},  \nonumber \\
\label{eq:app3}
\end{align}
where $k=\frac{2\pi\frac{J_{eff}^2N_I}{h}D_{\downarrow}(E) }{v_{\uparrow}^{ch}(E)}$. Under the assumption of  constant density of states over the range of energy between $\mu_S$ and $\mu_D$, $k$ can be taken as a constant.
Similar derivations can be made to show that 
\begin{equation}
T_{\downarrow\uparrow}^{sfb}(E)=kT_{\downarrow}(E)\{1-F_{-\frac{3}{2}}\}
\label{eq:appref1}
\end{equation}
 for $G>\frac{e^2}{h}$.\\
 \indent We call the parameter $k$ the \emph{spin-flip transmission coefficient}, denoted by $T_{\uparrow \downarrow}^f$ and $T_{\downarrow \uparrow}^f$ ($T_{\uparrow \downarrow}^b$ and $T_{\downarrow \uparrow}^b$) for spin-flip scattering at the QPC from  forward  propagating up-spin  channel terminating in the drain contact to forward propagating down-spin channel terminating in the drain contact and forward propagating down-spin channel originating in the source contact to forward propagating up-spin channel terminating in the drain contact (forward propagating up-spin channel terminating in the drain contact  to backward propagating down-spin channel originating in the drain contact  and forward propagating down-spin channel terminating in the drain contact to backward propagating up-spin channel originating in the drain contact) respectively. The coefficients $T_{\uparrow \downarrow}^f$ and $T_{\downarrow \uparrow}^f$  cause a change in the total charge  current at the drain  since the scattering occurs to a forward propagating channel. The coefficients $T_{\uparrow \downarrow}^b$ and $T_{\downarrow \uparrow}^b$, however, also cause a change in the total output current since the scattering occurs  to backward propagating states.  It should be noted that the above equations are only valid at the QPC in the energy range between $\mu_S$ and $\mu_D$ ($\mu_S>\mu_D$). For $E<\mu_D$, both the forward propagating and backward propagating channels are filled and hence spin-flip scattering cannot occur giving $T_{\uparrow \downarrow}^f$ = $T_{\downarrow \uparrow}^f=T_{\uparrow \downarrow}^b$= $T_{\downarrow \uparrow}^b=0$
\section{Non-Equilibrium Green's function formalism for calculation of current through a potential energy barrier.} \label{appendix4}
We calculate the conductance of the device by  calculating the direct  transmission coefficients $T_{\uparrow}$ and $T_{\downarrow}$ of the system using the non-equilibrium Green's function (NEGF) formalism assuming that the edge states of the Landau levels can be modeled as a quasi $1-D$ ballistic conductor. Such method of modeling has been shown to  accurately match the experimental results \cite{salahuddin,aono}. We model the region of the QPC as a smooth Gaussian energy barrier \citep{qpcmodel},
\[
U_O=150e^{-\frac{(x-x_o)^2}{2\sigma^2}}meV,
\]
  with $\sigma=22nm$ and $x_o$ is the narrowest region of the QPC that determines the transmissivity of the channels.

In case of ballistic transport in nano devices, the generalized equations  for Green's function  and scattering matrices  are  given by the equations:
\begin{gather} 
G_{\uparrow(\downarrow)}(E_x)=[E_xI-H_{\uparrow (\downarrow)}-U-\Sigma_{\uparrow (\downarrow)}(E_z)]^{-1} \nonumber \\
\Sigma_{\uparrow (\downarrow)}(E_x)=\Sigma_{L\uparrow(\downarrow)}(E_x)+\Sigma_{R\uparrow(\downarrow)}(E_x) \nonumber \\
A_{\uparrow(\downarrow)}(E_x)=i[G_{\uparrow (\downarrow)}(E_x)-G_{\uparrow (\downarrow)}^{\dagger}(E_x)] \nonumber \\
\Gamma_{L (R)\uparrow (\downarrow)}(E_x)=[\Sigma_{L (R)\uparrow (\downarrow)}(E_x)-\Sigma_{L (R)\uparrow (\downarrow)}^{\dagger}(E_x)], \nonumber \\
 \label{eq:negf}  
\end{gather}
where $H_{\uparrow (\downarrow)}$ is the discretized device Hamiltonian matrix in $1-D$ constructed using  the effective mass approach \cite{qtransport}.
\[
H_{\uparrow(\downarrow)}=H_0+U_{\uparrow (\downarrow)},
\]
 where $H_0$ is the Hamiltonian matrix in the absence of an externally applied potential at the QPC. $U_{\uparrow (\downarrow)}$ is the minimum potential energy of the electrons for the up-spin (down-spin) channel respectively given by \eqref{eq:overhausereffect}. $U$ is the additional electronic potential energy due to an externally applied voltage and $\Sigma_{L\uparrow (\downarrow)}(E_x)$ and $\Sigma_{R\uparrow (\downarrow)}(E_x)$  describe the   coupling and scattering of electronic wavefunctions  due to left and right contacts respectively for the up-spin (down-spin) channel. In the above sets of equations, $E_x$ is the free variable denoting the electronic energy  along the transport direction. $A_{\uparrow (\downarrow)}(E_x)$ is the  $1-D$  spectral function and $\Gamma_{\uparrow(\downarrow)}(E_x)$ is the broadening matrix at  energy $E_x$ for the up-spin (down-spin) electrons. \\
\indent The minimum potential energy of the up-spin and down-spin channels ($U_{\uparrow}$ and $U_{\downarrow}$) at the QPC are modeled by   potential barriers of the form \cite{Imamoglu}
  \begin{gather}
  U_{\uparrow}=U_O-\underbrace{\frac{g_e \mu_B B}{2}}_{\frac{Zeeman}{field}}+\underbrace{(A_{eff}^{Ga}+A_{eff}^{As})\frac{F_I}{2}}_{\frac{Overhauser}{field}} \nonumber  \\
  U_{\downarrow}=U_O+\overbrace{\frac{g_e \mu_B B}{2}}^{}-\overbrace{(A_{eff}^{Ga}+A_{eff}^{As})\frac{F_I}{2}}, \nonumber \\
  \label{eq:overhausereffect}
\end{gather}
where $A_{eff}^{Ga}$ and $A_{eff}^{As}$ are the effective hyperfine constants for Ga and As respectively and has the respective value of $A_{eff}^{Ga}=42\mu eV$ and $A_{eff}^{As}=46 \mu eV$.  \\
\indent The electron and hole densities per unit length at point $j$ are given by the electron and hole correlation functions.
\[
n_{j\uparrow (\downarrow)}=\int\frac{[G^n_{\uparrow(\downarrow)}(E_x)dE_x]}{2\pi a}
\]
\[
p_{j\uparrow (\downarrow)}=\int\frac{[G^p_{\uparrow(\downarrow)}(E_x)dE_x]}{2\pi a},
\]
where $'a'$ is the lattice constant. In the above equations, $G^n_{\uparrow(\downarrow)}(E_x)$ and $G^p_{\uparrow(\downarrow)}(E_x)$ are the electron and hole correlation functions given by:
\begin{gather}
G^n_{\uparrow (\downarrow)}(E_x)=G_{\uparrow (\downarrow)}(E_x)\Sigma^{in}_{\uparrow (\downarrow)}(E_x)G^{\dagger}_{\uparrow (\downarrow)}(E_x) \nonumber \\
G^p_{\uparrow (\downarrow)}(E_x)=G_{\uparrow (\downarrow)}(E_x)\Sigma^{out}_{\uparrow (\downarrow)}(E_x)G^{\dagger}_{\uparrow (\downarrow)}(E_x), \nonumber \\
 \label{eq:correlation} 
\end{gather}
$ \Sigma^{in}(E_x)$ and $ \Sigma^{out}(E_x)$ are the in-scattering and the  out-scattering functions which model the rate of scattering of the  electrons and holes respectively from the contact to the device.
\begin{gather}
  \Sigma^{in}_{\uparrow (\downarrow)}(E_x)=\Sigma^{in}_{S\uparrow (\downarrow)}(E_x)+\Sigma^{in}_{D\uparrow (\downarrow)}(E_x) \nonumber \\
   \Sigma^{out}_{\uparrow (\downarrow)}(E_x)=\Sigma^{out}_{S\uparrow (\downarrow)}(E_x)+\Sigma^{out}_{D\uparrow (\downarrow)}(E_x), \nonumber \\
   \label{eq:sig}
  \end{gather}
   where the subscript $'S'$ and  $'D'$ denote the influence of source contact and drain contact  respectively on the scattering matrices. The in-scattering and out-scattering functions are dependent on the contact quasi-Fermi distribution functions as:
   \begin{multline}
 \Sigma^{in}(E_x)=\underbrace{\Gamma_S(E_x)f_S(E_x)}_{\frac{left-contact}{inflow}}  
 +\underbrace{\Gamma_D(E_x)f_D(E_x)}_{\frac{right-contact}{inflow}} \nonumber  \\
 \end{multline}
 \begin{multline}
 \Sigma^{out}(E_x)=\underbrace{\Gamma_S(E_x)\Big\{ 1-f_S(E_x)\Big \} }_{\frac{left-contact}{outflow}} \\ +\underbrace{\Gamma_D(E_x)\Big \{ 1-f_D(E_x)\Big \}}_{\frac{right-contact}{outflow}}, \nonumber  \\
 \label{eq:sig1}
 \end{multline}
where $f_{S(D)}$ denote the quasi-Fermi distribution of source(drain) contact. \\
\indent The direct transmission coefficients  are given by:
\begin{equation}
T_{\uparrow (\downarrow)}(E)=Trace\left[ \Gamma_{L\uparrow(\downarrow)}G_{\uparrow(\downarrow)} \Gamma_{R\uparrow(\downarrow)}G^{\dagger}_{\uparrow(\downarrow)}\right].
\end{equation}
The current that flows directly through the QPC from the  up-spin (down-spin) edge channel originating in the source contact to the up-spin (down-spin) edge channel terminating in the drain contact  (without electronic  spin-flip) is given by:
\begin{eqnarray}
I_{\uparrow (\downarrow)}=\frac{q}{h} \int T_{\uparrow (\downarrow)}(E_x)\{ f_L(E_x)-f_R(E_x) \}dE_x. \nonumber \\
\label{eq:currentnegf}
\end{eqnarray}

\section{Derivation of $\left[ \Gamma\right]$} \label{appendix2}

In case, where the electronic spin-flip rate is limited by the supply of electrons from the source, $\Gamma_{\uparrow\downarrow}$ and $\Gamma_{\downarrow\uparrow}$ must be related to $|I^{sf}_{\uparrow\downarrow}|$ and $|I^{sf}_{\downarrow\uparrow}|$ respectively. We show that  $\Gamma_{\uparrow\downarrow}=C_1\left|I^{sf}_{\uparrow\downarrow}\right|$ and $\Gamma_{\downarrow\uparrow}=C_2|I^{sf}_{\downarrow\uparrow}|$ and solve for $C_1$ and $C_2$.  We start our derivation from equation \eqref{eq:flip1}. The total spin-flip current $I^{sf}$ at the QPC can be written as the sum of up-to-down electronic spin-flip current and down-to-up electronic spin-flip current at the QPC. 
\color{black}
\small
\begin{eqnarray}
I^{sf}=\int I^{sf}(E)dE=qN_I \times \Big[-\Gamma_{\downarrow\uparrow}F_{\frac{3}{2}}+\Big \{\Gamma_{\uparrow\downarrow}-\Gamma_{\downarrow\uparrow}\Big \}F_{\frac{1}{2}} \nonumber \\  +\Big \{\Gamma_{\uparrow\downarrow}
-\Gamma_{\downarrow\uparrow}\Big\}F_{-\frac{1}{2}} 
+\Gamma_{\uparrow\downarrow}F_{-\frac{3}{2}}  \Big] \nonumber  \\
=qN_I \Big( \Big[\Gamma_{\uparrow\downarrow}\Big\{ F_{\frac{1}{2}}+ F_{-\frac{1}{2}}+ F_{-\frac{3}{2}}\Big\}  
-  \Gamma_{\downarrow\uparrow}\Big\{ F_{\frac{3}{2}}+ F_{\frac{1}{2}}+ F_{-\frac{1}{2}}\Big\} \Big]\Big) \nonumber
\end{eqnarray}
\normalsize
\begin{equation}
=\overbrace{qN_I \times \Gamma_{\uparrow\downarrow}\Big\{ 1-F_{\frac{3}{2}}\Big\} }^{|I^{sf}_{\uparrow \downarrow}|}
-\overbrace{qN_I \times \Gamma_{\downarrow\uparrow}\Big\{1- F_{-\frac{3}{2}}\Big\} }^{|I^{sf}_{\downarrow\uparrow}|}. \nonumber
\end{equation}
Let us consider the situation where $I^{sf}_{\downarrow\uparrow}=0$. In this case, $\Gamma_{\downarrow\uparrow}=0$.

\begin{align}
I^{sf} &=|I^{sf}_{\uparrow\downarrow}| \nonumber \\ 
 & = qN_I \times \Gamma_{\uparrow\downarrow}\Big\{ 1-F_{\frac{3}{2}}\Big\} 
-qN_I \times \Gamma_{\downarrow\uparrow}\Big\{1- F_{-\frac{3}{2}}\Big\}  \nonumber \\
&=qN_I\Gamma_{\uparrow\downarrow} \Big\{1-F_{\frac{3}{2}}\Big\} \nonumber \\
 \Gamma_{\uparrow\downarrow} &=\frac{|I^{sf}_{\uparrow\downarrow}|}{qN_I\Big\{1-F_{\frac{3}{2}}\Big\}}.
\end{align}
\color{black}

In a similar way, setting $I^{sf}_{\uparrow\downarrow}=0$, we get 
\begin{equation}
\Gamma_{\downarrow\uparrow}=\frac{|I^{sf}_{\downarrow\uparrow}|}{qN_I\Big\{1-F_{-\frac{3}{2}}\Big\}}.
\end{equation}
The transition matrix $\left[\Gamma\right]$ is then given by:

\begin{widetext}
\begin{equation}
\left [ \Gamma \right ] =\begin{bmatrix}
-C_2|I_{\downarrow\uparrow}^{sf}| & C_1|I_{\uparrow\downarrow}^{sf}|  & 0 & 0 \\
C_2|I_{\downarrow\uparrow}^{sf}| & -\left (C_2|I_{\downarrow\uparrow}^{sf}|+ C_1|I_{\uparrow\downarrow}^{sf}| \right ) & C_1|I_{\uparrow\downarrow}^{sf}|&  0 \\
0 & C_2|I_{\downarrow\uparrow}^{sf}| & -\left (C_2|I_{\downarrow\uparrow}^{sf}|+ C_1|I_{\uparrow\downarrow}^{sf}| \right ) & C_1|I_{\uparrow\downarrow}^{sf}|   \\
0 & 0 & C_2|I_{\downarrow\uparrow}^{sf}| & -C_1|I_{\uparrow\downarrow}^{sf}|.
\end{bmatrix},
\label{eq:rewrite}
\end{equation}
\end{widetext}
with $C_1=\frac{1}{qN_I\Big\{1-F_{\frac{3}{2}}\Big\}}$ and $C_2=\frac{1}{qN_I\Big\{1-F_{-\frac{3}{2}}\Big\}}$.

\section{Derivation of $\left[\Gamma_{NMR}\right]$ (Interaction between nuclear spin levels due to an externally applied RF field).} \label{appendix3}
When the nuclear spin levels are perturbed by an externally applied RF field with energy  corresponding approximately to the difference in energy between the nuclear spin levels,  the nuclei  undergoes a periodic oscillatory transition between the consecutive spin states (Rabi oscillations \cite{APL_Edge_Scatt,NMR_book}). However due to inhomogeneities and lack of coherence  in the applied RF field, the periodic  oscillatory transitions decays along with an exponential decay in the nuclear polarization. Such decay time is of the order of $100\mu s$ \cite{APL_Edge_Scatt_ORIG,APL_Edge_Scatt}. We model the temporal evolution of the nuclear polarization without taking into account such periodic oscillatory transition and calculate the occupancy of the nuclear density of states without  bookkeeping of the correlation terms. \\
\indent Let us assume that the applied RF frequency is $\omega$. In steady state, the rate of  transition between the $s^{th}$ and the $(s+1)^{th}$ nuclear spin levels is given by \cite{NMR_book}:

\[
N_s(\xi)P_{s+1}(\xi+\hslash \omega)r_{s\rightarrow s+1}=N_{s+1}(\xi+\hslash \omega)P_s(\xi)r_{s+1\rightarrow s} 
\]
\begin{multline*}
\Rightarrow N_I^2D_s(\xi)F_s(\xi)D_{s+1}(\xi+\hslash \omega)\{ 1-F_{s+1}(\xi+\hslash \omega)\}r_{s\rightarrow s+1}\\=N_I^2D_{s+1}(\xi+\hslash\omega)F_{s+1}(\xi+\hslash\omega)D_s(\xi)\{1-F_s(\xi)\}r_{s+1\rightarrow s}
\end{multline*}

\[
\Rightarrow \frac{F_s(\xi)}{1-F_s(\xi)}\frac{1-F_{s+1}(\xi+\hslash \omega)}{F_{s+1}(\xi+\hslash \omega)}=\frac{r_{s+1\rightarrow s}}{r_{s \rightarrow s+1}}.
\]
In the above equations, $D_s(\xi)$, $N_s(\xi)$,  $P_s(\xi)$  and $F_s(\xi)$ denote the normalized density of states of the nuclear state with spin $'s'$ at energy $\xi$, density of occupied nuclear states  with spin $'s'$ at energy $\xi$, density of vacant nuclear states  with spin $'s'$ at energy $\xi$ and fraction of the density of states which is occupied at energy $\xi$. $r_{s\rightarrow s+1}$ denote the rate at which the nuclei can uundergo a transition from the nuclear state  with  spin $'s'$ to the nuclear state with spin $'s+1'$. In general, $r_{s\rightarrow s+1}=cN$ for stimulated absorption  and $r_{s+1\rightarrow s}=c(N+1)$ for stimulated$+$spontaneous emission. $N$ is the probability of occupancy  of the bosonic density of states for photons with energy $\hslash\omega$ in equilibrium at temperature $T$
 given by $N=\frac{1}{e^{\frac{\hslash \omega}{k_BT}}-1}$ and $c$ is a constant of proportionality. Putting these values in the above equation, we get;
\[
\frac{F_s(\xi)}{1-F_s(\xi)}\frac{1-F_{s+1}(\xi+\hslash \omega)}{F_{s+1}(\xi+\hslash \omega)}=\frac{N+1}{N}=e^{\frac{\hslash \omega}{k_BT}}.
\]
When the nuclear spins are irradiated by an externally applied RF field of sufficient power,  $N+1 \approx N$. Hence, 
\[
F_s(\xi)\approx F_{s+1}(\xi+\hslash\omega).
\]
If the initial occupancy of the $s^{th}$ and $(s+1)^{th}$ levels be denoted by $F_s^0$ and $F_{s+1}^0$, then
\begin{eqnarray}
F_s(\xi)=\frac{D_s(\xi)F_s^0(\xi)+D_{s+1}(\xi+\hslash\omega)F_{s+1}^0(\xi+\hslash\omega)}{D_s(\xi)+D_{s+1}(\xi+\hslash\omega)}\nonumber \\=F_{s+1}(\xi+\hslash\omega). \nonumber \\
\end{eqnarray}
We assume that under the influence of the perturbing RF frequency the decay of the difference in occupancy between the nuclear spin levels occurs with a time constant $\tau_{NMR}$. Hence, the rate of decay of nuclear spin polarization due to RF frequency perturbation for the case discussed above is given by:
\begin{widetext}
\begin{equation}
\small
\begin{split}
\frac{d}{dt}\{D_s(\xi)F_s(\xi)\}  =\frac{1}{\tau_{NMR}}\frac{\pi \eta}{2}\Big \{ &D_{s+1}(\xi+\hslash \omega)D_s(\xi) F_{s+1}(\xi+\hslash \omega)\{1-F_s(\xi)\}-D_{s+1}(\xi+\hslash \omega)D_s(\xi)\{1-F_{s+1}(\xi+\hslash \omega)\}F_s(\xi)\Big\} \\
&=\frac{1}{\tau_{NMR}}\frac{\pi \eta}{2}\Big\{ D_{s+1}(\xi+\hslash \omega)D_s(\xi)\{ F_{s+1}(\xi+\hslash \omega)-F_{s}(\xi)\}\Big\} \\
\Rightarrow \frac{d}{dt}&\{F_s(\xi)\}  = \frac{1}{\tau_{NMR}}\frac{\pi \eta}{2}\Big\{ D_{s+1}(\xi+\hslash \omega)\{ F_{s+1}(\xi+\hslash \omega)-F_{s}(\xi)\}\Big\}
\end{split}
\end{equation}
Similarly,
\[
\frac{dF_{s+1}(\xi)}{dt}=\frac{1}{\tau_{NMR}}\frac{\pi \eta}{2}\Big \{{D_{s}(\xi-\hslash\omega)\{F_{s}(\xi-\hslash\omega)-F_{s+1}(\xi)}\}\Big \}.
\]

In the above equations, the factor $\frac{\pi \eta}{2}$ acts as a normalization constant, $\eta$ being the broadening of the nuclear density of states.  For a nuclear system with more than two spin levels, the nuclear energy level can interact directly with the levels which  energetically lie immediately above and below the level, that is, a nuclear spin level with spin $'s'$ can interact directly with the levels with spins $'s+1'$ and $'s-1'$. The equation for time evolution occupancy  of the $s^{th}$ nuclear spin density of states in such a case is given by:

\small
\begin{align}
\frac{dF_s(\xi)}{dt}=\frac{1}{\tau_{NMR}}\frac{\pi \eta}{2}\Big \{{D_{s+1}(\xi+\hslash\omega)\{F_{s+1}(\xi+\hslash\omega)-F_s(\xi)\}+D_{s-1}(\xi-\hslash\omega)\{F_{s-1}(\xi-\hslash\omega)-F_s(\xi)}\Big \}.
\end{align}
\normalsize
 For a system with four nuclear spin levels, the time evolution of the occupancy of the nuclear spin density of states   under the influence of an externally applied RF field (neglecting correlation) is hence given by:
\[
\left [ \frac{dF(\xi)}{dt} \right ]_{NMR} =diag \left(\frac{1}{\tau_{NMR}} \frac{\pi \eta}{2}\times [\Gamma_{NMR}(\xi) ]\times [F_{NMR}(\xi,\hslash \omega)]\right),
\]
where
\scriptsize
\begin{eqnarray}
\Gamma_{NMR}(\xi)=\begin{bmatrix}
-{D_{\frac{1}{2}}(\xi-\hslash \omega)} & {D_{\frac{1}{2}}(\xi-\hslash \omega)}  & 0 & 0 \\
{D_{\frac{3}{2}}(\xi+\hslash \omega)} & -\{D_{\frac{3}{2}}(\xi+\hslash\omega)+D_{-\frac{1}{2}}(\xi-\hslash \omega)\} & {D_{-\frac{1}{2}}(\xi-\hslash \omega)} &  0 \\
0 & {D_{\frac{1}{2}}(\xi+\hslash\omega)} & -\{D_{\frac{1}{2}}(\xi+\hslash \omega)+D_{-\frac{3}{2}}(\xi-\hslash\omega)\} & {D_{-\frac{3}{2}}(\xi-\hslash\omega)}  \\
0 & 0 & {D_{-\frac{1}{2}}(\xi+\hslash\omega)} & -{D_{-\frac{1}{2}}(\xi+\hslash\omega)}
\end{bmatrix} \nonumber \\
\end{eqnarray}
\begin{eqnarray}
F_{NMR}(\xi,\hslash\omega)=
\begin{bmatrix}
F_{\frac{3}{2}}(\xi) & F_{\frac{3}{2}}(\xi+\hslash \omega) & 0 & 0 \\
F_{\frac{1}{2}}(\xi-\hslash \omega) & F_{\frac{1}{2}}(\xi) & F_{\frac{1}{2}}(\xi+\hslash \omega) &  0 \\
0 &  F_{-\frac{1}{2}}(\xi-\hslash \omega) & F_{-\frac{1}{2}}(\xi) & F_{-\frac{1}{2}}(\xi+\hslash \omega) \\
0 & 0 & F_{-\frac{3}{2}}(\xi-\hslash \omega) & F_{-\frac{3}{2}}(\xi)
\end{bmatrix} \nonumber \\
\end{eqnarray}
 
\end{widetext}

\bibliography{Hyperfine_QH_REF}% Produces the bibliography via BibTeX.
\end{document}